\documentclass[aps,prd,nopacs,nofootinbib,notitlepage]{revtex4-1}

\usepackage{amsfonts}
\usepackage{amsmath}
\usepackage{mathtools}
\usepackage{xcolor}
\usepackage{bm}
\usepackage{graphicx}
\usepackage{fancybox}
\usepackage{comment}
\usepackage{multirow}
\usepackage{tipa}
\definecolor{refs}{RGB}{245,156,74}
\usepackage[colorlinks=true,hyperfootnotes=true,citecolor=cyan]{hyperref}
\usepackage{enumitem}
\usepackage{comment}
\usepackage{subcaption}
\usepackage{esint}

\newcommand{\be}{\begin{equation}}
\newcommand{\ee}{\end{equation}}
\newcommand{\ba}{\begin{eqnarray}}
\newcommand{\ea}{\end{eqnarray}}
\newcommand{\bs}{\begin{subequations}}
\newcommand{\es}{\end{subequations}}

\newcommand{\bbe}{\boldsymbol{\mathrm{e}}}

\newcommand{\bdelta}{\boldsymbol{\delta}}

\newcommand{\bomega}{\boldsymbol{\omega}}

\newcommand{\bA}{\boldsymbol{A}}

\newcommand{\bF}{\boldsymbol{F}}

\newcommand{\bH}{\boldsymbol{H}}
\newcommand{\bh}{\boldsymbol{h}}

\newcommand{\bk}{\boldsymbol{k}}

\newcommand{\bL}{\boldsymbol{L}}
\newcommand{\bM}{\boldsymbol{M}}
\newcommand{\bO}{\boldsymbol{O}}

\newcommand{\bR}{\boldsymbol{R}}
\newcommand{\bT}{\boldsymbol{T}}
\newcommand{\bt}{\boldsymbol{t}}
\newcommand{\bu}{\boldsymbol{u}}

\newcommand{\bX}{\boldsymbol{X}}
\newcommand{\bx}{\boldsymbol{x}}
\newcommand{\bY}{\boldsymbol{Y}}

\newcommand{\bdiff}{\boldsymbol{\mathrm{d}}}
\newcommand{\bDiff}{\boldsymbol{\mathrm{D}}}

\newcommand{\lp}{\left(}
\newcommand{\rp}{\right)}
\newcommand{\lb}{\left[}
\newcommand{\rb}{\right]}

\newcommand{\nn}{\nonumber}

\newcommand{\0}{0$^\text{th}$}
\newcommand{\1}{1$^\text{st}$}
\newcommand{\2}{2$^\text{nd}$}

\newcommand{\+}{ \prescript{+}{}}
\newcommand{\viff}{\bdiff\tilde{\varphi}}

\begin{document}

\title{Black holes in Lorentz gauge theory}

\date{\today}

\author{Tomi S. Koivisto}
\email{tomi.koivisto@ut.ee}
\address{Laboratory of Theoretical Physics, Institute of Physics, University of Tartu, W. Ostwaldi 1, 50411 Tartu, Estonia}
\address{National Institute of Chemical Physics and Biophysics, R\"avala pst. 10, 10143 Tallinn, Estonia}
\author{Luxi Zheng}
\email{luxy.zheng@ut.ee}
\address{Laboratory of Theoretical Physics, Institute of Physics, University of Tartu, W. Ostwaldi 1, 50411 Tartu, Estonia}

\begin{abstract}

Black hole solutions are explored in the Lorentz gauge theory of gravity. The fields of the theory are the gauge potential in the adjoint and a scalar in the fundamental representation of the Lorentz group, a metric tensor then emerging as a composite field in a symmetry-broken phase. Three distinct such phases of the theory are considered.  
In SO(3) phase, the fundamental field is identified with a generalised Painlev{\'e}-Gullstrand-Lema{\^i}tre coordinate time. In the static spherically symmetric case it is a stealth scalar, and the general vacuum solution is then parameterised by two constants, one related to the black hole mass and the other to an observer. Also, formulations of pregeometric first order electromagnetism are considered in order to construct a consistent realisation of a charged black hole. In SO(1,2) phase of the theory, the Schwarzschild solution is a configuration wherein the fundamental field is real outside and imaginary inside the horizon. In this phase the field can be associated with an effective radial pressure resulting in additional singularities and asymptotic non-flatness. Finally, a symmetry-broken phase which would correspond to solutions in an alternative attempt at a Lorentz gauge theory is shown to be incompatible with black holes. 
   
\end{abstract}

\maketitle

\tableofcontents

\section{Introduction}

General Relativity (GR) describes gravity in terms of a dynamical metric, and its remarkable prediction is that the metric can become singular in a black hole \cite{taylor}.
The description is spectacularly succesful, as highlighted by the recent detection of gravitational 
waves and the rapidly improving understanding of the astrophysical black holes in our universe \cite{Sedda:2024duh}. However, the acceleration of the universe and some tensions in the cosmological data have also motivated studies into possible modifications of the metric dynamics \cite{DiValentino:2024wgi}, as well as into alternative geometrical descriptions of gravity in terms of different fields \cite{CANTATA:2021ktz}. 
In fact, a long-standing foundational problem is, why there exists a metric at all. GR does not allow a non-degenerate metric, and the lack of a zero-metric  ``ground state'' has often been pointed out as the reason for various theoretical problems with GR, in particular, its non-renormalisability as a quantum theory. Arguably, this is the strongest theoretical motivation for developing alternative geometrical viewpoints on gravity \cite{Krasnov:2020lku}. 

Notably, Pleba{\'n}ski's formulation of GR put forward a chiral description of spacetime \cite{Plebanski:1977zz}.  Based on the reducibility of the Lorentz group, the vacuum Einstein equations are encoded solely into one chiral sector. The dynamical fields of the theory are a triple of 2-forms plus the chiral connection. No metric is introduced by hand, but by imposing certain conditions on the 2-forms they can be interpreted as pairs of wedged coframes that give rise to a metric. This elegant formulation underlies some of the prominent current approaches to quantum gravity \cite{Engle:2023qsu}. It is even possible to further eliminate the triple of 2-forms in favour of the chiral Lorentz connection, resulting in a pure SU(2)-connection formulation of GR \cite{Krasnov:2011pp}. However, the coupling of matter to gravity is not straightforward in these formulations, though it is understood that matter fields are the source of the other chiral sector of the curvature \cite{Capovilla:1991qb}. 

A chiral description of gravity can also be arrived at via gauging the global Lorentz invariance in a special-relativistic parameterised field theory \cite{Koivisto:2022uvd}. In the resulting {\it Lorentz gauge theory of gravity}, the field parameterising the coordinates of the Minkowski space in Special Relativity can be identified as a canonical clock field, as first proposed by Z\l{}o\'snik {\it et al} \cite{Zlosnik:2018qvg}. In this way, a possible solution to the problems of time is in-built into the Lorentz gauge theory. The coframe for a composite metric is simply the Lorentz-covariant derivative of the clock field. Thus, the only fields in the theory are the Lorentz connection and the clock field. Instead of an additional triple of 2-forms as in Pleba{\'n}ski's original formulation, the metric is thus constructed from the degrees of freedom in the anti-self dual half of the connection, which is already implied by construction. There is precisely the required 3$\times$4 degrees of freedom in the SU(2) connection to span a generic Bartels frame (aka spatial Dreibein), and it is not necessary to impose constraints with Lagrange multipliers. Spinor matter and Yang-Mills gauge fields in a first order formulation are naturally incorporated into the theory \cite{Gallagher:2023ghl}. 

This motivates further investigations of the Lorentz gauge theory. A Hamiltonian analysis of the generalised class of theories including arbitrary linear combinations of the self-dual and antiself-dual curvatures in the action
was carried out recently, and it was found that only the chiral action (involving only either one of the twin curvatures) is singled out as the one resulting in an extension to GR whereas, quite interestingly, the action for the total curvature describes a topological theory in the sense that it possesses no local degrees of freedom \cite{Nikjoo:2023flm}. The latter topological version could be more tractable as a quantum theory, and its chiral projection then yields the desired Lorentz gauge theory. This might be exploited in the eventual construction of the quantum gravity theory\footnote{In perhaps some analogy to the spin foam approach wherein a topological BF theory is deformed by the so called simplicity constraints that basically impose the Pleba{\'n}ski's triple of 2-forms to a wedged pair of coframes  \cite{Engle:2023qsu}. A different, dynamical perspective will be proposed at (\ref{dymaxion}).}. It is at least superficially a promising feature of the Lorentz gauge theory that the action can be written in a ``curvature-squared'' form in a rather direct analogy to the usual Yang-Mills theories describing the Standard Model interactions. 

However, it can be prudent to begin with a more detailed study of the classical solutions of the theory. This far, there are only some preliminary considerations pertaining to the cosmology of the Lorentz gauge theory \cite{Gallagher:2021tgx,Koivisto:2023epd}. The cosmological solutions illustrate the constitution of spacetime as described above: the clock field, dubbed the {\it khronon}, can be identified with the cosmic time in a given coordinate system, and in the limit of Minkowski space, the Lorentz gauge field strength remains non-vanishing in one chiral sector that supports the Bartels frame for space. The Riemann curvature of the metric, which becomes non-zero for expanding cosmological solutions, is encoded in the Lorentz gauge field strength of the opposite chirality. There can be effective energy density associated with the khronon field, which behaves as rotationless dust. This ``dust of time'' is a special case of Brown-Kuchar dust, which specifies a material frame and can thus provide a convenient standard of time for (canonical quantum) gravity \cite{Brown:1994py,Husain:2011tm,Giesel:2012rb}. Such an energy component can appear\footnote{In the reverse way, a similar dust degree of freedom can result from restoring the space-time symmetry into a non-covariant theory \cite{Magueijo:2024zxz},
or simply by varying a non-covariant action without time-reparametrisation invariance \cite{Kaplan:2023wyw},  see also \cite{Casadio:2024bdb}.} (as a limiting case) in various models wherein Lorentz symmetry is broken spontaneously by additional fields in the gravitational sector \cite{Jacobson:2000xp, Arkani-Hamed:2003pdi,Blas:2009qj,Lim:2010yk,Blanchet:2024mvy}. 

In this paper, we explore static spherically symmetric solutions in the Lorentz gauge theory. As we restrict to static solutions, they do not allow the addition of the dust energy, which can be seen as a special case of the no-go theorem derived in \cite{Izumi:2009ry}. In Section \ref{theorysec} we review the action formulation and field equations of the theory, highlighting its property as the chiral projection of a topological gauge theory. 
The spherically symmetric spacetime geometry is constructed in Section \ref{sphericalsec}, and it is found  
that the khronon field can be identified as the time coordinate of the Lema{\^i}tre coordinates, and that the emergent geometry obeys the Birkhoff-Jebsen theorem. Next, the solution is generalised to the electromagnetic vacuum, for which purpose we have to couple the Lorentz gauge theory with a pregeometric version of Maxwell's theory. In Section \ref{electrosec} we discuss alternative versions and construct explicitly the electrically and magnetically charged Reissner-Nordstr\"om solutions. 
A completely different class of black hole solutions is taken into consideration in Section \ref{radialsec}, where the symmetry-breaking field, taking space-like values, could be called the {\it radion}. In this case, new solutions exist though with singular features. In Section \ref{alternativesec} we comment on yet one more class of possible solutions and their relevance to a proposal for a different SO(1,3) theory of gravity in the literature. Section \ref{conclusionsec} is the conclusion. In the Appendix \ref{extrasym} we supplement the introductory discussion below and in the Appendix \ref{geodesics} we briefly review known results for coordinate systems adapted to generic geodesics.  

\section{Lorentz gauge theory}
\label{theorysec}

The fields of the theory are a dimensionless scalar $\phi^a$ in the fundamental representation of the Lorentz group\footnote{Perhaps more accurately, the $\phi^a$ represents the Lorentz torsor, which is realised via the shift symmetry discussed below.}, and the canonical connection 1-form $\bomega^a{}_b$ in the adjoint representation. The connection defines the covariant exterior derivative  $\bDiff$ and its curvature is given as $\bR^a{}_b = \bdiff\bomega^a{}_b + \bomega^a{}_c\wedge\bomega^c{}_b$. We thus work on an SO(1,3) bundle over a 4-dimensional manifold $M$. 

\subsection{The action}
\label{action}

The action is to be posited as an integral of a polynomial 4-form functional of the fields. In terms of the connection as the only field, the most general action then includes only two parameters,
\bs
\be \label{tqft}
I_{(0)} = \int_M\lp g_1\eta_{ab}\eta^{cd} + g_2\epsilon_{ab}{}^{cd}\rp \bR^a{}_c\wedge\bR^b{}_d\,.  
\ee  
The symbols $\eta_{ab}$ and $\epsilon_{ab}{}^{cd}$ denote the two invariants of the Lorentz algebra, and our sign conventions are $\eta_{ab}=\eta^{ab}=\text{diag}(-,+,,+,+)$, and $\epsilon_{01}{}^{23}=-\epsilon^{01}{}_{23}=1$. The action (\ref{tqft}) represents a topological field theory which contains no local degrees of freedom in the bulk of $M$, as is readily seen by performing a partial integration using the Bianchi identity $\bDiff\bDiff\bDiff = 0$ and the Stokes theorem, 
\be
I_{(0)} = \oint_{\partial M}\lp g_1\eta_{ab}\eta^{cd} + g_2\epsilon_{ab}{}^{cd}\rp\bomega^{a}{}_{c}\wedge\bR^b{}_d\,.  
\ee
\es
Topological field theories, which can be interpreted as integral representations of topological invariants (in this case, Donaldson polynomials) are of great interest due to their mathematical properties \cite{Birmingham:1991ty}.  As quantum field theories containing a finite number of degrees of freedom with simple dynamics, they may even be solved exactly in the perturbative regime. It seems therefore an appealing approach to quantum gravity to consider that GR could emerge from a topological field theory by some symmetry breaking unleashing the bulk degrees of freedom \cite{Mielke:2017nwt}. Current attempts towards cosmological applications include \cite{Agrawal:2020xek,Kehagias:2021smx}. An interesting recent study indeed considers the theory (\ref{tqft}), positing a symmetry breaking onto an $s$-exact boundary added to (\ref{tqft}) in the BRST formalism\footnote{In Ref. \cite{Sadovski:2024uhg}, see also \cite{Junqueira:2016hlu}, a doublet of Lorentz-adjoint 2-forms plus a metric is introduced in the construction of the boundary and the coframe is then inserted by an explicit symmetry breaking. From the Lorentz gauge theory perspective, a more minimal scheme could be very interesting (involving the single scalar $\phi^a$, presumably a component of a BRST doublet).} \cite{Sadovski:2024uhg}.

Once we take into account the $\phi^a$, it appears that the number of possible polynomial actions becomes infinite, as one could construct arbitrary polynomials functions of the argument $\phi^2 = \eta_{ab}\phi^a\phi^b$. However, we require the additional global shift symmetry of the action. This means that the theory cares only about changes (naturally, in Lorentz-covariant terms) in the field $\phi^a$, whereas the absolute value of $\phi^a$ is unphysical. Thus, a shift in the field $\phi^a \rightarrow \phi^a + \xi^a$, by any constant $\bDiff\xi^a=0$ should be irrelevant. Then there are but three independent 4-forms that we may add to (\ref{tqft}). In addition to the quartic 4-form (here and in the following indices will be raised and lowered with $\eta_{ab}$) 
\be \label{Ilambda}
I_{(4)} = \lambda\int\epsilon_{abcd}\bDiff\phi^a\wedge\bDiff\phi^b\wedge\bDiff\phi^c\wedge\bDiff\phi^d\,,   
\ee  
we can only write down the two-parameter theory
\be \label{Iphi}
I_{(2)} = g_3\int\bDiff\phi^a\wedge\bDiff\phi^b\wedge\bR_{ab}  + g_4\int\epsilon_{abcd}\bDiff\phi^a\wedge\bDiff\phi^b\wedge\bR^{cd}\,. 
\ee 
If the shift symmetry is not required to be exact, but the action is allowed to be changed by a boundary term, then the three further terms can be taken into account,
\bs
\ba
\tilde{I}_{(2)} & = & \int\lp \tilde{g}_3\phi_a\phi_c\eta^{bd} + \tilde{g}_4\phi^e\phi_a\epsilon_{ce}{}^{bd}\rp\bR^a{}_b\wedge\bR^c{}_d\,, \label{tildeI2} \\
\tilde{I}_{(4)} & = & 3\tilde{\lambda}\int\epsilon_{abcd}\bR^a{}_e\phi^e\phi^b\wedge\bDiff\phi^c\wedge\bDiff\phi^d\,. 
\ea
\es
Note that these are equivalent to the previous terms up to boundary terms, i.e. we have $\tilde{I}_{(2)} \overset{b}{=} I_{(2)}$ and  $\tilde{I}_{(4)} \overset{b}{=} I_{(4)}$ if we remove the tildes from the couplings constants.  Comparison of the expression (\ref{tildeI2}) with the point of departure (\ref{tqft})  shows that the dynamics of the Lorentz gauge theory, encoded in (\ref{Iphi}), can emerge from the topological field theory by a deformation of the inner product.  

Further, the Hamiltonian analysis in Ref.\cite{Nikjoo:2023flm} concluded that the theory (\ref{Iphi}) is devoid of local degrees of freedom when $g_3=0$, and the reason was traced to the enchanced symmetry
\bs
\ba
\phi^a & \rightarrow & \phi^a + \xi^a\,, \quad \text{where} \quad \bDiff\xi^a = 0\,, \\
\omega^{ab} & \rightarrow & \xi^{ab}\bdiff\phi^2\,, \quad \text{where} \quad \xi^{ab}=-\xi^{ba}\,. \label{extra}
\ea
\es
We verify the presence of the extra symmetry (\ref{extra}) in Appendix \ref{extrasym}, and check that in fact it applies to the general action $I=I_{(0)} + I_{(2)} + I_{(4)}$ (obviously one could add $+\tilde{I}_{(2)} + \tilde{I}_{(4)}$) if one sets $g_3=0$. The desired extension of GR is obtained by choosing the $g_3$ and $g_4$ to correspond to the maximally chiral projection of the curvature, 
\bs
\label{actions}
\be
I_+  =  \int \bDiff\phi^a\wedge\bDiff\phi^b\wedge\+\bR_{ab}\,, \label{primeplus}
\ee
or alternatively,
\be
\tilde{I}_+  =  \int \+\bR^b{}_a\wedge\bR_{cb} \phi^a\phi^c\,.
\ee
\es  
However, in Ref.\cite{Gallagher:2023ghl} it was found that the $I_+$ rather than the $\tilde{I}_+$ predicts the unique correct results for physical charges, and here we have noted that this can be related to the exactness of the shift symmetry. 

The chiral projection explicitly reads
\be \label{projection}
\+\bR^a{}_b = \frac{1}{2}\lp \delta^a_c\delta^d_b - \frac{i}{2}\epsilon^{ad}{}_{bc}\rp\bR^{c}{}_d\,,
\ee  
which shows that the formulation of the theory requires the complexification of the Lorentz group. Whilst this may require some extra technical care in some cases, for the purposes of the present article we can be non-chalant about some factors of $i$. 
In fact, we've have in mind that fundamentally, the theory is rather based on the Euclidean group, in which case the $i$ does not appear in projections such as (\ref{projection}), but instead one imaginary field is required to recover the emergent pseudo-Riemannian geometry, and this is naturally the khronon i.e. the ``Higgs scalar'' in the symmetry-broken phase $\phi^a \sim \tau\delta^a_0$. Recently, it has been argued that an analytic continuation from Euclidean spacetime is a way to render Standard Model quantum field theory well-defined in Minkowski space, and a chiral realisation was suggested that hints at relations of the symmetries of the Standard Model to the geometry of a Euclidean twistor space \cite{Woit:2021bmb}. Interesting perspectives to gravity are known from twistor geometry that also become available only in the Euclidean signature \cite{Krasnov:2020lku}. However, here we shall not pursue those lines and will stick to notation in the conventional signature.

To conclude our discussion, the theory $I=I_{(0)}+I_{(2)}+I_{(4)}$ has five dimensionless coupling constants and seems to lack any local dynamics when $g_3=0$. 
Setting $g_4=-i/4$ and promoting the constant $g_3=a/2$ into an interpolating parameter\footnote{In the context of loop quantum gravity, the coefficient of the Holst term analogous to the first term in (\ref{dymaxion}) is the celebrated Barbero-Immirzi parameter \cite{BarberoG:1994eia,Holst:1995pc}. Here the roles of the two terms are in some sense switched, as usually the Holst term does not contribute to the dynamics (due to absence of torsion). Since $a$ is the coefficient of the dual curvature in some analogy with the QCD axion, it could be called the gravaxion (if not the dymaxion due to all the dynamics hinging on nonvanishing $a$).}, the quadratic piece (\ref{Iphi}) reads  
\be \label{dymaxion}
I_{(2)} = \frac{1}{2}\int \bDiff\phi_a\wedge\bDiff\phi^b\wedge\lp a\delta^a_c\delta^d_b - \frac{i}{2}\epsilon^{ad}{}_{bc}\rp\bR^{c}{}_d\,. 
\ee
It would seem natural to consider $a$ as a field, whose expectation value is $a=0$ in the topological phase, and $a= 1$ in the dynamical (extended chiral GR) phase, $I_{(2)} = I_+$. It could be a hint about the precise mechanism for the primordial emergence of a vacuum that the Minkowski spacetime in this theory is a nontrivial instanton in the topological $a=0$ phase, but has identically zero instanton number in the dynamical $a=1$ phase. For the rest of this article, we work in the latter phase and set $a=1$. 

\subsection{The dynamics}

Let $\bbe^a \equiv \bDiff\phi^a$. This short-hand notation makes transparent that the covariant derivative of the scalar will eventually become the coframe, the ``soldering form'' that maps objects from the SO(1,3) fibres to the tangent space of the manifold $M$. We should clarify that in fact we should consider $m_P\bbe^a = \bDiff\phi^a$, introducing this way the Planck mass $m_P$ in order to obtain a dimensionless coframe. However, in the dynamical equations the gravitational coupling $\sim 1/m_P^2$ can be absorbed into the definition of the source currents, and we do this in order to ease the notation throughout the article. 

The field equations of theory are obtained by varying either of the two actions (\ref{actions}) wrt to the fields $\phi^a$ and the $\bomega^{ab}$. We take into account that these fields might be sourced by matter fields, and parameterise the sources by the 3-forms $\bt^a$ and $\bO^{ab}$, respectively. We obtain   
\bs
\ba
\bDiff\lp\+{}\bR^a{}_b\wedge\bbe^b - \bt^a\rp & = & 0\,, \label{first} \\
\frac{1}{2}\bDiff\+{}\lp\bbe^{[a}\wedge\bbe^{b]}\rp & = & \phi^{[a}\+{}\bR^{b]} + \bO^{ab}\,. \label{second}
\ea
\es
By formally integrating the first equation (\ref{first}) and inserting the solution into the second equation (\ref{second}), we can rewrite the above as  
\bs
\ba
\+{}\bR^a{}_b\wedge\bbe^b & = & \bt^a + \bM^a \quad \text{where} \quad \bDiff\bM^a = 0\,, \label{field1} \\
\frac{1}{2}\bDiff\+{}\lp\bbe^{[a}\wedge\bbe^{b]}\rp & = & \phi^{[a}\bM^{b]} + \bO^{ab}\,.  \label{field2}
\ea
\es
The first equation (\ref{field1}) is apparently identical to the Einstein field equation in the self-dual first order formulation, in the presence of an additional source term $\bM^a$, 
\be
-\frac{i}{2}\epsilon_{abcd}\+\bR^{bc}\wedge\bbe^b =  \bt^a + \bM^a\,.
\ee  
This is because the self-dual curvature, by definition (\ref{projection}), satisfies 
\be
\star \+\bR^{ab} \equiv \frac{1}{2}\epsilon_{abcd}\+\bR^{cd} = i\+\bR^{ab}\,.
\ee
However, besides the additional source term the structure of the theory is also modified by the property that the $\bbe^a \equiv \bDiff\phi^a$ depends as well upon the Lorentz connection.

To clarify this structure, it is useful to fix the gauge $\phi^a = \phi\delta^a_0$. Then the {\it khronon} field $\phi$ gives a canonical 4-velocity and in terms of the adapted time coordinate $\tau=\phi$ one can simply regard $\bbe^0 = \bdiff \phi = \bdiff \tau$. 
The Bartels coframe is constructed as $\bbe^I = \bomega^I{}_0\phi$, where here and in the following we will denote the spatial indices with upper case Latin letters $I$, $J$, $K$. Omitting here the details of the derivations, one can deduce that  \cite{Koivisto:2023epd}
\begin{itemize}
\item Minimal coupling $\bO^{ab}=0$ and (\ref{field2}) results in $\bM^a = \bM\delta^a_0$.
\item Conservation of the spatial components $\bDiff \bM^I = 0$ results in $\bM = M\star\bbe^0$.
\end{itemize}
So, we obtain the set of field equations
\bs
\label{fieldequations}
\ba
\+{}\bR^0{}_I\wedge\bbe^I & = & \bt^0 + M\star\bbe^0 \quad \text{where} \quad \bdiff\lp M\star\bbe^0\rp = 0\,, \label{Field1}  \\ 
\+{}\bR^I{}_a\wedge\bbe^a & = & \bt^I\,, \label{Field2} \\
\bDiff\+{}\lp\bbe^{[a}\wedge\bbe^{b]}\rp & = & 0\,. \label{Field3}
 \ea
 \es
 Solving some connection coefficients from (\ref{Field3}) and plugging into (\ref{Field1},\ref{Field2}) gives us the Einstein equations, effectively sourced by a pure energy component specified by a single function $M$. This
 kind of energy field, which is constant in a given comoving volume, can be identified as rotationless dust \cite{Brown:1994py,Husain:2011tm,Giesel:2012rb}. This is perhaps the simplest instance of a dust component, similar to which have been introduced in GR for the sole purpose of providing a reference frame and thus a useful phenomenological approach to the problems of time encountered in canonical quantum gravity. In the Lorentz gauge theory, the existence of the canonical reference frame is deduced from the structure of the composite metric geometry emerging in a symmetry-broken phase. The metric line element of GR is now constructed as $\bdiff s^2 = \bdiff \phi^2 +\delta_{IJ}\bbe^I\bbe^J$. 
 
 In the rest of this article we study the architecture of a static and spherically symmetric spacetime as built from $\phi^a$ and $\omega^{ab}$.  In Section \ref{radialsec} we will consider also an alternative symmetry breaking where the spacelike $\phi^a\phi_a>0$ plays the role of a {\it radion} field $\phi^a$, and in Section \ref{alternativesec} briefly comment on yet one more alternative prescription.


\section{The SO(3) synchronous phase}
\label{sphericalsec}

A static spherically symmetric metric is most often considered in the coordinates
\bs
\label{schwarz}
\be \label{sline}
\bdiff s^2 = -f^2(r)\bdiff t^2 + g^2(r)\bdiff r^2 + r^2\lp \bdiff \theta^2 + \sin^2{\theta}\bdiff \varphi^2\rp\,,
\ee  
corresponding to the diagonal (co)tetrad
\be \label{stetrad}
\bbe^a = \delta^a_0 f\bdiff t + \delta^a_1 g\bdiff r  + \delta^a_2 r\bdiff \theta + \delta^a_3 r\sin\theta \bdiff \varphi\,.  
\ee
\es
Looking for a solution to $\bDiff\tau^a = \bbe^a$, one quickly finds it inconvenient, at least in the synchronous symmetry-breaking phase $\tau^I=0$. Besides, the Schwarzschild coordinates are singular already at the horizon. From this perspective, Lema{\^i}tre coordinates, for which $-f^2(r)\bdiff t^2 + g^2(r)\bdiff r^2 = -\bdiff \tau^2 + F^2(\rho,\tau) \bdiff \rho^2 $ with a suitable definition $\tau$ and $\rho$ for some function $F(\rho)$, would appear to be natural. This form of the metric would be associated with the connection described by the Levi-Civita 1-form (in this article we denote quantities in the metric geometry with a ring)  
\ba \label{metricomega}
\mathring{\bomega}^0{}_1 & = & \dot{F}\bdiff\rho\,, \quad 
\mathring{\bomega}^0{}_2  =  \dot{r}\bdiff\theta\,, \quad
\mathring{\bomega}^0{}_3  =  \dot{r}\sin\theta\bdiff\varphi\,, \nn \\ 
\mathring{\bomega}^1{}_2 & = & -\frac{r'}{F}\bdiff\theta\,, \quad 
\mathring{\bomega}^1{}_3  =  -\frac{{r}'}{F}\sin\theta\bdiff\varphi\,, \quad
\mathring{\bomega}^2{}_3  =  -\cos\theta\bdiff\varphi\,,
\ea
from which one may compute the Riemann curvature
\bs
\label{metricriemann}
\ba
\mathring{\bR}^0{}_1 & = & \ddot{F}\bdiff\tau\wedge\bdiff\rho\,, \\
\mathring{\bR}^0{}_2 & = & \ddot{r}\bdiff\tau\wedge\bdiff\theta + \lp \dot{r}'-\dot{F}r'\rp \bdiff\rho\wedge\bdiff\theta\,, \\
\mathring{\bR}^0{}_3 & = & \ddot{r}\sin\theta\bdiff\tau\wedge\bdiff\varphi + \lp \dot{r}'-\dot{F}r'\rp \sin\theta\bdiff\rho\wedge\bdiff\varphi\,, \\
\mathring{\bR}^1{}_2 & = & \lp\frac{\dot{F}r'}{F^2} - \frac{\dot{r}'}{F}\rp\bdiff\tau\wedge\bdiff\theta + \lp \frac{F'r'}{F^2} - \frac{r''}{F} + \dot{F}\dot{r}\rp\bdiff\rho\wedge\bdiff\theta\,, \\
\mathring{\bR}^1{}_3 & = & \lp\frac{\dot{F}r'}{F^2} - \frac{\dot{r}'}{F}\rp\sin\theta\bdiff\tau\wedge\bdiff\varphi + \lp \frac{F'r'}{F^2} - \frac{r''}{F} + \dot{F}\dot{r}\rp\sin\theta\bdiff\rho\wedge\bdiff\varphi\,.
\ea
\es
However, we cannot assume that the solution for the gauge potential $\bomega^{a}{}_b$ in the Lorentz gauge theory would coincide with the metric connection 
$\mathring{\bomega}^a{}_b$ in (\ref{metricomega}). To find solutions of the Lorentz gauge theory, we will assume the synchronous symmetry-breaking phase of the Lorentz group SO(1,3)/SO(3) such that the field $\phi^a$ (called the khronon field $\tau^a$ in this and the following Section) can be identified with the time coordinate $\tau$, and then use the field equations (\ref{fieldequations}) to solve for the Lorentz gauge potential $\bomega^a{}_b$.  

Historically, both P. Painlev{\'e}  and A. Gullstrand had discovered spherically symmetric non-diagonal solutions to A. Einstein's theory in the beginning of 1920's terms of the coordinates $(\tau,r,\theta,\varphi)$, though not recognising the relation of their solutions to K. Schwarzschild's earlier solution by  a coordinate transformation which was not clarified until 1933 in G. Lema{\^i}tre's article (which considered embedding the Schwarzschild solution in a de Sitter universe and featured metrics in the diagonal form (\ref{lemaitre}))  \cite{Unruh:2014aka}. 

\subsection{The symmetry breaking}
\label{synchronous}

The SO(1,3) phase is defined as $\phi^2<0$. The gauge choice $\tau^a = \tau\delta^a_0$ is sometimes called the ``time gauge'', but here we refer to it as the synchronous symmetry breaking. We also avoid the ambiguous term ``frame'' in this context. 

To arrive at the convenient coordinates $(\tau,\rho)$, for which $-f^2(r)\bdiff t^2 + g^2(r)\bdiff r^2 = -\bdiff \tau^2 + F^2(\rho,\tau) \bdiff \rho^2$,
we begin with an Ansatz introducing the functions $h(r)$ and $k(r)$ s.t.
\bs
\label{cchange}
\ba
\bdiff \tau & = & \bdiff t + h(r)\bdiff r\,, \\
\bdiff \rho & = & \bdiff t + k(r)\bdiff r\,.
\ea
\es
The solutions are\footnote{The two branches of solutions are adapted to outward and inward travelling radial geodesics, respectively. Considering a black hole, one can see that null geodesics either originate or terminate at the horizon, respectively, but together the two branches cover the region $r>0$, and maximal extension can be considered in terms of $\tau_\pm$ \cite{Unruh:2014aka,Lemos:2020qxk}.}
\be \label{solution}
h  = \pm\frac{Fg}{f}\,, \quad k =  \pm\frac{g}{Ff}\,, \quad \text{with} \quad F = \sqrt{1-f^2}\,.
\ee 
Thus, the line element (\ref{sline}) is written in Lema{\^i}tre coordinates as
\bs
\be \label{lemaitre}
\bdiff s^2 = -\bdiff \tau^2 +\lb 1 - f^2(r(\tau,\rho))\rb \bdiff \rho^2 + r^2(\tau,\rho)\lp \bdiff \theta^2 + \sin^2{\theta}\bdiff \varphi^2\rp\,,
\ee
corresponding to the diagonal (co)tetrad
\be
\bbe^a = \delta^a_0 \bdiff \tau + \delta^a_1 F\bdiff \rho  + \delta^a_2 r\bdiff \theta + \delta^a_3 r\sin\theta \bdiff \varphi\,.  
\ee
\es
We then easily obtain the electric components of the connection in the phase $\tau^a = \tau\delta^a_0$,
\be \label{electric}
\bomega^1{}_0 = \frac{F}{\tau}\bdiff\rho\,, \quad 
\bomega^2{}_0 = \frac{r}{\tau}\bdiff\theta\,, \quad 
\bomega^3{}_0 = \frac{r\sin\theta}{\tau}\bdiff\varphi\,.
\ee
The time component of torsion $\bT^0 = \bDiff\bDiff \phi^0$ vanishes identically, as it should in the synchronous phase. The notation for $\tau$ as the time coordinate and the \0 component of the khronon in the SO(3) phase is justified. A price to pay is, as indicated in (\ref{lemaitre}), that we have to express the Schwarzschild radius $r$ in the Lema{\^i}the coordinates $r=r(\tau,\rho)$. 

Let us consider this in more details. The relations (\ref{cchange}) together with the solution (\ref{solution}) lead to the differential equation,
\be \label{rhotau}
\frac{\pm \bdiff r}{\sqrt{1-f^2}} = \bdiff \rho - \bdiff \tau\,.
\ee
This can be solved given the metric function $f$. There exists an analytic solution for the Schwarzschild-Nordstr\"om-de Sitter case, but dropping the de Sitter part the expression
remains unwieldy enough as will be seen shortly. For explicit expressions, we therefore specialise to the case
\bs
\label{example}
\be \label{s-nr}
f^2(r) = 1 - \frac{m_S}{4\pi m_P^2 r} + \frac{q^2}{8\pi m_P^2 r^2}\,,
\ee
with the mass parameters $m_P$, $m_S$ and the charge parameter $q$.
Integrating (\ref{rhotau}) we obtain
\be \label{rhophi}
\pm (\rho - \tau) = \frac{2\sqrt{2\pi}m_P}{3m_S^2}\lp q^2 + m_S r\rp\sqrt{2m_S r - q^2} + \text{constant}\,.
\ee
We note that the Lema{\^i}tre metric becomes singular $F=0$ at $r=q^2/2m_S$, but this is always beyond not only the event but also the 
Cauchy horizon. We will return to this point in Section \ref{RNsolution}.  
Let us now invert (\ref{rhophi}) as a function of $z=\pm (\rho-\tau - \text{constant})$. We obtain 
\ba \label{rnsolu}
r(z)  & = &  \frac{1}{4m_P} Z  - \frac{q^2}{2m_S} + \frac{q^4 m_P}{m_S^2} Z^{-1}\,,
\ea
where we used the shorthand by introducing a dimensionless function $Z(z)$ where
\be
Z^3(z) = \frac{2m_P}{m_S}\lb  9 m_S^2 z^2 +  3z \sqrt{9m_S^4 z^2 + 8\pi m_P q^6 }  + 4\pi\lp\frac{m_P}{m_S}\rp^2  q^6 \rb\,. 
\ee
\es
This calculation shows that there exists an analytic solution in the SO(3) phase. This solution becomes more tractable in the charge-free limit 
\be
r \quad \overset{q \rightarrow 0}{=} \quad \lp \frac{3z}{2}  \rp^{\frac{2}{3}}\lp \frac{m_S}{4\pi m_P^2}\rp^{\frac{1}{3}}\,.
\ee
The example (\ref{example}) will be the relevant solution in the next Section wherein we will couple electromagnetism to gravity.  

Next, we continue the construction of a generic static spherically symmetric geometry in terms of the Lorentz gauge potential. 
A generic the spherically symmetric spin connection can be parameterised in terms of 12 functions of 2 variables \cite{Hohmann:2019fvf},
\bs
\label{educated}
\ba
\bomega^0{}_1 & = & S_3\bdiff \tau + S_4\bdiff\rho\,, \\
\bomega^0{}_2 & = & S_9\bdiff\theta - S_{15}\sin\theta\bdiff\varphi\,, \\
\bomega^0{}_3 & = & S_{15}\bdiff\theta + S_9\sin\theta\bdiff\varphi\,, \\
\bomega^1{}_2 & = & S_{10}\bdiff \theta + S_{16}\sin\theta\bdiff\varphi\,, \\
\bomega^1{}_3 & = & -S_{16}\bdiff\theta + S_{10}\sin\theta\bdiff\varphi\,, \\ 
\bomega^2{}_3 & = & -S_{17}\bdiff \tau  - S_{18}\bdiff\rho - \cos\theta\bdiff\varphi\,.
\ea
\es
We've already deduced from (\ref{electric}) the coefficients $S_3=S_{15}=0$, $S_4=F/\tau$ and $S_9=r/\tau$. There remain the 4 coefficients which we rename as $S_{10}=A$, $S_{16}=iB$, $S_{17}=iC$ and $S_{18}=iD$. So,
\bs
\label{Tparam}
\ba
\bomega^0{}_1 & = & \frac{F}{\tau}\bdiff\rho\,, \quad 
\bomega^0{}_2 = \frac{r}{\tau}\bdiff\theta\,, \quad 
\bomega^0{}_3 = \frac{r\sin\theta}{\tau}\bdiff\varphi\,, \\
\bomega^1{}_2 & = &  A\bdiff \theta + iB\sin\theta\bdiff\varphi\,, \quad
\bomega^1{}_3  =  -iB\bdiff\theta + A\sin\theta\bdiff\varphi\,, \quad
\bomega^2{}_3  =  -iC\bdiff \tau  - iD\bdiff\rho - \cos\theta\bdiff\varphi\,.
\ea
\es 
The boost field strength of this connection is
\bs
\label{boost}
\ba
\bR^0{}_1 & = & \lp \frac{\dot{F}}{\tau} - \frac{F}{\tau^2}\rp\bdiff\tau\wedge\bdiff\rho - 2\frac{iB r}{\tau}\sin\theta\bdiff\theta\wedge\bdiff\varphi\,, \\
\bR^0{}_2 & = & \lp \frac{\dot{r}}{\tau}- \frac{r}{\tau^2}\rp\bdiff\tau\wedge\bdiff\theta - \frac{iCr}{\tau}\sin\theta\bdiff\tau\wedge\bdiff\varphi 
+\frac{1}{\tau}\lp r' + {FA}\rp\bdiff\rho\wedge\bdiff\theta + \frac{i}{\tau}\lp FB - rD\rp\sin\theta\bdiff\rho\wedge\bdiff\varphi\,, \\
\bR^0{}_3 & = & \frac{irC}{\tau}\bdiff\tau\wedge\bdiff\theta +\lp \frac{\dot{r}}{\tau}- \frac{r}{\tau^2}\rp\sin\theta\bdiff\tau\wedge\bdiff\varphi
+ \frac{i}{\tau}\lp rD -FB\rp \bdiff\rho\wedge\bdiff\theta +  \frac{1}{\tau}\lp r' + FA\rp\sin\theta\bdiff\rho\wedge\bdiff\varphi\,,
\ea
and the rotation field strength is 
\ba
\bR^1{}_2 & = & \lp \dot{A} - BC\rp\bdiff\tau\wedge\bdiff\theta + i\lp \dot{B} - AC\rp\sin\theta\bdiff\tau\wedge\bdiff\varphi \nn \\
                     & + &    \lp A' - BD + \frac{Fr}{\tau^2}\rp\bdiff\rho\wedge\bdiff\theta 
                     +  i\lp B' - AD\rp\sin\theta\bdiff\rho\wedge\bdiff\varphi -iB\cos\theta\bdiff\theta\wedge\bdiff\varphi\,, \\
\bR^1{}_3 & = & -i\lp \dot{B} - AC\rp \bdiff\tau\wedge\bdiff\theta + \lp \dot{A} -BC\rp\sin\theta \bdiff\tau\wedge\bdiff\varphi \nn \\
                  & - & i\lp B' - AD \rp\bdiff\rho\wedge\bdiff\theta
                  + \lp A' - BD + \frac{Fr}{\tau^2}\rp\sin\theta\bdiff\rho\wedge\bdiff\varphi\,, \\ 
\bR^2{}_3 & = & i\lp {C}' - \dot{D}\rp \bdiff\tau\wedge\bdiff\rho + \lp \frac{r^2}{\tau^2} - A^2 + B^2 + 1\rp\sin\theta\bdiff\theta\wedge\bdiff\varphi\,.    
\ea
\es
Already at the level of kinematics, we see that the Lorentz gauge curvature (\ref{boost}) is different from the metric Riemann 2-form (\ref{metricriemann}).  
Next, we consider the dynamics of the Lorentz gauge and then check the vacuum solutions for the 4 remaining functions.

\subsection{The field equations}

In the synchronous phase, it is convenient to consider the system in terms of the two SO(3)-valued 2-forms, the torsion $\bT^I$ and the self-dual field strength $\bF^I$. These satisfy the relations
\bs
\ba
\bT^I & = & \bDiff\bbe^I = \bDiff\bDiff \tau^I = \bR^{0I}\tau\,, \label{bartelstorsion} \\
\bF^I & = & \bdiff\bA^I - \epsilon^I{}_{JK}\bA^J\wedge\bA^K = -i\+\bR^{0I} = -\epsilon^I{}_{JK}\+\bR^{JK}\,.
\ea
\es
In terms of these 2-forms, we can rewrite field equations (\ref{fieldequations}) in the form
\bs
\ba
\bF_I\wedge\bbe^I & = & M\star\bbe^0 + \bt^0\,, \label{teom1} \\
\bF^I\wedge\bbe^0 +  i\epsilon^I{}_{JK}\bF^J\wedge\bbe^K& = & - \bt^I\,, \label{teom2} \\
\bT^I\wedge\bbe^0 - i\epsilon^I{}_{JK}\bT^J\wedge\bbe^K & = & 2i\bO^{0I}\,. \label{ceom}
\ea
\es
This form of the system is completely general. 

When specialising to spherical symmetry, the self-dual gauge potential assumes the form
\bs
\ba
\bA^1 & = & - \frac{i}{2}C\bdiff\tau-\frac{i}{2}\lp \frac{F}{\tau} + D\rp\bdiff\rho -\frac{1}{2}\cos\theta\bdiff\varphi\,, \\
\bA^2 & = & - \frac{i}{2}\lp \frac{r}{\tau}- B\rp \bdiff\theta -\frac{1}{2}A\sin\theta \bdiff\varphi  \,, \\
\bA^3 & = &  \frac{1}{2}A\bdiff\theta - \frac{i}{2}\lp \frac{r}{\tau}- B\rp\sin\theta\bdiff\varphi\,,
\ea
\es
which is associated with a field strength whose components are given as
\bs
\label{selfdual}
\ba
\bF^1 & = & -\frac{i}{2}\lp \frac{\dot{F}}{\tau} - \frac{F}{\tau^2} - C' + \dot{D}\rp \bdiff\tau\wedge\bdiff\rho + 
\frac{1}{2}\lb 1 - A^2 + \lp \frac{r}{\tau} - B \rp^2 \rb\sin\theta\bdiff\theta\wedge\bdiff\varphi\,, \\
\bF^2 & = & -\frac{i}{2}\lp \frac{\dot{r}}{\tau} - \frac{r}{\tau^2} + AC - \dot{B}\rp\bdiff\tau\wedge\bdiff\theta
- \frac{1}{2}\lb \dot{A} + \lp\frac{r}{\tau} -B\rp C\rb\sin\theta\bdiff\tau\wedge\bdiff\varphi \nn \\
& - & \frac{i}{2}\lb \frac{r'}{\tau} - B' + A\lp\frac{F}{\tau}+D\rp\rb\bdiff\rho\wedge\bdiff\theta 
- \frac{1}{2}\lb A' + \lp \frac{r}{\tau} - B\rp\lp\frac{F}{\tau}+D\rp\rb\sin\theta\bdiff\rho\wedge\bdiff\varphi\,, \\
\bF^3 & = & \frac{1}{2}\lb \dot{A} + \lp \frac{r}{\tau} -B\rp C\rb\bdiff\tau\wedge\bdiff\theta 
- \frac{i}{2}\lp \frac{\dot{r}}{\tau} - \frac{r}{\tau^2} + AC - \dot{B} \rp\sin\theta\bdiff\tau\wedge\bdiff\varphi \nn \\
& + & \frac{1}{2}\lb A' + \lp\frac{r}{\tau}-B\rp\lp\frac{F}{\tau}+D\rp\rb\bdiff\rho\wedge\bdiff\theta
- \frac{i}{2}\lb \frac{1}{\tau}\lp r' + FA\rp - B' + AD\rb\sin\theta\bdiff\rho\wedge\bdiff\varphi\,.  
\ea
\es
We note that there is a solution for a teleparallel connection,
\be \label{teleparallel}
A=\pm 1, \quad B=\frac{r}{\tau}\,, \quad C=0\,, \quad D=-\frac{F}{\tau}  \quad \Rightarrow \quad \bF^I=0\,.
\ee
The torsion of the Bartels frame we can read directly from the boost components of the Lorentz field strength (\ref{boost}) as seen from the relation (\ref{bartelstorsion}). 

Having determined the 2-forms $\bT^I$ and $\bF^I$, we can consider the EoM's.  
The 3 components of the connection EoM (\ref{ceom}) are
\bs
\ba
-2i\lb \lp \dot{r} - \frac{r}{\tau} + B \rp \bdiff\tau + \lp r' + FA\rp \bdiff\rho\rb\wedge  r \sin\theta\bdiff\theta\wedge\bdiff\varphi & = & i\tau \bM^1 + 2i\bO^{01}\,, \\
\lb \lp r' + FA - rFC\rp\bdiff\theta + i\lp r\dot{F} + \dot{r}F - 2F\frac{r}{\tau} + FB - rD\rp \sin\theta\bdiff\varphi\rb \wedge\bdiff\tau\wedge\bdiff\rho & = & 
 i\tau \bM^2 + 2i\bO^{02}\,, \\
 \lb-i\lp r\dot{F} + \dot{r}F - 2\frac{r}{\tau}F + FB - rD\rp\bdiff\theta + \lp r' + FA - rFC\rp\sin\theta\bdiff\varphi\rb\wedge\bdiff\tau\wedge\bdiff\rho
 & = &  i\tau \bM^3 + 2i\bO^{03}\,.
\ea
\es 
If there are no matter fields with spin currents, the RHS sources are zero. Then we find the solution
\be \label{ABCDsolution}
A = - \frac{r'}{F}\,, \quad B = \frac{r}{\tau} - \dot{r}\,, \quad C = 0\,, \quad D = \dot{F} - \frac{F}{\tau}\,.
\ee
Thus, in the absence of spin currents of matter, the spherically symmetric solutions are determined by the one free function $F$. It has to be solved from the equations sourced by the energy-momentum of matter. 
The energy equation (\ref{teom1}) is
\be \label{hamilton}
\frac{1}{2}\lb 1-A^2 + \frac{2rA'}{F} +  \lp \frac{r}{\tau} - B\rp \lp \frac{3r}{\tau} - B + \frac{2rD}{F}\rp\rb \star\bbe^0
+ r\lb \dot{A} + \lp \frac{r}{\tau} - B\rp C\rb \star \bbe^1 =  -r^2\lp \bM\star\bbe^0 + \bt^0\rp\,, \\
\ee
and the 3 momentum equations (\ref{teom2}) are
\bs
\label{momentum}
\ba
\frac{r}{2}\lb \frac{r'}{\tau} - B' + A\lp\frac{F}{\tau}+D\rp\rb  \star\bbe^0 & + & \nn \\  \frac{1}{2}\lb 1 - A^2 + \frac{2r\dot{r}}{\tau} - \frac{2r^2}{\tau^2} + \lp \frac{r}{\tau}-B\rp^2 
- 2 rAC - 2r\dot{B}\rb \star\bbe^1 & = & r^2\bt^1\,, \label{mom1} \\
-\frac{1}{2}\lb \dot{F}\frac{r}{\tau} + A' + \lp\frac{r}{\tau}-B\rp D - rC' + r\dot{D} + F\lp \frac{\dot{r}}{\tau} - \frac{r}{\tau^2} - AC - \dot{B} - \frac{B}{\tau}\rp\rb\star\bbe^2 & + &  \nn \\
\frac{i}{2}\lb \frac{r'}{\tau} + AD - B' + F\lp \dot{A} + \frac{A}{\tau} - BC + \frac{rC}{\tau}\rp\rb\star\bbe^3 
& = & r F\bt^2\,, \label{mom2} \\
-\frac{i}{2}\lb \frac{r'}{\tau} + AD - B' + F\lp \dot{A} + \frac{A}{\tau} - BC + \frac{rC}{\tau}\rp\rb\star\bbe^2 & - & \nn \\
\frac{1}{2}\lb \dot{F}\frac{r}{\tau} + A' + \lp\frac{r}{\tau}-B\rp D - rC' + r\dot{D} + F\lp \frac{\dot{r}}{\tau} - \frac{r}{\tau^2} - AC - \dot{B} - \frac{B}{\tau}\rp\rb\star\bbe^3 & = & r F\bt^3\,. \label{mom3}
\ea
\es
For generality, we have not implemented the solution (\ref{ABCDsolution}) in the these field equations since (\ref{ABCDsolution}) assumed matter without spin. However, for any spherically symmetric configuration of source matter fields, the angular components of the field equations, (\ref{mom2}) and (\ref{mom3}), are equivalent.  

\subsection{Vacuum solution}

It is not difficult to find the solutions in the absence of both geometrical $\bM^a=0$ and material $\bt^a=\bO^{ab}=0$ sources. 
The coefficients of the spin connection in this case were determined at (\ref{ABCDsolution}). Then any 1 of the 4 equations (\ref{hamilton},\ref{momentum}) 
suffices the determine the solution. Consider e.g. the energy constraint (\ref{hamilton}). Its 2 independent components now reduce to
\bs
\ba
1-\lp \frac{r'}{F}\rp^2 - \frac{2r}{F} \lp \frac{r'}{F}\rp' + \dot{r}\lp \dot{r} + \frac{2r\dot{F}}{F} \rp & = & 0\,, \label{eka}  \\
\lp \frac{r'}{F}\rp' & = & 0\,. \label{toka}
\ea
\es
The \2 equation (\ref{toka}) sets the connection coefficient $A=A_0$ to a constant and shows that $r' = -A_0F$. Plugging this into the \1 equation (\ref{eka}) yields the \1 order differential equation
\be \label{eka2}
1-A_0^2 + A_0^2 F(r)\lp F(r) + 2r\frac{\partial F(r)}{\partial r}\rp = 0\,,
\ee
whose general solution is
\be \label{vacuumsolution}
F(r) = \pm \sqrt{\frac{m_S}{4\pi m_P^2 r} + 1 - A_0^{-2} }\,,
\ee
where we absorbed the integration constant into $m_S/4\pi m_P^2$. It is then straightforward to show that this solution is consistent with the rest of the field equations, i.e. all
the 4 independent 3-form components appearing in the (\ref{momentum}) are zero. 3 of them are identically zero regardless of the form of $F$, given only (\ref{ABCDsolution}) and $r'=-A_0F$, but one of components gives the \2 order differential equation $r\ddot{F}+\dot{r}\dot{F}+\ddot{r}F=0$. This is equivalent to 
\be \label{kolkki}
F(r)\frac{\partial^2 F(r)}{\partial r^2} + \lp\frac{\partial F(r)}{\partial r}\rp^2 + 2\frac{F(r)}{r}\frac{\partial F(r)}{\partial r} = 0\,, 
\ee
whose solution has 2 integration constants which are fixed by matching with (\ref{vacuumsolution}). At this point, we note that the system is not compatible with a non-zero integration constant $M \neq 0$. Such would introduce a non-vanishing RHS in the equation (\ref{eka}) and thus modify the solution (\ref{vacuumsolution}), but then it would no longer satisfy the equation (\ref{kolkki}). This is a manifestation of a no-go theorem for rotationless dust density in static spherically symmetric configuration in GR (which can be circumvented in special electrically counterpoised cases \cite{PhysRev.72.390,Acena:2023yly}) that has been observed, e.g. in the contexts of Ho{\v r}ava-Lifshitz gravity \cite{Izumi:2009ry} and mimetic matter \cite{Gorji:2020ten}.  

Choosing $A_0=\pm 1$ we recover the Schwarzschild solution. The naive guess for the metric
of the generic solution,
\bs
\label{Alines}
\be \label{Aline}
\bdiff s^2 \overset{?}{=} -\lp \frac{1}{A_0^2} - \frac{m_S}{4\pi m_P^2 r}\rp\bdiff t^2 + \lp\frac{1}{A_0^2} - \frac{m_S}{4\pi m_P^2 r}\rp^{-1}\bdiff r^2 + r^2\lp \bdiff \theta^2 + \sin^2{\theta}\bdiff \varphi^2\rp\,,
\ee  
is {\it not} gauge-equivalent to the usual case, as confirmed by checking that the curvature scalar of the above metric $\mathring{R}\overset{?}{=}2(1-A_0^2)/r^2$ is not independent of $A_0$. 
Rescaling the coordinates, $\tilde{t} = -A_0^{-1}t$, $\tilde{r}=-A_0r $, and the mass parameter $\tilde{m}_S=-A_0m_S$, we obtain the line element with the conformal rescaling of the angular part, 
\be \label{Aline2}
\bdiff s^2  \overset{?}{=} -\lp 1 - \frac{\tilde{m}_S}{4\pi m_P^2 \tilde{r}}\rp\bdiff \tilde{t}^2 + \lp 1 - \frac{\tilde{m}_S}{4\pi m_P^2 \tilde{r}}\rp^{-1}\bdiff \tilde{r}^2 + \frac{\tilde{r}^2}{A_0^2}\lp \bdiff \theta^2 + \sin^2{\theta}\bdiff \varphi^2\rp\,,
\ee  
\es
which shows that the spatial slices are curved. We cannot rescale the angular parameterisation to obtain the standard Schwarzschild form. This would seem to invalidate the Birkhoff-Jebsen theorem in the synchronous phase of Lorentz gauge theory. However, the line element (\ref{Alines}) does {\it not} correctly describe the solution in the $(t,r,\theta,\varphi)$ coordinates. 
One can check that the curvature scalar of the Lema{\^i}tre line element vanishes identically $\mathring{R}=0$ regardless of $A_0$ for any (\ref{vacuumsolution}). The resolution is that the relation of the Schwarzschild and Lema{\^i}tre charts we established in Section \ref{synchronous} needs to be reconsidered for generic $A_0$. The revised transformation then brings the Lema{\^i}tre coordinates for any $A_0$ to the standard Schwarzschild form, reconciling the curvature scalars $\mathring{R}=0$ in both coordinates. Thus the Birkhoff-Jebsen theorem is reinstated in the synchronous phase of Lorentz gauge theory. Nevertheless, the parameter $A_0$ of the spin connection coefficient turns out to have a physical significance. 

To clarify all this, recall that the solution to the radial geodesic equation in the Schwarzschild spacetime is given by the 4-velocity
\be \label{observer1}
\lp \frac{\bdiff t}{\bdiff\tau},  \frac{\bdiff r}{\bdiff\tau},  \frac{\bdiff \theta}{\bdiff\tau},  \frac{\bdiff \varphi}{\bdiff\tau}\rp = \lp \frac{e}{f^2}, \sqrt{e^2 - f^2}, 0, 0\rp\,,
\ee 
where $e$ is a constant number characterising the initial state of a freely-falling particle infinitely far from the black hole. The particle is at rest for $e=1$ and has a non-zero velocity towards the black hole if $e>1$; the case $e<1$ one may interpret as the particle being initially at rest at some finite $r$. The derivation and interpretation of time-like geodesics is reviewed in more detail in the Appendix \ref{geodesics}. 
The perspective of (\ref{observer1}) suggests the generalisation of the relations (\ref{cchange}) in the more general spherically symmetric case,
\bs
\ba
\bdiff \tau & = & e\bdiff t + h(r) \bdiff r\,, \\
\bdiff \rho & = & \bdiff t + e k(r) \bdiff r\,.  
\ea
\es
The solution (\ref{solution}) is then generalised to 
\be \label{solution2}
h  = \pm\frac{Fg}{f}\,, \quad k =  \pm\frac{g}{Ff}\,, \quad \text{with} \quad F = \sqrt{e^2-f^2}\,.
\ee
Thus, we obtain the generalised Lema{\^ i}tre coordinates adapted to freely-falling observers with arbitrary initial conditions. 
Indeed, in the new coordinates, the 4-velocity (\ref{observer1}) reads simply 
\be \label{observer2}
\lp \frac{\bdiff \tau}{\bdiff\tau},  \frac{\bdiff \rho}{\bdiff\tau},  \frac{\bdiff \theta}{\bdiff\tau},  \frac{\bdiff \varphi}{\bdiff\tau}\rp = \lp 1, 0, 0, 0\rp\,.
\ee 
These solutions emerged naturally from the Lorentz gauge theory wherein the spin connection coefficient $A_0$ is determined as $A_0=e$ for the family of observers whose velocity is characterised by the constant $e$. Now we should clarify the minor technical detail that the relation of the two constants seems different when comparing (\ref{vacuumsolution}) and (\ref{solution2}): this is just because for $e \neq 1$ one should take into account the generalisation of (\ref{rhotau}) as $\bdiff r = \pm F(e\bdiff\rho - \bdiff\tau)$ which results in the slight modification of (\ref{eka2}) and its solution (\ref{vacuumsolution}):
\be
1-A_0^2 + F(r)\lp F(r) + 2r\frac{\partial F(r)}{\partial r}\rp = 0 \quad \Rightarrow \quad 
F(r) = \pm \sqrt{\frac{m_S}{4\pi m_P^2 r} - 1 + A_0^2} \equiv \pm \sqrt{e^2 - f^2}\,, \quad \text{where } e=A_0\,.
\ee  
Thus, the two integration constants $A_0$ and $m_S$ both have a physical interpretation. 

Already the case $m_S=0$ is interesting, as it shows that the limit of Minkowski space requires a non-trivial choice of $A_0^2 \neq 1$. As $m_S \rightarrow 0$, the function $F \rightarrow  \sqrt{A_0^2-1}$. Thus, for $m_S=0$ the Lema{\^ i}tre from of the metric is degenerate in the standard case $e^2=A_0^2=1$. The relation to the regular Minkowski limit of the Schwarzschild metric breaks down for $F=0$ since for a static observer $\bdiff r=0$. Instead, let us consider the (pre)geometry of the gauge fields in the case that $m_S=0$, $A^2_0 > 1$. We obtain a slight generalisation of the solution (\ref{teleparallel}) s.t. $A=A_0$, $B=r/\tau \mp \sqrt{A_0^2-1}$, $C=0$, $D = -\sqrt{A_0^2-1}$, which still corresponds to vanishing self-dual field strength $\bF^I=0$. However, the torsion of the Bartels frame is non-trivial,
\bs
\ba
\bT^1 & = & -\frac{\sqrt{A_0^2-1}}{\tau}\bdiff\tau\wedge\bdiff\rho - 2i\lp\frac{r}{\tau} \mp  \sqrt{A_0^2-1}\rp\sin\theta\bdiff\theta\wedge\bdiff\varphi\,, \\
\bT^2 & = & \lp\pm \sqrt{A_0^2-1} - \frac{r}{\tau}\rp\bdiff\tau\wedge\bdiff\theta + i \lp 2\sqrt{A_0^2-1}\frac{r}{\tau} \pm 1 \mp A_0^2\rp\sin\theta\bdiff\rho\wedge\bdiff\varphi\,, \\
\bT^3 & = & \lp\pm \sqrt{A_0^2-1} - \frac{r}{\tau}\rp\sin\theta\bdiff\tau\wedge\bdiff\varphi - \lp 2\sqrt{A_0^2-1}\frac{r}{\tau} \pm 1 \mp A_0^2\rp\bdiff\rho\wedge\bdiff\theta\,.
\ea
\es
Along the geodesics, we can set the spin connection coefficient $B=0$ and the above expressions simplify as half of their terms will vanish. The case of $A_0=\pm \sqrt{2}$ corresponds to the standard normalisation of the radial coordinate, i.e. $\bdiff s^2 = -\bdiff\tau^2 + \bdiff\rho^2 + r^2\bdiff\theta^2 + r^2\sin^2\theta\bdiff\varphi^2$. The torsion is regular for $A_0^2=1$, as expected since this choice brings about merely a coordinate issue in the relation of more conventional charts to the (generalised) Lema{\^ i}tre chart.    

Now we return from the Minkowski limit to the black hole spacetime and finally solve for the fundamental field $\tau$.  
Integrating from (\ref{cchange}) in the case $A_0 = 1$, we find an explicit expression for the khronon in the standard Schwarzschild coordinates,
\bs
\label{tausolutions}
\be \label{tausolution}
A_0=1: \quad \tau(r,t) = t - t_0 \pm \sqrt{2}\lp \sqrt{r_Sr} - r_S\tanh^{-1}{\sqrt{\frac{r}{r_S}}}\rp\,, 
\ee
where $r_S=m_S/(4\pi m_P^2)$. 
The divergence of the expression at the horizon reflects the coordinate singularity of the Schwarzschild coordinates. Following a geodesic, the coordinate $t$ dives to (minus) infinity at $r_S$, cancelling the divergence of $\tanh^{-1}{\sqrt{\frac{r}{r_S}}}$ as $\tau$ evolves smoothly. We note that the expression holds for arbitrary branches of the 
function $\tanh^{-1}(x)$ and by default is chosen to be real though the symmetry of the theory allows arbitrary shifts of the constant $t_0$. Choosing e.g. the principal branch would amount to the imaginary recalibration $t_0 \rightarrow t_0  \mp i \pi r_S/\sqrt{2}$ above the horizon. The same holds for the generic solutions, which have somewhat more complicated expressions. In the case $0<A_0<1$ we obtain
\be
0<A_0<1: \quad \tau(r,t) = t - t_0 \pm \lb r F(r) - 2A_0 r_S\tanh^{-1}\lp\frac{F(r)}{A_0}\rp + \frac{1- 2A_0^2}{\sqrt{1-A_0^2}} r_S\tanh^{-1}\lp\frac{F(r)}{\sqrt{1-A_0^2}}\rp\rb\,.   
\ee 
Note that the case $A_0<1$ can only be straightforwardly extended to cover the manifold for $r<r_S/(1-A_0^2)$, since beyond that $F(r)$ becomes imaginary, as does the square brackets in the expression above. Physically meaningful analytic extensions might be suggested by the gauge theory of the complexified Lorentz group, as they might for the negative-energy case $A_0<0$, but we will not pursue this direction in the present article. The case $A_0>1$ covers all $r>0$. Then the khronon field is expressed in the Schwarzschild coordinates as
\be
A_0>1: \quad \tau(r,t) = t - t_0 \pm \lb rF(r) - 2A_0 r_S \tanh^{-1}\lp\frac{A_0}{F(r)}\rp + \frac{2A_0^2-1}{\sqrt{A_0^2-1}}r_S\log{\lp2\sqrt{\frac{r}{r_S}}\lp \sqrt{A_0^2-1} - F(r)\rp\rp}\rb\,.  
\ee
\es
The solution (\ref{tausolutions}) can be interpreted as a {\it stealth} scalar field solution, for which the field profile is determined by the equations of motion but wherein the field itself does not contribute an energy-momentum source to gravity, e.g. \cite{Maeda:2012tu,Sotiriou:2015pka,Gorji:2020ten}.

\begin{figure}[t!]
\includegraphics[width=0.5\linewidth,keepaspectratio]{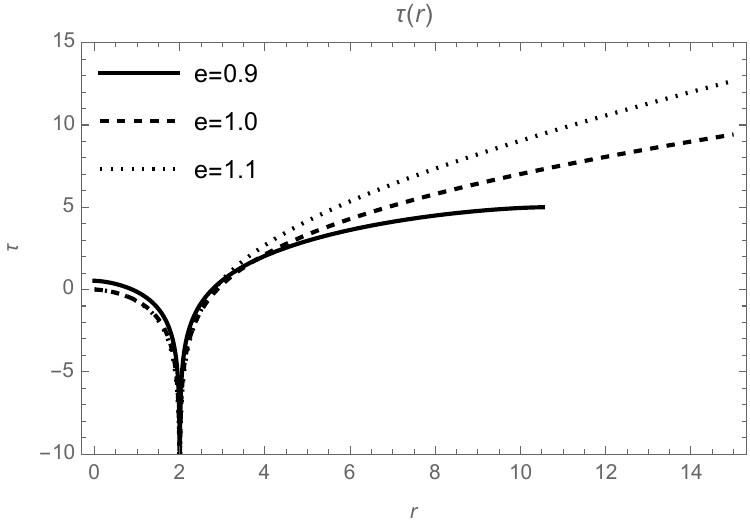} 
\caption{The radial profile of the khronon $\tau(r)$ in units of $1/m_S$ for some representatives of the three different types of geodesics in the case of a Schwarszchild black hole with the mass $m_S$.}
\end{figure}

\section{Electromagnetism}
\label{electrosec}

In order to proceed beyond vacuum solutions, we need to couple additional fields to Lorentz gauge theory. The source fields must then admit a pregeometric description, by which we simply mean that the matter theory should remain well-defined even in the absence of spacetime geometry, and in particular, without assuming an a priori metric tensor. Recent candidate theories of pregeometric Yang-Mills fields have been considered by Gallagher {\it et al} \cite{Gallagher:2022kvv,Gallagher:2024haq}. Here we shall restrict to the Abelian case of electromagnetism. The electromagnetically charged black hole solution is constructed explicitly, providing, to our knowledge, the first example of an exact solution to a pregeometric first order Yang-Mills theory. We will consider three alternative formulations of the theory, and show that the solution does not exist in one them. 

The fields in a pregeometric \1 order electromagnetic action functional 
\be \label{electroaction}
I  =  \int \bH^{ab}\wedge\lp\star\bH_{ab} - \eta_{ab}\bF\rp \quad \text{where} \quad \bH^{ab} = \bh^a\wedge\bbe^b\,, 
\ee
are the electromagnetic gauge potential and the additional coframe $\bh^a$. 
However, the field equations obtained by the variation wrt these 2 fields, respectively,
\bs
\label{homot}
\ba
\bdiff\bH &= & 0\,, \label{inhomo} \\
2\star \bH_{ab}\wedge\bbe^b - \bF\wedge\bbe_a & = & 0\,, \label{homo}
\ea
\es
do not explicitly feature the gauge potential, and thus it suffices for our present purposes to consider the system in terms of the $\bh^a$ and the physical field strength $\bF$. 
A rationale of the \1 order formulation that the field strength and the field excitation should be considered as independent degrees of freedom, will be made yet more apparent in Section \ref{altfor} wherein we'll consider the excitation field $h^{ab}=h^{[ab]}$ in the adjoint representation of the Lorentz group s.t. $\bh^a=h^a{}_b\bbe^b$.
The \1 Eq.(\ref{inhomo}) represents a pregeometric generalisation of the inhomogeneous Maxwell equations (in the absence of sources) where the usual field excitation $\ast\bF$ will be obtained as the solution $\bH = \ast \bF$ to the equation (\ref{homo}) in the geometric phase. There are alternative formulations to (\ref{electroaction}) which can allow dynamical solutions that modify the constitutive relation of the Maxwell theory s.t. $\bH \neq \ast \bF$, giving rise to new phenomenology. We will consider also such an alternative later in Section \ref{altfor}. 

First, we will explore spherically symmetric electromagnetic field in the formulation (\ref{inhomo}). In this Section we will work using the standard radial coordinate $r$. 
The generic electromagnetic field strength has the form
\bs
\label{EMandDUAL}
\be \label{EMfield}
\bF  =  \frac{q_e(r)}{r^2}\bdiff t \wedge\bdiff r - q_b(r)\sin\theta\bdiff \theta \wedge\bdiff \varphi 
  =  -\frac{q_e(r)}{r^2}F\bdiff \tau \wedge\bdiff \rho - q_b(r)\sin\theta\bdiff \theta \wedge\bdiff \varphi\,,
\ee
whose dual field strength is 
\be \label{dual}
\ast\bF =  -\frac{q_b(r)}{r^2}\bdiff t \wedge\bdiff r + q_e(r)\sin\theta\bdiff \theta \wedge\bdiff \varphi\,.
\ee
\es
Strictly speaking, the field strengths are not spherically symmetric in the presence of a non-zero magnetic field $q_b/r^2 \neq 0$. Nevertheless, as will be shown in detail below, the field strengths (\ref{EMandDUAL}) are compatible with spherically symmetric spacetime geometry. We have taken $q_e=q_e(r)$ and $q_b=q_b(r)$ to be arbitrary radial functions, so that the Ansatz (\ref{EMfield}) can be considered as the most generic one, but the notation anticipates that in vacuum they will be enforced to be constants $q_e'=q_m'=0$ by the EoMs (\ref{homot}). 


\subsection{Hyperframe formulation}

Let us rewrite the coframe (\ref{lemaitre}) using (\ref{cchange}),
\bs
\label{eejahoo}
\be
\bbe^0 = \bdiff t + h(r)\bdiff r\,, \quad \bbe^1 = F(r)\lp \bdiff t + k(r)\bdiff r\rp\,, \quad \bbe^2 = r\bdiff\theta\,, \quad \bbe^3 = r\sin\theta\bdiff\varphi\,, 
\ee
and make an Ansatz for the hypercoframe
\be \label{hoo}
\bh^0 = a(r)\bdiff t + b(r)\bdiff r\,, \quad \bh^1 = c(r)\bdiff t + d(r)\bdiff r\,, \quad \bh^2 = n(r)\bdiff\theta + m(r)\sin\theta\bdiff\varphi \,, \quad \bh^3 = s(r)\bdiff\theta + p(r)\sin\theta\bdiff\varphi\,.
\ee
\es
The configuration is simple enough that with a more clever Ansatz, one could more directly guess that the solution will be (\ref{EMsolution}), but it can be useful to go through a more detailed derivation. The most generic Ansatz would have 16 instead of just 8 functions in (\ref{hoo}), but the form of the electromagnetic field (\ref{EMfield}) suggests that (\ref{hoo}) is sufficient.  

Comparing with the spacetime area element $\bbe^a\wedge\bbe^b$, the antisymmetric components of $\bH^{ab}$ in (\ref{electroaction}) span some kind of hybrid area element. However, 
the $\bH^{ab}$ has also symmetric components. We readily find the diagonal components from (\ref{eejahoo}),  
\ba \label{diagonals}
\bH^{00}  =  \lp ah - b \rp\bdiff t\wedge\bdiff r\,, \quad
\bH^{11}  =     F\lp ck - d\rp\bdiff t\wedge\bdiff r \,, \quad
\bH^{22}  =  -m r\sin\theta\bdiff\theta\wedge\bdiff\varphi\,, \quad
\bH^{33}  =  s r\sin\theta\bdiff\theta\wedge\bdiff\varphi\,,
\ea
and obtain the trace
\be \label{hootrace}
\bH = - \lp ah - b  - Fck + Fd\rp \bdiff t\wedge\bdiff r  - \lp m - s\rp r\sin\theta\bdiff\theta\wedge\bdiff\varphi\,. 
\ee
We note that though none of the diagonal components (\ref{diagonals}) appears in the homogeneous equation (\ref{homo}), nevertheless that very equation should determine the trace (\ref{hootrace}) correctly s.t. the inhomogeneous equation (\ref{inhomo}) then describes the dynamics of the Maxwell theory. This pregeometric dynamical structure is quite non-trivial and it is worthwhile to check its workings in more detail.  

The 8 functions in (\ref{hoo}) are determined by the EoM (\ref{homo}). Computing first
\bs
\label{Hnolla}
\ba
\star\bH_{01} & = & \bH^{[23]} = \frac{1}{2}\lp n+p\rp r \sin\theta\bdiff\theta\wedge\bdiff\varphi\,, \\
\star\bH_{02} & = & -\bH^{[13]} = - \frac{1}{2}Fs\lp \bdiff t + k\bdiff r\rp\wedge\bdiff\theta   
-\frac{1}{2}\lp cr+Fp\rp\sin\theta\bdiff t\wedge\bdiff\varphi - \frac{1}{2}\lp dr + Fkp\rp\sin\theta\bdiff r\wedge\bdiff\varphi\,, \\ 
\star\bH_{03} & = & \bH^{[12]} = +\frac{1}{2}\lp cr + Fn\rp\bdiff t\wedge\bdiff\theta + \frac{1}{2}\lp dr + Fkn\rp\bdiff r\wedge\bdiff\theta
+ \frac{1}{2}Fm\lp\bdiff t+ k\bdiff r\rp\sin\theta \wedge\bdiff\varphi\,,
\ea
\es
we can obtain the 0-component of the EoM,
\ba \label{eq0}
2\lb F\lp n+p\rp +rc \rb r\sin\theta\bdiff t\wedge\bdiff\theta\wedge\bdiff\varphi
& + & 2\lb Fk\lp n+p\rp +rd \rb r\sin\theta\bdiff r\wedge\bdiff\theta\wedge\bdiff\varphi \nn \\
=  q_b\lp \bdiff t + h\bdiff r\rp\wedge \sin\theta\bdiff\theta\wedge\bdiff\varphi\,.
\ea 
Using further
\bs
\ba
\star\bH_{10} & = & -\star\bH_{01}\,, \\ 
\star\bH_{12} & = & \bH^{[03]} = \frac{1}{2}s\bdiff t\wedge\bdiff\theta + \frac{1}{2}hs \bdiff r\wedge\bdiff\theta + \frac{1}{2}\lp ar +p\rp\sin\theta\bdiff t\wedge\bdiff \varphi 
+ \frac{1}{2}\lp br +h p\rp\sin\theta\bdiff r\wedge\bdiff \varphi \,, \\
\star\bH_{13} & = & -\bH^{[02]} = -\frac{1}{2}\lp ar+n\rp\bdiff t\wedge\bdiff \theta - \frac{1}{2}\lp br + hn\rp\bdiff r\wedge\bdiff\theta - \frac{1}{2}m\sin\theta\bdiff t\wedge\bdiff\varphi 
 - \frac{1}{2}mh\sin\theta\bdiff r\wedge\bdiff\varphi\,,
\ea
\es
we obtain the 1-component of iso-khronon EoM,
\ba \label{eq1}
- 2\lp n+p + ra\rp  r\sin\theta\bdiff t\wedge\bdiff\theta\wedge\bdiff\varphi
& - & 2\lb h\lp n+p\rp +rb \rb r\sin\theta\bdiff r\wedge\bdiff\theta\wedge\bdiff\varphi \nn \\
=  F q_b \lp \bdiff t + k\bdiff r\rp\wedge \sin\theta\bdiff\theta\wedge\bdiff\varphi\,.
\ea 
Then, using 
\bs
\ba
\star\bH_{20} & = & -\star\bH_{02}\,, \\ 
\star\bH_{21} & = & -\star\bH_{12}\,, \\ 
\star\bH_{23} & = & \bH^{[01]} = \frac{1}{2}\lb F\lp ak-b\rp -ch + d \rb\bdiff t \wedge\bdiff r\,,
\ea
\es
we obtain
\be \label{eq2}
2Fs\lp k-h\rp \bdiff t\wedge\bdiff r\wedge\bdiff\theta +
2\lb F\lp akr  -br -ph + pk\rp -crh+dr\rb \sin\theta\bdiff t\wedge\bdiff r\wedge\bdiff\varphi = \frac{q_e}{r}\bdiff t\wedge\bdiff r\wedge\bdiff\theta\,. 
\ee
Finally, noting that $\star\bH_{3a}=-\star\bH_{a3}$, we can use the previous expressions to deduce that
\be \label{eq3}
-2\lb F\lp akr+nk-nh-br\rp  - crh+dr\rb \bdiff t\wedge\bdiff r\wedge\bdiff\theta
-2Fm\lp k -h \rp \sin\theta\bdiff t\wedge\bdiff r\wedge\bdiff\varphi =  \frac{q_e}{r}\sin\theta\bdiff t\wedge\bdiff r\wedge\bdiff\varphi\,.
\ee
The 4 components of the homogeneous equation (\ref{homo}) are (\ref{eq0}), (\ref{eq1}), (\ref{eq2}), (\ref{eq3}), respectively, and they are solved by 
\be
a = \frac{b}{k} = Fc = \frac{Fd}{h} = - \frac{Fq_b}{2r^2}\,, \quad m = -s = -\frac{q_e}{2r}\,, \quad n=p=0\,. 
\ee
Plugging this solution into (\ref{hootrace}) and checking against (\ref{dual}) confirms that $\bH=\ast \bF$, as expected. We now have
\bs
\be \label{EMsolution}
\bh^0 = -\frac{q_b}{2r^2}\bbe^1\,, \quad \bh^1 = -\frac{q_b}{2r^2}\bbe^0\,, \quad \bh^2 = -\frac{q_e}{2r^2}\bbe^3\,, \quad
\bh^3 = \frac{q_e}{2r^2}\bbe^2\,,
\ee
so that finally, the excitation has the two independent non-vanishing components
\be \label{final}
h^{01} = -\frac{q_b}{2r^2}\,, \quad h^{23} = -\frac{q_e}{2r^2}\,, 
\ee
\es
the counterparts of the magnetic and the electric field strengths, respectively. 

Having at hand the (hyper)coframe (\ref{EMsolution}), we could define an electrified-magnetised (hyper)metric tensor $\bh^a\otimes\bh_a$. It would correspond to a line element 
\be
\bdiff \tilde{s}^2 = \frac{q^2_b}{4r^4}\lp - f^{-2}\bdiff r^2 + f^2\bdiff t^2\rp + \frac{q_e^2}{4r^2}\lp \bdiff \theta^2 + \sin^2\theta\bdiff\varphi^2\rp\,,   
\ee
describing a curved (hyper)spacetime wherein $r$ would play the role of a time coordinate. When the charges are equal, this metric is obtained from the spacetime metric by switching the time versus space -like properties of the coordinates $t$ and $r$ and by perfoming a conformal rescaling by the factor $\sim r^{-4}$.    
For amusement, we could compute various further geometrical quantities, e.g. the Riemann curvature $\tilde{\bR}^{ab}$ of the (hyper)metric, the (hyper)non-metricity $\bDiff\tilde{g}_{ab}$ and the (hyper)torsion $\tilde{\bT}^a = \bDiff\bh^a$ of the hyperframe wrt the Lorentz connection. 

\subsection{Alternative formulations}
\label{altfor}

The \1 order formulation of electromagnetism in terms of the hyperframe $\bh^a$ points towards possibly interesting connections with
other bimetric theories \cite{Isham:1971gm,Israelit:1986ez,BeltranJimenez:2012sz,Schmidt-May:2015vnx,Castro:2016rvj,Krssak:2017nlv,Blixt:2023qbg}. However, it is somewhat superfluous to introduce the $\bh^a$ as a fundamental field in a theory of pure electromagnetism. As an idea of the formulation is to consider the field excitation as an independent variational degree of freedom, 
it is natural is to regard the scalar $h^{ab}$ as the fundamental field instead of the hyperframe. This is also the more economical  point of departure, the $h^{ab}=-h^{ba}$ in the adjoint of the Lorentz group carrying the same number 6 of free components as the $\ast \bF$. 

Thus, rather than (\ref{electroaction}), we consider 
\be \label{electroaction2}
I  =  \int \bH^{ab}\wedge\lp\star\bH_{ab} - \eta_{ab}\bF\rp \quad \text{where} \quad \bH^{ab} = h^a{}_c\bbe^c\wedge\bbe^b\,,
\ee
and the field equations
\bs
\ba
\bdiff\bH &= & 0\,, \label{inhomo2} \\
2\bbe_{[a}\star \bH_{b]c}\wedge\bbe^c - \bF\wedge\bbe_{a}\wedge\bbe_{b} & = & 0\,. \label{homo2}
\ea
\es
The 4-form equation (\ref{homo2}) has now only 6 independent components. To find the solution for $h^{ab}$ associated with the
electromagnetic field (\ref{EMfield}), we could start with the simplification of the Ansatz (\ref{hoo}) by the antisymmetry, and obtain the solution very easily. However, as we know that (\ref{homo2}) implies $h_{ab}\bbe^a\wedge\bbe^b = \ast\bF$, the solution (\ref{final}) can be directly read off from the dual (\ref{dual}).  The action (\ref{electroaction2}) should be considered the fundamental version of the theory. 

A modified gauge theory has been considered motivated by the analogy with the gauge gravity theory, s.t. $\bh^a = \bDiff h^a$, in terms of an {\it iso-khronon} scalar field $h^a$ in the fundamental representation of the
Lorentz group. The modified action principle
\be \label{electroaction3}
I  =  \int \bH^{ab}\wedge\lp\star\bH_{ab} - \eta_{ab}\bF\rp \quad \text{where} \quad \bH^{ab} = \bDiff h^a\wedge\bbe^b\,,
\ee
however, does not reproduce the dynamics of Maxwell electromagnetism but results in a completely different theory. The field equations
\bs
\ba
\bdiff\bH &= & 0\,, \label{inhomo3} \\
\bDiff\lp 2\star \bH_{ab}\wedge\bbe^b - \bF\wedge\bbe_a\rp & = & 0\,, \label{homo3}
\ea
\es
now impose only the covariant derivative of the previous inhomogeneous equation. The equation itself $2\star \bH_{ab}\wedge\bbe^b -=\bF\wedge\bbe_a$ cannot be satisfied in general, because the $h^a$ does not carry enough degrees of freedom. This can be demonstrated explicitly in the spherically symmetric case at hand. With an Ansatz $h^a=h^a(\tau,\rho)$, we get
\bs
\label{ehdot}
\ba
\bDiff h^0 & = &  \dot{h}^0\bbe^0 + \lp \frac{(h^0)'}{F} + \frac{h^1}{\tau}\rp\bbe^1 + \frac{h^2}{\tau}\bbe^2  + \frac{h^3}{\tau}\bbe^3  \nn \\
& = & \bh^0 \quad \Rightarrow \quad \dot{h}^0 = 0\,, \quad  \frac{(h^0)'}{F} + \frac{h^1}{\tau} = -\frac{q_b}{2r^2}\,, \quad h^2 = h^3 = 0\,.  \label{ehto0}
\ea
where in the second line we have taken the solution (\ref{EMsolution}). Taking into account the result $h^2=h^3=0$, 
in order to reproduce the spatial components of the hypercoframe, we should then have
\ba
\bDiff h^1 & = & \dot{h}^1\bbe^0 + \lp \frac{(h^1)'}{F} + \frac{h^0}{\tau}\rp\bbe^1 = -\frac{q_b}{2r^2}\bbe^0\,, \label{ehto1} \\
\bDiff h^2 & = & \lp \frac{h^0}{\tau}+\frac{r' h^1}{rF}\rp\bbe^2 - i\lp \frac{1}{\tau}-\frac{\dot{r}}{r}\rp h^1\bbe^3  =  -\frac{q_e}{2r^2}\bbe^3\,, \label{ehto2} \\
\bDiff h^3 & = & i\lp \frac{1}{\tau}-\frac{\dot{r}}{r}\rp h^1\bbe^2  +  \lp \frac{h^0}{\tau}+\frac{r' h^1}{rF}\rp\bbe^3  = \frac{q_e}{2r^2}\bbe^2\,. \label{ehto3}
\ea 
\es
The last components of both the (\ref{ehto1}) and the  (\ref{ehto3}) vanish, which implies that $h^1=h(\tau) r$ for some function $h(\tau)$. From the time-independence of $h^0$ in (\ref{ehto0}) we then deduce that $h \sim 1/\tau$, and that the relation $r'=-A_0F$ must hold. However, then $h^0$ must be a constant and the system (\ref{ehdot}) becomes impossible to satisfy for non-zero charges $q_e$ and $q_b$. We see that the theory (\ref{electroaction3}) cannot mimic the spherically symmetric configuration (\ref{EMfield}) described by the standard electromagnetic theory.

Thus, this case study illustrates the fact noticed already in Refs.\cite{Gallagher:2023ghl,Gallagher:2024haq}, that the iso-khronon theory (\ref{electroaction3}) does not possess the solutions of the standard electromagnetic theory. This type of a pregeometric theory is therefore not a viable candidate description of the standard model interactions, though presents a curious vector field theory that could in principle be phenomenologically interesting. A satisfactory \1 order pregeometric extension of the standard model interactions is achieved with the action of the type (\ref{electroaction2}) in terms of the excitation field $h^{ab}$ valued in the adjoint of the Lorentz group.


\subsection{The Reissner-Nordstr{\"o}m solution}
\label{RNsolution}

The energy current contributed by the action (\ref{electroaction2}) (equivalently (\ref{electroaction})) is
\be
\bt_a = 2\star\bH_{ab}\wedge\bh^a - \bF\wedge\bh_a\,,
\ee
which assumes a very simple form for the solution (\ref{EMsolution})\,,
\be
\bt_0  =  \frac{q^2}{2r^4}\star\bbe_0\,, \quad  
\bt_1  =  \frac{q^2}{2r^4}\star\bbe_1\,, \quad  
\bt_2  =  \frac{q^2}{2r^4}\star\bbe_2\,, \quad  
\bt_3  =  -\frac{q^2}{2r^4}\star\bbe_3\,.   
\ee
Inserting this as the source in the field equations (\ref{hamilton}) and (\ref{momentum}), we find that the homogeneous components of these equations are satisfied identically
given (\ref{ABCDsolution}) and $r'=-A_0F(r)$, where $A_0$ is again a constant. The homogeneous components of (\ref{hamilton}) and (\ref{mom1}) both yield the same equation
\bs
\label{yht}
\be
1-A_0^2 + F^2(r) + 2rF(r)F'(r) =  \frac{q^2}{r^2}\,, \label{yht1}
\ee  
and the homogeneous components of (\ref{mom2}) and (\ref{mom3}) are also degenerate and both equivalent to
\be
2F(r)F'(r)+r(F'(r))^2+rF(r)F''(r) = -\frac{q^2}{r^3}\,. \label{yht2}
\ee
\es
Equation (\ref{yht1}) does not feature the constant $A_0$, unlike (\ref{yht1}) which does. However, (\ref{yht2}) is a second order differential equation and its solution contains
an additional integration constant. Choosing the constant suitably, both the equations (\ref{yht}) are solved by
\be
F^2(r) = A_0^2 - 1 + \frac{r_S}{r} - \frac{q^2}{r^2}\,. 
\ee 
The function $F(r)$ vanishes $F(r_0)=0$ at 
\be
r_0 = \frac{1}{2\lp A_0^2-1\rp}\lp \sqrt{4\lp A_0^2 -1 \rp q^2 + r_S^2} - r_S\rp\,,  
\ee
and becomes pure imaginary below $r<r_0$. The $r_0$ is always beyond not only the outer event horizon at $r_+$ but also the inner Cauchy horizon at $r_-$ where
\be
r_\pm = \frac{1}{2}\lp r_S \pm \sqrt{r_S^2-4q^2}\rp\,. 
\ee
By increasing the parameter $A_0$ related to the initial condition of a free-falling object whose proper time along a geodesic is measured directly by the khronon $\tau$, the range of real and positive-definite $F(r)$ can be extended to arbitrarily small $r$. The termination of Painlev{\'e}-Gullstrand description of the Reissner-Nordstr{\"o}m geometry has been recognised as a consequence of the Missner-Sharp mass becoming negative \cite{Faraoni:2020ehi} (see also \cite{Herrero:2010yt,Fazzini:2023scu}). Indeed, the Missner-Sharp mass coincides with the canonical energy which does become arbitrarily negative as one approaches the center of a  Reissner-Nordstr{\"o}m black hole \cite{Gomes:2022vrc}. At $r_0=0$, the energy that attracts an object on a geodesic trajectory disappears. Thus the geodesics may bounce at $r_0$, and in GR the Painlev{\'e}-Gullstrand (as well as the Lema{\^i}tre) coordinate system breaks down at this point. However, in the Lorentz gauge theory it is not obvious whether any fatal discontinuity occurs at the radius $r=r_0$. As the function $F$ becomes pure imaginary, the spacetime metric changes signature, and there appear two time-like directions instead of one. This hints at a ``bitemporal''  (c.f. ``atemporal'' \cite{Capozziello:2024ucm}) avoidance of the singularity, and as we will conclude in Section \ref{conclusionsec}, might merit further investigation. 

\section{The SO(1,2) radial phase}
\label{radialsec}

In the conventional Schwarzschild-like coordinate system (\ref{sline}) it is natural to consider the symmetry breaking pattern SO(1,3)/SO(1,2), such that the radial
field is described as the radion $\phi^a = \phi(r)\delta^1_0$. The coframe (\ref{stetrad}) is then reproduced by the connection coefficients
\bs
\label{radialSSB}
\be
\bomega^0{}_1 = \frac{f}{\phi}\bdiff t\,, \quad \bomega^1{}_2 = -\frac{r}{\phi}\bdiff\theta\,, \quad \bomega^1{}_3 = -\frac{r\sin\theta}{\phi}\bdiff\varphi\,,
\ee
when the radion is given by the function
\be
\phi(r) = \int^r {g(x)}\bdiff x\,. 
\ee
\es
As an example, the Schwarzschild solution is described by 
\be \label{phi_S}
\phi(r) = \pm \sqrt{r\lp r- r_S\rp} \pm \frac{r_S}{2}\log{\lb \frac{4r}{r_S}\lp 1+ \sqrt{1-\frac{r_S}{r}}\rp -2\rb} + \phi_0\,.  
\ee
First, we note that if $m_S=0$, the radion is simply (anti)linear in the radius, $\phi(r)=\pm r+\phi_0$. 
Choosing $\phi_0=\mp r_S\log{2}/2$, the radion vanishes at the horizon and becomes pure imaginary inside the horizon. Then, at the singularity
$r \rightarrow 0$ the radion $\phi \rightarrow \pm r_S\log{i}/2=ir_S(1/2+k)\pi$ is a multiple of $i\pi$ given by the integer $k$ depending on the branch.  Asymptotically at $r \rightarrow \infty$ we have $\phi(r) \rightarrow \pm (r + \log{4r/r_S})$. 


\begin{figure}[t!]
\begin{subfigure}[h]{0.45\linewidth}
\includegraphics[width=1.0\linewidth,keepaspectratio]{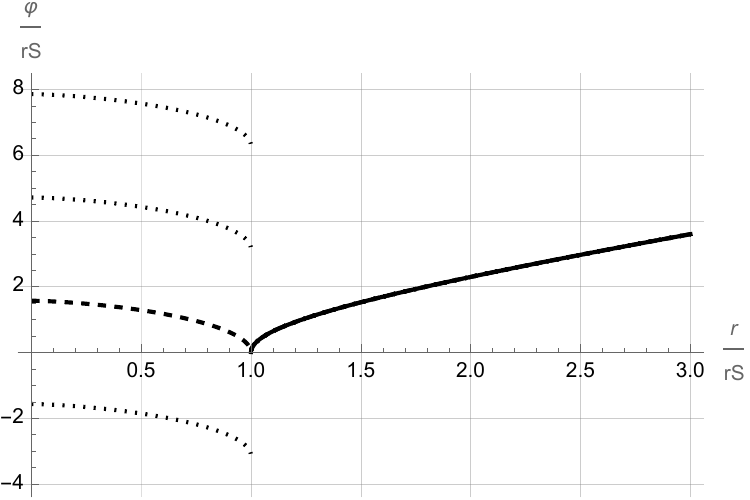} 
\caption{The radion field profile. Below $r<r_S$ the field is imaginary. Thus, the solid line is the Re($\phi$),
and the dashed line is the Im($\phi$).}
\end{subfigure}
\hfill
\begin{subfigure}[h]{0.45\linewidth}
\includegraphics[width=1.0\linewidth,keepaspectratio]{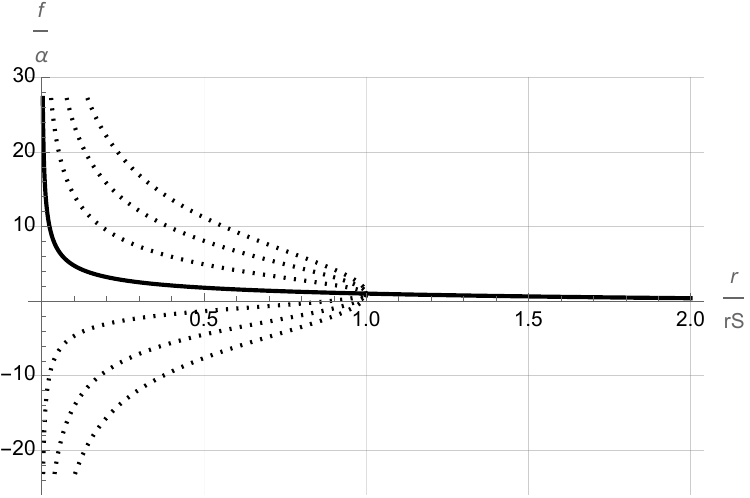}
\caption{The lapse function $f(r)$ as a function of the radius. Though alternative branches of solutions depicted as dotted lines are continuous,
they are not smooth.}
\end{subfigure}
\caption{The solution $\beta=0$ which is well-defined everywhere though singular at $r=0$ and $r=r_0$.}
\label{betafigure}
\end{figure}

Let us consider the generic spherically symmetric system in the configuration (\ref{radialSSB}). According to (\ref{educated}), we shall now parameterise the remaining three connection coefficients in terms of four $r$-dependent functions, as
\be
\bomega^0{}_2 = A(r)\bdiff\theta - iB(r)\viff\,, \quad \bomega^0{}_3 = iB(r)\bdiff\theta + A(r)\bdiff\tilde{\varphi}\,, \quad \bomega^2{}_3 = iC(r)\bdiff t + iD(r)\bdiff r - \tan\theta\bdiff\tilde{\varphi}\,.
\ee
Note that despite we use the same symbols as previously, these functions are different from those in (\ref{Tparam}), and we introduced the shorthand $\bdiff\tilde{\varphi} = \sin\theta\bdiff\varphi$.  Also, in this section the prime will denote derivative wrt $r$. The curvature of the connection is
\bs
\ba
\bR^0{}_1 & = & -\lp \frac{f'}{\phi} - \frac{f\phi' }{\phi^2} \rp \bdiff t\wedge\bdiff r + 2\frac{iBr}{\phi}\bdiff\theta\wedge\bdiff\tilde{\varphi}\,, \\
\bR^0{}_2 & = & -\lp \frac{fr}{\phi^2} + BC\rp\bdiff t\wedge\bdiff\theta  + iAC\bdiff t\wedge\bdiff\tilde{\varphi} + \lp A' - BD\rp\bdiff r\wedge\bdiff \theta - i\lp B' - AD\rp\bdiff r\wedge\viff\,, \\
\bR^0{}_3 & = & -iAC\bdiff t\wedge\bdiff\theta - \lp\frac{fr}{\phi^2} + BC\rp\bdiff t\wedge\viff + i\lp B'-AD\rp\bdiff r\wedge\bdiff\theta + \lp A' - BD\rp\bdiff r\wedge\viff\,, \\
\bR^1{}_2 & = & \frac{fA}{\phi}\bdiff t\wedge\bdiff\theta   - i\lp \frac{fB}{\phi } + \frac{rC}{\phi}\rp \bdiff t\wedge\viff + \lp\frac{r\phi'}{\phi^2} - \frac{1}{\phi}\rp \bdiff r\wedge\bdiff\theta - \frac{irD}{\phi}\bdiff r\wedge\viff\,, \\  
\bR^1{}_3 & = &i\lp \frac{fB}{\phi } + \frac{rC}{\phi}\rp \bdiff t\wedge\bdiff\theta + \frac{fA}{\phi}\bdiff t\wedge\viff + \frac{irD}{\phi}\bdiff r\wedge\bdiff\theta
+ \lp\frac{r\phi'}{\phi^2} - \frac{1}{\phi}\rp \bdiff r\wedge\viff\,, \\
\bR^2{}_3 & = & -iC'\bdiff t\wedge\bdiff r + \lp 1 - \frac{r^2}{\phi^2} + A^2 - B^2  \rp\bdiff\theta\wedge\viff\,. 
\ea
\es
The self-dual connection $\bA^I = -i\+{}\bomega^{0I}$ is 
\bs
\ba
\bA^1 & = & -\frac{i}{2}\lp \frac{f}{\phi} - C\rp\bdiff t + \frac{i}{2}D\bdiff r - \frac{1}{2}\tan\theta\viff\,, \\ 
\bA^2 & = & -\frac{i}{2}A\bdiff\theta + \frac{1}{2}\lp \frac{r}{\phi} - B\rp\viff\,, \\
\bA^3 & = & -\frac{1}{2}\lp \frac{r}{\phi} - B\rp\bdiff\theta - \frac{i}{2}A\viff\,,
\ea
\es
and its field strength is
\bs
\ba
\bF^1 & = & \frac{i}{2}\lp \frac{f'}{\phi} - \frac{f\phi'}{\phi^2} - C'\rp\bdiff t\wedge\bdiff r + \frac{1}{2}\lb 1 + A^2 - (\frac{r}{\phi}-B )^2\rb\bdiff\theta\wedge\viff\,, \\
\bF^2 & = & \frac{i}{2}\lp \frac{f}{\phi}-C\rp\lp\frac{r}{\phi} -B\rp\bdiff t\wedge\bdiff\theta \nn  -  \frac{A}{2}\lp\frac{f}{\phi}-C\rp\bdiff t\wedge\viff \nn \\
& - & \frac{i}{2}\lb A' + \lp \frac{r}{\phi}-B\rp D\rb\bdiff r\wedge\bdiff\theta - \frac{1}{2}\lp \frac{r\phi'}{\phi^2} - \frac{1}{\phi} + B' - AD\rp\bdiff r\wedge\viff\,, \\
\bF^3 & = & \frac{A}{2}\lp\frac{f}{\phi}-C\rp\bdiff t\wedge\bdiff\theta +\frac{i}{2}\lp\frac{f}{\phi}-C\rp\lp\frac{r}{\phi}-B\rp\bdiff t\wedge\viff \nn \\ & + & \frac{1}{2}\lp \frac{r\phi'}{\phi^2}-\frac{1}{\phi} + B' - AD\rp\bdiff r\wedge\bdiff\theta 
- \frac{i}{2}\lb A' + \lp\frac{r}{\phi} - B\rp D\rb\bdiff r\wedge\viff\,.
\ea
\es
Due to the different symmetry breaking configuration, the radial torsion now vanishes. We have
\bs
\ba
\bT^0 & = & -\lp f' - f\log{\phi}' \rp \bdiff t\wedge\bdiff r + 2iBr\bdiff\theta\wedge\bdiff\tilde{\varphi}\,, \\
\bT^1 & = & 0\,, \\
\bT^2 & = &  -fA\bdiff t\wedge\bdiff\theta + i\lp fB + rC\rp \bdiff t\wedge\viff + \lp 1 - r\log{\phi}' \rp \bdiff r\wedge\bdiff\theta + irD\bdiff r\wedge\viff\,, \\
\bT^3 & = & -i\lp {fB} + {rC}\rp \bdiff t\wedge\bdiff\theta - {fA}\bdiff t\wedge\viff - {irD}\bdiff r\wedge\bdiff\theta
+ \lp 1 - r\log{\phi}'\rp \bdiff r\wedge\viff\,. 
\ea
\es
The six components of the connection field equations now read explicitly 
\bs
\ba
\bT^0\wedge\bbe^1 + i\bT^2\wedge\bbe^3 -i\bT^3\wedge\bbe^2 & = & 2\bO^{01}\,, \label{con1} \\
\bT^0\wedge\bbe^2 - \bT^2\wedge\bbe^0  + i\bT^3\wedge\bbe^1 & = & 2\bO^{02}\,, \label{con2} \\
\bT^0\wedge\bbe^3  - i\bT^2\wedge\bbe^1- \bT^3\wedge\bbe^0  & = & 2\bO^{03}\,, \\
- i\bT^0\wedge\bbe^3-\bT^2\wedge\bbe^1  + i\bT^3\wedge\bbe^0 & = & 2\bO^{12}\,, \\
i\bT^0\wedge\bbe^2 - i\bT^2\wedge\bbe^3 - \bT^3\wedge\bbe^1 & = & 2\bO^{13}\,, \\
 - i\bT^0\wedge\bbe^1 + \bT^2\wedge\bbe^3 - \bT^3\wedge\bbe^2 & = & 2\bO^{23}\,.
\ea
\es
Combining the first three with the last three imposes $\prescript{-}{}\bO^{ab}=0$. Due to spherical symmetry, the equations sourced by $\bO^{02}$ and $\bO^{03}$ (equivalently, the equations sourced by the $\bO^{02}$ and $\bO^{03}$) are equivalent. Thus, out of the six there remain two independent equations. Let us choose the two first ones,
(\ref{con1}) and (\ref{con2}),
\bs
\ba
i\bO^{01} & = & \lp \frac{1}{r\phi'} - \frac{1}{\phi} + \frac{B}{r}\rp\star\bbe^0 + \frac{A}{r}\star\bbe^1\,, \\
2\bO^{02} & = & i\lp \frac{A}{r} - \frac{D}{\phi'}\rp\star\bbe^2 - \lp \frac{f'}{f\phi'}+\frac{1}{r\phi'} - \frac{2}{\phi} + \frac{B}{r}
+ \frac{C}{f}\rp\star\bbe^3\,.
\ea
\es  
Assuming there are no spinning sources, the solutions are given by
\be \label{nospin}
A = 0\,, \quad B = \frac{r}{\phi} - \frac{\partial r}{\partial \phi}\,, \quad
C = \frac{f}{\phi} -   \frac{\partial f}{\partial \phi}\,, \quad D= 0\,. 
\ee
The connection coefficients $B$ and $C$ measure the non-linearity of the radius $r$ and the lapse function $f$ as
functions of the radion $\phi$, respectively. The energy equation is
\be \label{efe1a}
-\bt^0 = \lb \frac{1}{2\phi^2} + \frac{1}{2r^2}\lp 1 + A^2 - B^2\rp + \frac{B}{r\phi} 
- \frac{1}{r\phi'}\lp \frac{1}{\phi} - B' + AD\rp\rb\star\bbe^0 + \frac{A}{r}\lp \frac{1}{\phi} - \frac{C}{f}\rp\star\bbe^2
\ee
Without an energy source, $\bt^0=0$, and upon plugging in (\ref{nospin}), this equation reduces to the differential equation
\be \label{efe1b}
r\phi'' + \frac{1}{2}\phi'\lb (\phi')^2 - 1\rb =0\,,
\ee
whose solution is given by (\ref{phi_S}). The radial component of the field equation is
\be \label{efe2a}
\bt^1 + \bM^1 = \frac{1}{r\phi'}\lb A' + \lp\frac{r}{\phi}-B\rp D\rb\star\bbe^0 
+\frac{1}{r}\lb \frac{1}{2r}\lp 1 + A^2 - \lp \frac{r}{\phi} - B\rp^2\rp -\lp\frac{1}{\phi}-\frac{C}{f}\rp\lp\frac{r}{\phi}-B\rp\rb\star\bbe^1\,. 
\ee
Let us again assume no material sources $\bt^1=0$ but take into account the integration form $\bM^1 = f^{-1}r^{-2}\alpha/2\star\bbe^1$, where $\alpha$ is a dimensionless constant. Plugging in (\ref{nospin}) we radial momentum equation reduces to 
\be \label{efe2b}
rf' + \frac{1}{2}f +  \frac{1}{2}(\phi')^2\lp \alpha - f\rp = 0\,.  
\ee 
Using the radion profile (\ref{phi_S}), we find the solution
\be \label{lapse}
f  = \alpha \lb 1 + \sqrt{1-\frac{r_S}{r}}\log{\lp \sqrt{\frac{r}{r_S}} -\sqrt{\frac{r}{r_S}-1} \rp}\rb + \beta\sqrt{1-\frac{r_S}{r}}\,,
\ee
where $\beta$ is an integration constant. 
At $r=r_S$, we have $f=\alpha$. Inside the horizon, the $\alpha$-term remains real but the $\beta$-term becomes pure imaginary. 
The angular components of the field equations give
\bs
\label{angulars}
\ba
\bt^2  & = &  
- \frac{1}{2}\lb\frac{\partial}{f\partial\phi}\lp \frac{f}{\phi}-C\rp +\frac{1}{r\phi'}\lp\frac{1}{\phi}-B'+AD\rp -\lp\frac{1}{\phi}-\frac{B}{r}\rp\frac{C}{f} - \frac{B}{r\phi}\rb\star\bbe^2 \\ 
& - &  \frac{i}{2r}\lb \frac{\partial A}{\partial \phi} + \lp \frac{r}{\phi}-B\rp\frac{D}{\phi'} + A\lp\frac{1}{\phi}-\frac{C}{f}\rp\rb\star\bbe^3 \\
 \bt^3  & = & 
 \frac{i}{2r}\lb \frac{\partial A}{\partial \phi} + \lp \frac{r}{\phi}-B\rp\frac{D}{\phi'} + A\lp\frac{1}{\phi}-\frac{C}{f}\rp\rb\star\bbe^2 \\ 
   & - &\frac{1}{2}\lb\frac{\partial}{f\partial\phi}\lp \frac{f}{\phi}-C\rp +\frac{1}{r\phi'}\lp\frac{1}{\phi}-B'+AD\rp -\lp\frac{1}{\phi}-\frac{B}{r}\rp\frac{C}{f} - \frac{B}{r\phi}\rb\star\bbe^3\,.  
\ea
\es 
Plugging in the connection (\ref{nospin}) and setting the sources to vanish $\bt^2=\bt^3=0$, Eq.(\ref{angulars}) boils down to one non-trivial equation
\be
r\phi' f'' +\lp \phi' - r\phi''\rp f '  - \phi'' f = 0\,.  
\ee
Given the radion $\phi=\phi(r)$, this is a homogeneous  second order differential equation for the function $f(r)$, the solution possessing two integration constants.
Remarkably, when using the radion profile (\ref{phi_S}), the solution emerges in the form (\ref{lapse}) wherein  the $\alpha$ and the $\beta$ are the two integration constants.
Thus, (\ref{lapse}) is indeed a consistent vacuum solution of the theory.

The solutions can be classified into three distinct cases according to whether $\alpha$, $\beta$ or neither of them is zero. The case $\alpha = 0$ gives the standard Schwarzschild solution. The case $\beta=0$ describes an exotic spacetime which is not asymptotically flat. The case that both $\alpha$ and $\beta$ are non-zero can be considered as a class of modifications of the usual Schwarzschild solution. However, the modification is only well-defined outside $r>r_S$ the Schwarzschild radius, since the solution becomes complex inside $r<r_S$ (or, choosing imaginary $\beta$, the solution is consistent only inside the horizon). In contrast, the two first cases are well-defined everywhere (though have singularities). We plot an example of a case with $\beta \neq 0$ in figure \ref{betafigure}.

\begin{figure}[t!]
\begin{subfigure}[h]{0.45\linewidth}
\includegraphics[width=1.0\linewidth,keepaspectratio]{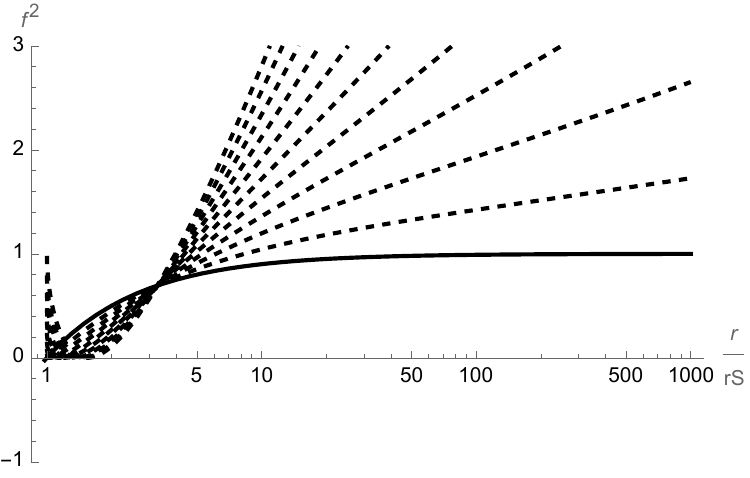} 
\caption{The lapse function $f^2(r)$ as a function of the radius $r$ in units of $r_S$ for cases $\alpha \le 0$. The solid line is $\alpha=0$,
and the dashed lines are for $\alpha=-0.1$, $\alpha=-0.2$, $\dots$, $\alpha=-1.0$.}
\end{subfigure}
\hfill
\begin{subfigure}[h]{0.45\linewidth}
\includegraphics[width=1.0\linewidth,keepaspectratio]{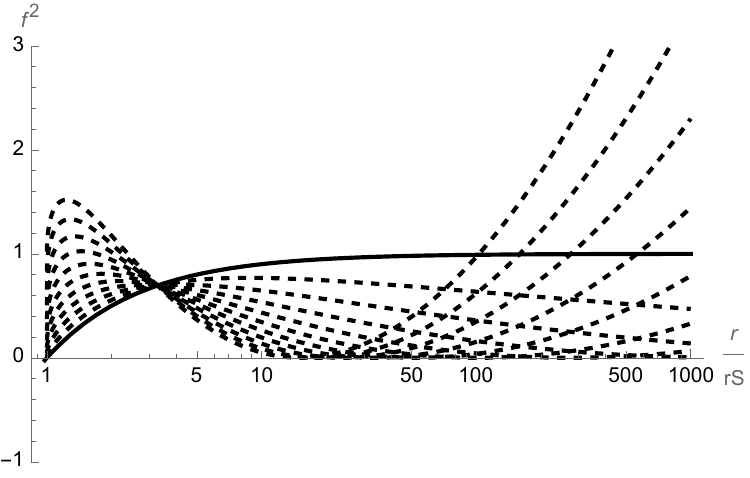}
\caption{The lapse function $f^2(r)$ as a function of the radius $r$ in units of $r_S$ for cases $\alpha \ge 0$. The solid line is $\alpha=0$,
and the dashed lines are for $\alpha=0.1$, $\alpha=0.2$, $\dots$, $\alpha=1.0$}
\end{subfigure}
\caption{The lapse function computed from (\ref{lapse}) for different values $\alpha$ with $\beta=1$. \label{lapsefigure}}
\label{ffigure}
\end{figure}

Let us consider in some detail the third case which represents a (rather drastic) modification of the Schwarzschild black hole. Firstly, we note that the $\alpha$-term becomes zero at approximately $r = r_0 \approx 3.72671753r_S$, so at this radius the lapse function $f(r_0)$ coincides with the lapse function of the standard solution. Assuming $\alpha<0$, then below the radius $r_S < r < r_0$ the lapse function is smaller than the usual one, and actually goes to zero at some $r>r_S$ depending on the $\alpha$. Thus, there is a bounce in the metric function $f^2(r)$ which occurs between $r_S$ and $r_0$.  Asymptotically at $r \gg r_0$,  the lapse function grows logarithmically, like 
\be
f(r \gg r_0) \rightarrow \frac{\alpha}{2}\lp 2 - \log{\frac{4r}{r_S}}\rp\,.   
\ee
When $\alpha>0$, the bounce in the metric function occurs at $r>r_0$. We plot these different cases in figure \ref{lapsefigure}. These singular solutions have been found also in the
context of mimetic gravity \cite{Gorji:2020ten,Khodadi:2024ubi}, see also \cite{Chen:2017ify,Nashed:2018qag,Chamseddine:2019fog,Nojiri:2024txy}. The new $r_0$-singularity was interpreted as a caustic singularity \cite{Gorji:2020ten}, and it was shown that the compact objects present in mimetic gravity do not reproduce the shadows observed by the Event Horizon Telescope \cite{Khodadi:2024ubi}. Thus, the exotic radion solutions in Lorentz gauge theory with $M \neq 0$ do not describe realistic astrophysics, but their existence is a theoretical curiosity. On the other hand, extensions of mimetic gravity could support regular black holes \cite{Han:2022rsx,Giesel:2024mps} and it could be interesting to study their possible relevance to ultraviolet extensions of the Lorentz gauge theory.  

\subsection{De Sitter space}

The caustic singularities found above are a generic feature of the spherically symmetric solutions in the space-like symmetry phase of the Lorentz gauge theory. 
We illustrate this by considering the generalisation of the de Sitter space solution in the theory. Therefore, we take into account a cosmological constant,
\be \label{Lambda}
\bt^a = \frac{3}{2}\Lambda\star\bbe^a\,. 
\ee
For convenience, we have rescaled the $\Lambda$ by the factor of 3. 
Using this simple source in the energy constraint (\ref{efe1a}), the equation (\ref{efe1b}) is modified to
\be \label{efe1c}
r\phi'' + \frac{1}{2}\phi'\lb (\phi')^2 - 1\rb = \frac{3}{2}(\phi')^3\Lambda r^2\,,
\ee
whose solution is given by the radion field
\be
\phi(r) = 
\frac{1}{\sqrt{\Lambda}}\arcsin{\lp\sqrt{\Lambda}r\rp} + \phi_0\,, 
\ee
which corresponds, as expected, to the metric function
\be
g^2 = \frac{1}{1-\Lambda r^2}\,. 
\ee
Next, using the source (\ref{Lambda}) in the radial momentum constraint (\ref{efe2a}) gives us the modification of the equation (\ref{efe2b}) as
\be \label{efe2b}
rf' + \frac{1}{2}f +  \frac{1}{2}(\phi')^2\lb \alpha +  \lp 3\Lambda r^2  -1\rp f\rb = 0\,,
\ee 
which is solved by
\be
f = \frac{\alpha}{2}\lp \sqrt{1-\Lambda r^2}\coth^{-1}\sqrt{1-\Lambda r^2} - 1\rp + \beta\sqrt{1-\Lambda r^2}\,,
\ee
where $\beta$ is an integration constant. As expected, the behaviour of the solutions is similar to the previously studied case, and three distinct classes of solutions can be considered.
The case $\alpha = 0$, $\beta \neq 0$, reproduces the usual de Sitter solution. The situation is reversed wrt the previously studied case of the black hole, wherein the radion became imaginary inside the horizon. Now, a branch of solutions for the radion $\phi$ can be chosen such that $\phi(r_{dS})=0$ vanishes at the horizon $r=r_{dS}=1/\sqrt{\Lambda}$ and becomes pure imaginary beyond the horizon $r>r_{dS}$. The case $\alpha \neq 0$, $\beta = 0$ also covers the full manifold. In this case,  $f/\alpha$ diverges at the origin, and the metric function $f^2$ has a bounce at $r = r_0 \approx 0.552434/\sqrt{\Lambda}$, similar to the caustic singularity found in the vacuum case. At $r=r_{dS}$ we have then $f^2 = \alpha^2/4$, and $f^2$ keeps again growing logarithmically with increasing $r$. Finally, the case that both $\alpha$ and $\beta$ is non-zero is well-defined only inside the de Sitter radius $r<r_{dS}$. 

\section{On an alternative Lorentz gauge theory}
\label{alternativesec}

We note that a different approach to a Lorentz gauge theory of gravity has also been introduced in the literature \cite{Wiesendanger:2018dzw} and black hole solutions have been recently considered also in that context \cite{Wiesendanger:2024dnp}. A main motivation for Wiesendanger's SO(1,3) theory is renormalisability. Since the only field in the theory has the mass dimension one, it may be possible to compose actions from this field and its derivatives which have the claim both for being renormalisable and for reducing to GR in the relevant (low energy) limit  \cite{Wiesendanger:2020lwa}. As such actions must be extremely non-polynomial, the aim is not a fundamental pregeometric theory but an effective, renormalisable theory of gravity. Interestingly, in terms of the number of fields the formulation of the theory is even more economical than ours, since there is no additional scalar (like $\phi^a$) besides the Lorentz connection (our $\bomega^{ab}$).  

One can consider eliminating the field $\phi^a$ by replacing it with the spacetime coordinate, like $\phi^a=(t,x,y,z) \equiv x^a$. This was actually considered in \cite{Koivisto:2022uvd}, wherein it was shown that at least the Minkowski solution exists in our theory for such a symmetry breaking pattern. The Minkowski solution is inequivalent to the flat spacetime -limits of all the solutions we have considered in this article, since in both the khronon and the radion scenarios the Minkowski space is associated with curvature and torsion, whereas the Minkowski solution in the $\phi^a=x^a$ phase is geometrically trivial \cite{Koivisto:2022uvd}. In that phase, the coframe is composed as
\be \label{wtetrad}
\bbe^a = \bdelta^a + \bomega^a{}_b x^b\,,
\ee  
and the Minkowski solution is then simply $\bomega^a{}_b=0$. In the following we will check whether more general spherically symmetric solutions, in particular black hole solutions, exist in the symmetry-breaking phase $\phi^a = x^a$. We believe these would essentially correspond to solutions in Wiesendanger's Lorentz gauge theory. There are various technical differences with respect Wiesendanger's formulation\footnote{There the $\bbe^a$ is not Lorentz-covariant, since the $x^b$ in (\ref{wtetrad}) is a Lorentz scalar and actually, the similar construction was proposed for the frame (and not the coframe).} but these should not be essential to our conclusions. In our theory, (\ref{wtetrad}) holds when assuming $x^a$ as the value of a fundamental field after a spontaneous symmetry breaking, whereas in Wiesendanger's theory there is no field but (\ref{wtetrad}) could be taken as a coordinate-dependent expression in terms of which one can build suitable action invariants. Regardless of the interpretation, however, (\ref{wtetrad}) would give the desired spacetime geometry solely in terms of the Lorentz connection.   

An important consequence of (\ref{wtetrad}) is that $\bbe^a x_a = \bx$, as follows directly from the antisymmetry of the Lorentz adjoint $\bomega^{ab}=-\bomega^{ba}$ \cite{Wiesendanger:2018dzw}. Thus, a consistent solution requires a coframe with this property, and in the problem at hand we require also that the coframe corresponds to spherically symmetric metrical geometry.  The coframe deduced in Ref.\cite{Wiesendanger:2024dnp} from a more general Ansatz is (the reason for the tilde-notation will be explained shortly) 
\bs
\label{wframe} 
\ba
\bbe^0 & = & A\bdiff t + C\bdiff r\,, \\
\tilde{\bbe}^I & = & \lp A-1\rp\frac{t}{r^2}x^I\bdiff t + D\lp \bdiff x^I - \frac{x^I}{r}\bdiff r\rp + \lp 1 + C\frac{t}{r}\rp\frac{x^I}{r}\bdiff r\,, \label{wbartel}
\ea
\es
where the three functions $A$, $C$ and $D$ can depend on both $t$ and $r=\sqrt{x_I x^I}$. One can easily check that $\bbe^a x_a = \bx$ and that $\bbe^a\otimes\bbe_a$ is indeed a metric in a spherically symmetric form. Now going back to (\ref{wtetrad}) we find that the coframe (\ref{wframe}) would ensue from the connection
\bs
\label{wconnection}
\ba
\bomega^{0I} & = & -\bomega^{I0} = \frac{A-1}{3x^I}\bdiff t + \frac{C}{r}\bdiff x^I\,, \\
\tilde{\bomega}^{IJ} & = & \lp A-1\rp \frac{t}{3}\lp\frac{x^I}{x_J r^2} - \frac{1}{3x_I x_J}\rp\bdiff t 
+ \lp 1- D+ C\frac{t}{r}\rp\lp\frac{x^I}{r^2}\bdiff x^J - \frac{1}{3x_J}\bdiff x^I\rp\,.   
\ea
\es
The problem is that this is not a Lorentz connection but has non-metricity,
\be
\tilde{\bDiff} \eta^{IJ} = 2\tilde{\bomega}^{(IJ)} = \lp A-1\rp \frac{t}{3}\lb \frac{1}{r^2}\lp \frac{x^I}{x_J} + \frac{x^J}{x_I} \rp - \frac{2}{3x_I x_J}\rb\bdiff t
+ \lp 1-D+C\frac{t}{r}\rp\lp \frac{2}{r^2}x^{(I}\bdiff x^{J)} - \frac{1}{3x_I}\bdiff x^J - \frac{1}{3x_J}\bdiff x^I\rp\,. 
\ee
This produces a coframe with the desired property $\bbe^a x_a = \bx$ by virtue of $\tilde{\bomega}_{(IJ)}x^I x^J=0$. However, the gauge field is not valued in the special orthogonal but rather the general linear group since though  $\tilde{\bomega}_{(IJ)}x^I x^J=0$, nevertheless $\tilde{\bomega}_{(IJ)} \neq 0$. If we just erase the non-metricity and consider the coframe obtained by the antisymmetrisation $\bomega^{IJ} \equiv \tilde{\bomega}^{[IJ]}$ (the tilde is now removed since $\bomega^{ab}$ is by construction a metric-compatible s.t. $\bDiff\eta_{ab}=0$), the resulting Bartels frame would be quite different from (\ref{wbartel}), 
\be
\bbe^I = \frac{1}{2}\lp A-1\rp t\lp \frac{x^I}{r^2} - \frac{1}{3x_I}\rp  \bdiff t + \lp D + C\frac{t}{r}\rp\bdiff x^I +  \frac{1}{2}\lp 1-D+C\frac{t}{r}\rp\lp \frac{x^I}{r}-\frac{r}{3x_I}\rp\bdiff r\,.  
\ee
This no longer describes a spherically symmetric geometry. For instance, the metric lapse function would read
\bs
\be
(\bbe^a\otimes\bbe_a)_{00} = -A^2 + \frac{1}{4}\lp A-1\rp^2 t^2 \lp \frac{1}{9x^2} + \frac{1}{9y^2} + \frac{1}{9z^2} -\frac{1}{r^2}\rp\,.   
\ee 
By fixing $A=1$, this can be made compatible with spherical symmetry, and then the metric shift function passes as well since it simplifies to
\be
(\bbe^a\otimes\bbe_a)_{0I} = -AC\frac{x_I}{r}\,.
\ee
However, the spatial components of the metric,
\ba
(\bbe^a\otimes\bbe_a)_{IJ} & = &  \lp D + C\frac{t}{r}\rp^2\delta_{IJ} +  \lp D + C\frac{t}{r}\rp\lp 1-D +  C\frac{t}{r}\rp\lp \frac{x_I x_J}{r^2}- \frac{x_I}{6x^J} - \frac{x_J}{6x^I}\rp \nn \\
& + & \lb \frac{1}{4}\lp 1-D +  C\frac{t}{r}\rp\lp  \frac{1}{9x^2} + \frac{1}{9y^2} + \frac{1}{9z^2} -\frac{1}{r^2}\rp - C^2\rb \frac{x_I x_J}{r^2}\,, 
\ea
\es
require us to set $D=1-Ct/r$. So, we are left with only one free function describing the geometry whose metric reduces to
\be
\bdiff s^2 = -\bdiff t^2 - 2C\bdiff t\bdiff r + \lp 1 - C^2 \rp \bdiff r^2 + r^2\lp\bdiff\theta^2 + \sin^2\theta\bdiff\varphi^2\rp\,. 
\ee
Redefining $\bdiff t \rightarrow \bdiff t + C\bdiff r$, we confirm that this is nothing but a Minkowski space. We conclude that the Minkowski space is the only spherically symmetric solution in the alternative Lorentz gauge theory (whilst dynamical spacetimes could perhaps be constructed in a GL(4) generalisation of the SO(1,3) gauge theory using the connection (\ref{wconnection})). On the other hand,  the symmetry breaking by $\phi^a=x^a$ does not lead to a viable phase of our SO(1,3) theory, which suggests that in the fundamental pregeometric theory the emergence of (even the Minkowski) spacetime requires non-trivial both torsion and curvature.  


\section{Conclusions}
\label{conclusionsec}

This article was an exploration of black holes in the Lorentz gauge theory. We discovered the first static spherically symmetric solutions existing in the fundamental SO(1,3) theory of spacetime and gravity, and by incorporating the U(1) interaction arrived at the first explicit solutions of the pregeometric first order electromagnetic theory.  

In the Lorentz gauge theory, the nature of spacetime is characterised by the symmetry-breaking field in the fundamental representation of the group. We considered the possibility of accommodating a black hole in three distinct kinds of spacetimes.   
\begin{itemize}
\item[-] Section \ref{sphericalsec}: in the synchronous phase, the {\it khronon} $\tau^a$ is a clock field. 
\item[-] Section \ref{radialsec}: in the SO(1,2) symmetry-breaking phase, the {\it radion} $\phi^a$ is space-like.
\item[-] Section \ref{alternativesec}: an alternative approach is to regard $x^a$ as the (Poincar{\'e}) {\it coordinate}. 
\end{itemize}
In the first case, the component $\tau^0=\tau$ is restricted to a stealth scalar field which does not contribute an effective energy density itself ($M=0$), but it turns out that the role of $\tau$ is to measure the proper time of a freely-falling time-like observer. The vacuum solution has two parameters, one which quantifies the mass of the black hole and one which quantifies the state of the observer. Whilst the solution for $\tau$ coincides with the time coordinates of the well-known Painlev{\'e}-Gullstrand and Lema{\^i}tre charts, the khronon is not an arbitrary coordinate but an elementary field of the theory, arguably realising the Cartan-geometric concept of ``lifting GR to Observer Space''  \cite{Gielen:2012fz} in an unprecedented depth.  The construction can be interesting to extend beyond spherical symmetry, and known results in GR suggest that it might be carried out at least for Kerr \cite{Doran:1999gb} and various other spacetimes \cite{Baines:2021jbj}. -The two alternative symmetry-breaking phases of the theory are curiosities whose possible physical relevance is less clear\footnote{Though not considered in this article, the theory admits also the null $\phi^2 =0$ phase giving rise to (when $\phi^a \neq 0$) a 2-dimensional space(time). An unexplored aspect also is a possible relation to phases of superfluids suggested by a formal analogy between the emergent coframe and the $\prescript{3}{}{\text{He}}$ order parameter, e.g. \cite{Makinen:2018ltj,Volovik:2020lwn}.}. The radion symmetry breaking with $M=0$ yields a rather neat description of the Schwarzschild black hole wherein the radion $\phi^a$ becomes imaginary beyond the horizon. When $M \neq 0$, exotic solutions emerge which are not asymptotically flat and which exhibit additional singularities, previously found in the context of mimetic gravity \cite{Gorji:2020ten,Khodadi:2024ubi}. The conclusion from the third case, employing the (Poincar{\'e}) coordinate parameterisation, is simply a no-go theorem. 

Whereas the above discussion concerned solutions in the different phases of the same gravitational theory, in Section \ref{electrosec} we studied the same Reissner-Nordstr{\"om} solution in three different versions of a pregeometric first order electromagnetic theory. The derivative order is reduced via introduction of a new field. 
\begin{itemize}
\item[-] Action (\ref{electroaction}): the field is a 16-component {\it hypercoframe} $\bh^a$.  
\item[-] Action (\ref{electroaction2}): the field is a 6-component {\it excitation} $h^{ab}$.
\item[-] Action (\ref{electroaction3}): the field is a 4-component {\it iso-khronon} $h^a$.
\end{itemize}
Our conclusion was that the satisfactory formulation is given in terms of the excitation $h^{ab}$ valued in the adjoint of the Lorentz group. As one expects, the dual of the scalar $\star h^{ab}$ corresponds to the Faraday tensor 
$\bF$ and the scalar $h^{ab}$ corresponds to the Hodge dual $\ast \bF$. The canonical Noether charges are given as the surface integrals $\oint h_{ab}\bDiff\phi^a\wedge\bDiff\phi^b$
\cite{Gallagher:2023ghl}. Evaluating this for our Reissner-Nordstr{\"om} solution, we get that the charge in an arbitrary spatial volume containing the singularity is equal to the electric parameter $q_e$. The magnetic parameter $q_b$ does not contribute to the Noether charge enclosed by a spatial surface, which provides one perspective to the asymmetry between an electric charge and a magnetic monopole. 

Finally, let us recapitulate some of the ``negative'' results of our explorations.
\begin{itemize}
\item The absence of non-singular solutions. All of the black hole solutions we found possess at least the $r=0$ singularity. Not only the (composite) spacetime metric, but also the fundamental fields exhibit divergent behaviour, and thus we cannot claim any black hole analogue of the non-singular ``khronogenetic'' cosmology.  
\item The absence of Reissner-Nordstr{\"o}m solutions\footnote{More precisely, we showed that the theory is not compatible with spherically symmetric electromagnetic field without a nontrivial vacuum excitation, and one can easily check that this conclusion holds in the approximation of neglecting gravity. We did not exclude the possibility of a 
Reissner-Nordstr{\"o}m spacetime geometry supported by a non-trivial vacuum excitation.} in the ``iso-khronon'' gauge theory. Since that version of the first order gauge theory cannot reproduce the static spherically symmetric field configuration mimicking a Coulomb field, the more general conclusion is that it is not a viable alternative to the standard Maxwell electromagnetism. 
\item The absence of Schwarzschild solutions in the alternative formulation of Lorentz gauge theory. The identification of the field $\phi^a$ with the spacetime coordinates is possible in the Minkowski spacetime, but not in a black hole spacetime. This might have repercussions in an alternative Lorentz gauge theory. 
\end{itemize}
The two latter ``negative'' results corroborate the claim for the uniqueness of the consistent first order action functional for gauge theories \cite{Gallagher:2023ghl}, pertaining, respectively, to the particle interactions and to the gravitational interactions. Both of the respective alternatives are too simple, in the sense that they reduce the field content too much to be able to still correctly capture the required physics. We demonstrated that the excitation $h^{ab}$ cannot be consistently reduced to the iso-khronon $h^a$ in pregeometric electromagnetism, and that the khronon $\phi$ cannot be simply replaced by the coordinate $x$ in pregeometric gravity. 

On the other hand, the first ``negative'' result points towards further investigations. We have mentioned the two complementary aspects of the theory calling for clarification, geometrical and topological. Quantum effects become important at extremely high curvatures, and could possibly result in the regularisation of geometry. The Bardeen and Hayward metrics are prototype regular black holes \cite{Hayward:2005gi}, and some such solutions which have relations to polymerized models and mimetic gravity \cite{Giesel:2024mps} could potentially be relevant in an ultraviolet extension of the Lorentz gauge theory. The problem of caustic singularities in mimetic gravity i.e. the rotationless dust model resurface in the Lorentz gauge theory when $M \neq 0$, which calls for a completion of the theory in the ultraviolet. We expect that the conformal symmetry of the more final unified gauge theory \cite{Koivisto:2019ejt} (whose twistorial structure as sketched in Section \ref{action} is reflected already in the current form of the theory) will be restored near the Planck scale, and hypothetically smoothen both the caustic and the black hole singularities.  

The complementary problem concerns not the regularisation of the geometry but its interpretation. One recent proposal is to avoid the black hole singularities in terms of {\it atemporality} \cite{Capozziello:2024ucm}. A dynamical mechanism was constructed for Lorentzian-Euclidean black holes such that the time coordinate turns imaginary inside the horizon, and the resolution of singularity was based on the non-prolongability of real-valued geodesics and accelerated orbits beyond the horizon. Whereas the construction of Ref.\cite{Capozziello:2024ucm} resorted to involved gluing and regularisation techniques, in Section \ref{radialsec} we witnessed the radion's analogous real-to-imaginary transition at the horizon as the straightforward prediction of the fundamental gauge theory in its SO(1,2) phase. Clearly, the Lorentz gauge theory would offer the natural framework for atemporality. In fact, as we recalled in the context of the Reissner-Nordstr{\"o}m solution, already standard GR hints at the possibility of the imaginary unit shielding from the singularity. An orbit towards a charged black hole hits such a barrier in coordinates adapted to the proper time along the orbit, and actually, even before that (regardless of the coordinates) the situation has become unclear. After crossing the inner Cauchy horizon, the orbit could in principle bounce back and enter into an entirely different region of spacetime outside both the horizons.       

Indeed, we have but initiated investigations into black hole solutions in the Lorentz gauge theory of gravity by demonstrating the existence of such solutions.
The maximal extension of the Schwarzschild metric has been found long since \cite{PhysRev.119.1743}, but to date its physical relevance remains completely obscure. It is fascinating to contemplate whether the chirality of spacetime in the Lorentz gauge theory could shed light on the potential reality of the regions III and IV in the Penrose-Carter diagram. 

\begin{acknowledgments}
This work was supported by the Estonian Research Council grant TK202 Center of Excellence ``Foundations of the Universe'', and benefitted from discussions with
P. Gallagher, L. Marzola and T. Z\l{}o\'snik. 
\end{acknowledgments}

\appendix 

\bibliography{KhrononBH}

\begin{thebibliography}{84}%
\makeatletter
\providecommand \@ifxundefined [1]{%
 \@ifx{#1\undefined}
}%
\providecommand \@ifnum [1]{%
 \ifnum #1\expandafter \@firstoftwo
 \else \expandafter \@secondoftwo
 \fi
}%
\providecommand \@ifx [1]{%
 \ifx #1\expandafter \@firstoftwo
 \else \expandafter \@secondoftwo
 \fi
}%
\providecommand \natexlab [1]{#1}%
\providecommand \enquote  [1]{``#1''}%
\providecommand \bibnamefont  [1]{#1}%
\providecommand \bibfnamefont [1]{#1}%
\providecommand \citenamefont [1]{#1}%
\providecommand \href@noop [0]{\@secondoftwo}%
\providecommand \href [0]{\begingroup \@sanitize@url \@href}%
\providecommand \@href[1]{\@@startlink{#1}\@@href}%
\providecommand \@@href[1]{\endgroup#1\@@endlink}%
\providecommand \@sanitize@url [0]{\catcode `\\12\catcode `\$12\catcode
  `\&12\catcode `\#12\catcode `\^12\catcode `\_12\catcode `\%12\relax}%
\providecommand \@@startlink[1]{}%
\providecommand \@@endlink[0]{}%
\providecommand \url  [0]{\begingroup\@sanitize@url \@url }%
\providecommand \@url [1]{\endgroup\@href {#1}{\urlprefix }}%
\providecommand \urlprefix  [0]{URL }%
\providecommand \Eprint [0]{\href }%
\providecommand \doibase [0]{http://dx.doi.org/}%
\providecommand \selectlanguage [0]{\@gobble}%
\providecommand \bibinfo  [0]{\@secondoftwo}%
\providecommand \bibfield  [0]{\@secondoftwo}%
\providecommand \translation [1]{[#1]}%
\providecommand \BibitemOpen [0]{}%
\providecommand \bibitemStop [0]{}%
\providecommand \bibitemNoStop [0]{.\EOS\space}%
\providecommand \EOS [0]{\spacefactor3000\relax}%
\providecommand \BibitemShut  [1]{\csname bibitem#1\endcsname}%
\let\auto@bib@innerbib\@empty
\bibitem [{\citenamefont {Taylor}\ and\ \citenamefont
  {Wheeler}(2001)}]{taylor}%
  \BibitemOpen
  \bibfield  {author} {\bibinfo {author} {\bibfnamefont {E.}~\bibnamefont
  {Taylor}}\ and\ \bibinfo {author} {\bibfnamefont {J.}~\bibnamefont
  {Wheeler}},\ }\href@noop {} {\emph {\bibinfo {title} {Exploring Black Holes:
  Introduction to General Relativity}}}\ (\bibinfo  {publisher}
  {Addison-Wesley},\ \bibinfo {address} {Boston},\ \bibinfo {year}
  {2001})\BibitemShut {NoStop}%
\bibitem [{\citenamefont {Sedda}\ \emph {et~al.}(2024)\citenamefont {Sedda},
  \citenamefont {Bortolas},\ and\ \citenamefont {Spera}}]{Sedda:2024duh}%
  \BibitemOpen
  \bibinfo {editor} {\bibfnamefont {M.~A.}\ \bibnamefont {Sedda}}, \bibinfo
  {editor} {\bibfnamefont {E.}~\bibnamefont {Bortolas}}, \ and\ \bibinfo
  {editor} {\bibfnamefont {M.}~\bibnamefont {Spera}},\ eds.,\ \href@noop {}
  {\emph {\bibinfo {title} {{Black Holes in the Era of Gravitational-Wave
  Astronomy}}}}\ (\bibinfo  {publisher} {Elsevier},\ \bibinfo {address}
  {Amsterdam},\ \bibinfo {year} {2024})\BibitemShut {NoStop}%
\bibitem [{\citenamefont {Di~Valentino}\ \emph {et~al.}(2024)\citenamefont
  {Di~Valentino}, \citenamefont {Perivolaropoulos},\ and\ \citenamefont
  {Levi~Said}}]{DiValentino:2024wgi}%
  \BibitemOpen
  \bibfield  {author} {\bibinfo {author} {\bibfnamefont {E.}~\bibnamefont
  {Di~Valentino}}, \bibinfo {author} {\bibfnamefont {L.}~\bibnamefont
  {Perivolaropoulos}}, \ and\ \bibinfo {author} {\bibfnamefont
  {J.}~\bibnamefont {Levi~Said}},\ }\href {\doibase 10.3390/universe10040184}
  {\  (\bibinfo {year} {2024}),\ 10.3390/universe10040184},\ \Eprint
  {http://arxiv.org/abs/2404.13981} {arXiv:2404.13981 [gr-qc]} \BibitemShut
  {NoStop}%
\bibitem [{\citenamefont {Akrami}\ \emph {et~al.}(2021)\citenamefont {Akrami}
  \emph {et~al.}}]{CANTATA:2021ktz}%
  \BibitemOpen
  \bibfield  {author} {\bibinfo {author} {\bibfnamefont {Y.}~\bibnamefont
  {Akrami}} \emph {et~al.} (\bibinfo {collaboration} {CANTATA}),\ }\href
  {\doibase 10.1007/978-3-030-83715-0} {\emph {\bibinfo {title} {{Modified
  Gravity and Cosmology}: {An Update by the CANTATA Network}}}},\ edited by\
  \bibinfo {editor} {\bibfnamefont {E.~N.}\ \bibnamefont {Saridakis}}, \bibinfo
  {editor} {\bibfnamefont {R.}~\bibnamefont {Lazkoz}}, \bibinfo {editor}
  {\bibfnamefont {V.}~\bibnamefont {Salzano}}, \bibinfo {editor} {\bibfnamefont
  {P.}~\bibnamefont {Vargas~Moniz}}, \bibinfo {editor} {\bibfnamefont
  {S.}~\bibnamefont {Capozziello}}, \bibinfo {editor} {\bibfnamefont
  {J.}~\bibnamefont {Beltr\'an~Jim\'enez}}, \bibinfo {editor} {\bibfnamefont
  {M.}~\bibnamefont {De~Laurentis}}, \ and\ \bibinfo {editor} {\bibfnamefont
  {G.~J.}\ \bibnamefont {Olmo}}\ (\bibinfo  {publisher} {Springer},\ \bibinfo
  {year} {2021})\ \Eprint {http://arxiv.org/abs/2105.12582} {arXiv:2105.12582
  [gr-qc]} \BibitemShut {NoStop}%
\bibitem [{\citenamefont {Krasnov}(2020)}]{Krasnov:2020lku}%
  \BibitemOpen
  \bibfield  {author} {\bibinfo {author} {\bibfnamefont {K.}~\bibnamefont
  {Krasnov}},\ }\href {\doibase 10.1017/9781108674652} {\emph {\bibinfo {title}
  {{Formulations of General Relativity}}}},\ Cambridge Monographs on
  Mathematical Physics\ (\bibinfo  {publisher} {Cambridge University Press},\
  \bibinfo {year} {2020})\BibitemShut {NoStop}%
\bibitem [{\citenamefont {Plebanski}(1977)}]{Plebanski:1977zz}%
  \BibitemOpen
  \bibfield  {author} {\bibinfo {author} {\bibfnamefont {J.~F.}\ \bibnamefont
  {Plebanski}},\ }\href {\doibase 10.1063/1.523215} {\bibfield  {journal}
  {\bibinfo  {journal} {J. Math. Phys.}\ }\textbf {\bibinfo {volume} {18}},\
  \bibinfo {pages} {2511} (\bibinfo {year} {1977})}\BibitemShut {NoStop}%
\bibitem [{\citenamefont {Engle}\ and\ \citenamefont
  {Speziale}(2023)}]{Engle:2023qsu}%
  \BibitemOpen
  \bibfield  {author} {\bibinfo {author} {\bibfnamefont {J.}~\bibnamefont
  {Engle}}\ and\ \bibinfo {author} {\bibfnamefont {S.}~\bibnamefont
  {Speziale}},\ }\enquote {\bibinfo {title} {{Spin Foams: Foundations}},}\ \
  (\bibinfo {year} {2023})\ \Eprint {http://arxiv.org/abs/2310.20147}
  {arXiv:2310.20147 [gr-qc]} \BibitemShut {NoStop}%
\bibitem [{\citenamefont {Krasnov}(2011)}]{Krasnov:2011pp}%
  \BibitemOpen
  \bibfield  {author} {\bibinfo {author} {\bibfnamefont {K.}~\bibnamefont
  {Krasnov}},\ }\href {\doibase 10.1103/PhysRevLett.106.251103} {\bibfield
  {journal} {\bibinfo  {journal} {Phys. Rev. Lett.}\ }\textbf {\bibinfo
  {volume} {106}},\ \bibinfo {pages} {251103} (\bibinfo {year} {2011})},\
  \Eprint {http://arxiv.org/abs/1103.4498} {arXiv:1103.4498 [gr-qc]}
  \BibitemShut {NoStop}%
\bibitem [{\citenamefont {Capovilla}\ \emph {et~al.}(1991)\citenamefont
  {Capovilla}, \citenamefont {Jacobson}, \citenamefont {Dell},\ and\
  \citenamefont {Mason}}]{Capovilla:1991qb}%
  \BibitemOpen
  \bibfield  {author} {\bibinfo {author} {\bibfnamefont {R.}~\bibnamefont
  {Capovilla}}, \bibinfo {author} {\bibfnamefont {T.}~\bibnamefont {Jacobson}},
  \bibinfo {author} {\bibfnamefont {J.}~\bibnamefont {Dell}}, \ and\ \bibinfo
  {author} {\bibfnamefont {L.~J.}\ \bibnamefont {Mason}},\ }\href {\doibase
  10.1088/0264-9381/8/1/009} {\bibfield  {journal} {\bibinfo  {journal} {Class.
  Quant. Grav.}\ }\textbf {\bibinfo {volume} {8}},\ \bibinfo {pages} {41}
  (\bibinfo {year} {1991})}\BibitemShut {NoStop}%
\bibitem [{\citenamefont {Koivisto}\ and\ \citenamefont
  {Zlosnik}(2023)}]{Koivisto:2022uvd}%
  \BibitemOpen
  \bibfield  {author} {\bibinfo {author} {\bibfnamefont {T.~S.}\ \bibnamefont
  {Koivisto}}\ and\ \bibinfo {author} {\bibfnamefont {T.}~\bibnamefont
  {Zlosnik}},\ }\href {\doibase 10.1103/PhysRevD.107.124013} {\bibfield
  {journal} {\bibinfo  {journal} {Phys. Rev. D}\ }\textbf {\bibinfo {volume}
  {107}},\ \bibinfo {pages} {124013} (\bibinfo {year} {2023})},\ \Eprint
  {http://arxiv.org/abs/2212.04562} {arXiv:2212.04562 [gr-qc]} \BibitemShut
  {NoStop}%
\bibitem [{\citenamefont {Z\l{}o\'snik}\ \emph {et~al.}(2018)\citenamefont
  {Z\l{}o\'snik}, \citenamefont {Urban}, \citenamefont {Marzola},\ and\
  \citenamefont {Koivisto}}]{Zlosnik:2018qvg}%
  \BibitemOpen
  \bibfield  {author} {\bibinfo {author} {\bibfnamefont {T.}~\bibnamefont
  {Z\l{}o\'snik}}, \bibinfo {author} {\bibfnamefont {F.}~\bibnamefont {Urban}},
  \bibinfo {author} {\bibfnamefont {L.}~\bibnamefont {Marzola}}, \ and\
  \bibinfo {author} {\bibfnamefont {T.}~\bibnamefont {Koivisto}},\ }\href
  {\doibase 10.1088/1361-6382/aaea96} {\bibfield  {journal} {\bibinfo
  {journal} {Class. Quant. Grav.}\ }\textbf {\bibinfo {volume} {35}},\ \bibinfo
  {pages} {235003} (\bibinfo {year} {2018})},\ \Eprint
  {http://arxiv.org/abs/1807.01100} {arXiv:1807.01100 [gr-qc]} \BibitemShut
  {NoStop}%
\bibitem [{\citenamefont {Gallagher}\ \emph {et~al.}(2024)\citenamefont
  {Gallagher}, \citenamefont {Koivisto}, \citenamefont {Marzola}, \citenamefont
  {Varrin},\ and\ \citenamefont {Zlosnik}}]{Gallagher:2023ghl}%
  \BibitemOpen
  \bibfield  {author} {\bibinfo {author} {\bibfnamefont {P.}~\bibnamefont
  {Gallagher}}, \bibinfo {author} {\bibfnamefont {T.~S.}\ \bibnamefont
  {Koivisto}}, \bibinfo {author} {\bibfnamefont {L.}~\bibnamefont {Marzola}},
  \bibinfo {author} {\bibfnamefont {L.}~\bibnamefont {Varrin}}, \ and\ \bibinfo
  {author} {\bibfnamefont {T.}~\bibnamefont {Zlosnik}},\ }\href {\doibase
  10.1103/PhysRevD.109.L061503} {\bibfield  {journal} {\bibinfo  {journal}
  {Phys. Rev. D}\ }\textbf {\bibinfo {volume} {109}},\ \bibinfo {pages}
  {L061503} (\bibinfo {year} {2024})},\ \Eprint
  {http://arxiv.org/abs/2311.07464} {arXiv:2311.07464 [hep-th]} \BibitemShut
  {NoStop}%
\bibitem [{\citenamefont {Nikjoo}\ and\ \citenamefont
  {Zlosnik}(2024)}]{Nikjoo:2023flm}%
  \BibitemOpen
  \bibfield  {author} {\bibinfo {author} {\bibfnamefont {M.}~\bibnamefont
  {Nikjoo}}\ and\ \bibinfo {author} {\bibfnamefont {T.}~\bibnamefont
  {Zlosnik}},\ }\href {\doibase 10.1088/1361-6382/ad1c84} {\bibfield  {journal}
  {\bibinfo  {journal} {Class. Quant. Grav.}\ }\textbf {\bibinfo {volume}
  {41}},\ \bibinfo {pages} {045005} (\bibinfo {year} {2024})},\ \Eprint
  {http://arxiv.org/abs/2308.01108} {arXiv:2308.01108 [gr-qc]} \BibitemShut
  {NoStop}%
\bibitem [{\citenamefont {Gallagher}\ and\ \citenamefont
  {Koivisto}(2021)}]{Gallagher:2021tgx}%
  \BibitemOpen
  \bibfield  {author} {\bibinfo {author} {\bibfnamefont {P.}~\bibnamefont
  {Gallagher}}\ and\ \bibinfo {author} {\bibfnamefont {T.}~\bibnamefont
  {Koivisto}},\ }\href {\doibase 10.3390/sym13112076} {\bibfield  {journal}
  {\bibinfo  {journal} {Symmetry}\ }\textbf {\bibinfo {volume} {13}},\ \bibinfo
  {pages} {2076} (\bibinfo {year} {2021})},\ \Eprint
  {http://arxiv.org/abs/2103.05435} {arXiv:2103.05435 [gr-qc]} \BibitemShut
  {NoStop}%
\bibitem [{\citenamefont {Koivisto}(2023)}]{Koivisto:2023epd}%
  \BibitemOpen
  \bibfield  {author} {\bibinfo {author} {\bibfnamefont {T.}~\bibnamefont
  {Koivisto}},\ }\href {\doibase 10.1142/S0219887824500403} {\bibfield
  {journal} {\bibinfo  {journal} {Int. J. Geom. Meth. Mod. Phys.}\ }\textbf
  {\bibinfo {volume} {20}},\ \bibinfo {pages} {2450040} (\bibinfo {year}
  {2023})},\ \Eprint {http://arxiv.org/abs/2306.00963} {arXiv:2306.00963
  [gr-qc]} \BibitemShut {NoStop}%
\bibitem [{\citenamefont {Brown}\ and\ \citenamefont
  {Kuchar}(1995)}]{Brown:1994py}%
  \BibitemOpen
  \bibfield  {author} {\bibinfo {author} {\bibfnamefont {J.~D.}\ \bibnamefont
  {Brown}}\ and\ \bibinfo {author} {\bibfnamefont {K.~V.}\ \bibnamefont
  {Kuchar}},\ }\href {\doibase 10.1103/PhysRevD.51.5600} {\bibfield  {journal}
  {\bibinfo  {journal} {Phys. Rev. D}\ }\textbf {\bibinfo {volume} {51}},\
  \bibinfo {pages} {5600} (\bibinfo {year} {1995})},\ \Eprint
  {http://arxiv.org/abs/gr-qc/9409001} {arXiv:gr-qc/9409001} \BibitemShut
  {NoStop}%
\bibitem [{\citenamefont {Husain}\ and\ \citenamefont
  {Pawlowski}(2011)}]{Husain:2011tm}%
  \BibitemOpen
  \bibfield  {author} {\bibinfo {author} {\bibfnamefont {V.}~\bibnamefont
  {Husain}}\ and\ \bibinfo {author} {\bibfnamefont {T.}~\bibnamefont
  {Pawlowski}},\ }\href {\doibase 10.1088/0264-9381/28/22/225014} {\bibfield
  {journal} {\bibinfo  {journal} {Class. Quant. Grav.}\ }\textbf {\bibinfo
  {volume} {28}},\ \bibinfo {pages} {225014} (\bibinfo {year} {2011})},\
  \Eprint {http://arxiv.org/abs/1108.1147} {arXiv:1108.1147 [gr-qc]}
  \BibitemShut {NoStop}%
\bibitem [{\citenamefont {Giesel}\ and\ \citenamefont
  {Thiemann}(2015)}]{Giesel:2012rb}%
  \BibitemOpen
  \bibfield  {author} {\bibinfo {author} {\bibfnamefont {K.}~\bibnamefont
  {Giesel}}\ and\ \bibinfo {author} {\bibfnamefont {T.}~\bibnamefont
  {Thiemann}},\ }\href {\doibase 10.1088/0264-9381/32/13/135015} {\bibfield
  {journal} {\bibinfo  {journal} {Class. Quant. Grav.}\ }\textbf {\bibinfo
  {volume} {32}},\ \bibinfo {pages} {135015} (\bibinfo {year} {2015})},\
  \Eprint {http://arxiv.org/abs/1206.3807} {arXiv:1206.3807 [gr-qc]}
  \BibitemShut {NoStop}%
\bibitem [{\citenamefont {Magueijo}(2024)}]{Magueijo:2024zxz}%
  \BibitemOpen
  \bibfield  {author} {\bibinfo {author} {\bibfnamefont {J.}~\bibnamefont
  {Magueijo}},\ }\href@noop {} {\  (\bibinfo {year} {2024})},\ \Eprint
  {http://arxiv.org/abs/2404.15809} {arXiv:2404.15809 [hep-th]} \BibitemShut
  {NoStop}%
\bibitem [{\citenamefont {Kaplan}\ \emph {et~al.}(2023)\citenamefont {Kaplan},
  \citenamefont {Melia},\ and\ \citenamefont {Rajendran}}]{Kaplan:2023wyw}%
  \BibitemOpen
  \bibfield  {author} {\bibinfo {author} {\bibfnamefont {D.~E.}\ \bibnamefont
  {Kaplan}}, \bibinfo {author} {\bibfnamefont {T.}~\bibnamefont {Melia}}, \
  and\ \bibinfo {author} {\bibfnamefont {S.}~\bibnamefont {Rajendran}},\
  }\href@noop {} {\  (\bibinfo {year} {2023})},\ \Eprint
  {http://arxiv.org/abs/2305.01798} {arXiv:2305.01798 [hep-th]} \BibitemShut
  {NoStop}%
\bibitem [{\citenamefont {Casadio}\ \emph {et~al.}(2024)\citenamefont
  {Casadio}, \citenamefont {Chataignier}, \citenamefont {Kamenshchik},
  \citenamefont {Pedro}, \citenamefont {Tronconi},\ and\ \citenamefont
  {Venturi}}]{Casadio:2024bdb}%
  \BibitemOpen
  \bibfield  {author} {\bibinfo {author} {\bibfnamefont {R.}~\bibnamefont
  {Casadio}}, \bibinfo {author} {\bibfnamefont {L.}~\bibnamefont
  {Chataignier}}, \bibinfo {author} {\bibfnamefont {A.~Y.}\ \bibnamefont
  {Kamenshchik}}, \bibinfo {author} {\bibfnamefont {F.~G.}\ \bibnamefont
  {Pedro}}, \bibinfo {author} {\bibfnamefont {A.}~\bibnamefont {Tronconi}}, \
  and\ \bibinfo {author} {\bibfnamefont {G.}~\bibnamefont {Venturi}},\
  }\href@noop {} {\  (\bibinfo {year} {2024})},\ \Eprint
  {http://arxiv.org/abs/2402.12437} {arXiv:2402.12437 [gr-qc]} \BibitemShut
  {NoStop}%
\bibitem [{\citenamefont {Jacobson}\ and\ \citenamefont
  {Mattingly}(2001)}]{Jacobson:2000xp}%
  \BibitemOpen
  \bibfield  {author} {\bibinfo {author} {\bibfnamefont {T.}~\bibnamefont
  {Jacobson}}\ and\ \bibinfo {author} {\bibfnamefont {D.}~\bibnamefont
  {Mattingly}},\ }\href {\doibase 10.1103/PhysRevD.64.024028} {\bibfield
  {journal} {\bibinfo  {journal} {Phys. Rev. D}\ }\textbf {\bibinfo {volume}
  {64}},\ \bibinfo {pages} {024028} (\bibinfo {year} {2001})},\ \Eprint
  {http://arxiv.org/abs/gr-qc/0007031} {arXiv:gr-qc/0007031} \BibitemShut
  {NoStop}%
\bibitem [{\citenamefont {Arkani-Hamed}\ \emph {et~al.}(2004)\citenamefont
  {Arkani-Hamed}, \citenamefont {Cheng}, \citenamefont {Luty},\ and\
  \citenamefont {Mukohyama}}]{Arkani-Hamed:2003pdi}%
  \BibitemOpen
  \bibfield  {author} {\bibinfo {author} {\bibfnamefont {N.}~\bibnamefont
  {Arkani-Hamed}}, \bibinfo {author} {\bibfnamefont {H.-C.}\ \bibnamefont
  {Cheng}}, \bibinfo {author} {\bibfnamefont {M.~A.}\ \bibnamefont {Luty}}, \
  and\ \bibinfo {author} {\bibfnamefont {S.}~\bibnamefont {Mukohyama}},\ }\href
  {\doibase 10.1088/1126-6708/2004/05/074} {\bibfield  {journal} {\bibinfo
  {journal} {JHEP}\ }\textbf {\bibinfo {volume} {05}},\ \bibinfo {pages} {074}
  (\bibinfo {year} {2004})},\ \Eprint {http://arxiv.org/abs/hep-th/0312099}
  {arXiv:hep-th/0312099} \BibitemShut {NoStop}%
\bibitem [{\citenamefont {Blas}\ \emph {et~al.}(2010)\citenamefont {Blas},
  \citenamefont {Pujolas},\ and\ \citenamefont {Sibiryakov}}]{Blas:2009qj}%
  \BibitemOpen
  \bibfield  {author} {\bibinfo {author} {\bibfnamefont {D.}~\bibnamefont
  {Blas}}, \bibinfo {author} {\bibfnamefont {O.}~\bibnamefont {Pujolas}}, \
  and\ \bibinfo {author} {\bibfnamefont {S.}~\bibnamefont {Sibiryakov}},\
  }\href {\doibase 10.1103/PhysRevLett.104.181302} {\bibfield  {journal}
  {\bibinfo  {journal} {Phys. Rev. Lett.}\ }\textbf {\bibinfo {volume} {104}},\
  \bibinfo {pages} {181302} (\bibinfo {year} {2010})},\ \Eprint
  {http://arxiv.org/abs/0909.3525} {arXiv:0909.3525 [hep-th]} \BibitemShut
  {NoStop}%
\bibitem [{\citenamefont {Lim}\ \emph {et~al.}(2010)\citenamefont {Lim},
  \citenamefont {Sawicki},\ and\ \citenamefont {Vikman}}]{Lim:2010yk}%
  \BibitemOpen
  \bibfield  {author} {\bibinfo {author} {\bibfnamefont {E.~A.}\ \bibnamefont
  {Lim}}, \bibinfo {author} {\bibfnamefont {I.}~\bibnamefont {Sawicki}}, \ and\
  \bibinfo {author} {\bibfnamefont {A.}~\bibnamefont {Vikman}},\ }\href
  {\doibase 10.1088/1475-7516/2010/05/012} {\bibfield  {journal} {\bibinfo
  {journal} {JCAP}\ }\textbf {\bibinfo {volume} {05}},\ \bibinfo {pages} {012}
  (\bibinfo {year} {2010})},\ \Eprint {http://arxiv.org/abs/1003.5751}
  {arXiv:1003.5751 [astro-ph.CO]} \BibitemShut {NoStop}%
\bibitem [{\citenamefont {Blanchet}\ and\ \citenamefont
  {Skordis}(2024)}]{Blanchet:2024mvy}%
  \BibitemOpen
  \bibfield  {author} {\bibinfo {author} {\bibfnamefont {L.}~\bibnamefont
  {Blanchet}}\ and\ \bibinfo {author} {\bibfnamefont {C.}~\bibnamefont
  {Skordis}},\ }\href@noop {} {\  (\bibinfo {year} {2024})},\ \Eprint
  {http://arxiv.org/abs/2404.06584} {arXiv:2404.06584 [gr-qc]} \BibitemShut
  {NoStop}%
\bibitem [{\citenamefont {Izumi}\ and\ \citenamefont
  {Mukohyama}(2010)}]{Izumi:2009ry}%
  \BibitemOpen
  \bibfield  {author} {\bibinfo {author} {\bibfnamefont {K.}~\bibnamefont
  {Izumi}}\ and\ \bibinfo {author} {\bibfnamefont {S.}~\bibnamefont
  {Mukohyama}},\ }\href {\doibase 10.1103/PhysRevD.81.044008} {\bibfield
  {journal} {\bibinfo  {journal} {Phys. Rev. D}\ }\textbf {\bibinfo {volume}
  {81}},\ \bibinfo {pages} {044008} (\bibinfo {year} {2010})},\ \Eprint
  {http://arxiv.org/abs/0911.1814} {arXiv:0911.1814 [hep-th]} \BibitemShut
  {NoStop}%
\bibitem [{\citenamefont {Birmingham}\ \emph {et~al.}(1991)\citenamefont
  {Birmingham}, \citenamefont {Blau}, \citenamefont {Rakowski},\ and\
  \citenamefont {Thompson}}]{Birmingham:1991ty}%
  \BibitemOpen
  \bibfield  {author} {\bibinfo {author} {\bibfnamefont {D.}~\bibnamefont
  {Birmingham}}, \bibinfo {author} {\bibfnamefont {M.}~\bibnamefont {Blau}},
  \bibinfo {author} {\bibfnamefont {M.}~\bibnamefont {Rakowski}}, \ and\
  \bibinfo {author} {\bibfnamefont {G.}~\bibnamefont {Thompson}},\ }\href
  {\doibase 10.1016/0370-1573(91)90117-5} {\bibfield  {journal} {\bibinfo
  {journal} {Phys. Rept.}\ }\textbf {\bibinfo {volume} {209}},\ \bibinfo
  {pages} {129} (\bibinfo {year} {1991})}\BibitemShut {NoStop}%
\bibitem [{\citenamefont {Mielke}(2017)}]{Mielke:2017nwt}%
  \BibitemOpen
  \bibfield  {author} {\bibinfo {author} {\bibfnamefont {E.~W.}\ \bibnamefont
  {Mielke}},\ }\href {\doibase 10.1007/978-3-319-29734-7} {\emph {\bibinfo
  {title} {{Geometrodynamics of Gauge Fields. On the Geometry of Yang-Mills and
  Gravitational Gauge Theories}}}},\ Mathematical Physics Studies\ (\bibinfo
  {publisher} {Springer},\ \bibinfo {year} {2017})\BibitemShut {NoStop}%
\bibitem [{\citenamefont {Agrawal}\ \emph {et~al.}(2020)\citenamefont
  {Agrawal}, \citenamefont {Gukov}, \citenamefont {Obied},\ and\ \citenamefont
  {Vafa}}]{Agrawal:2020xek}%
  \BibitemOpen
  \bibfield  {author} {\bibinfo {author} {\bibfnamefont {P.}~\bibnamefont
  {Agrawal}}, \bibinfo {author} {\bibfnamefont {S.}~\bibnamefont {Gukov}},
  \bibinfo {author} {\bibfnamefont {G.}~\bibnamefont {Obied}}, \ and\ \bibinfo
  {author} {\bibfnamefont {C.}~\bibnamefont {Vafa}},\ }\href@noop {} {\
  (\bibinfo {year} {2020})},\ \Eprint {http://arxiv.org/abs/2009.10077}
  {arXiv:2009.10077 [hep-th]} \BibitemShut {NoStop}%
\bibitem [{\citenamefont {Kehagias}\ and\ \citenamefont
  {Riotto}(2021)}]{Kehagias:2021smx}%
  \BibitemOpen
  \bibfield  {author} {\bibinfo {author} {\bibfnamefont {A.}~\bibnamefont
  {Kehagias}}\ and\ \bibinfo {author} {\bibfnamefont {A.}~\bibnamefont
  {Riotto}},\ }\href@noop {} {\  (\bibinfo {year} {2021})},\ \Eprint
  {http://arxiv.org/abs/2105.10669} {arXiv:2105.10669 [hep-th]} \BibitemShut
  {NoStop}%
\bibitem [{\citenamefont {Sadovski}\ and\ \citenamefont
  {Sobreiro}(2024)}]{Sadovski:2024uhg}%
  \BibitemOpen
  \bibfield  {author} {\bibinfo {author} {\bibfnamefont {G.}~\bibnamefont
  {Sadovski}}\ and\ \bibinfo {author} {\bibfnamefont {R.~F.}\ \bibnamefont
  {Sobreiro}},\ }\href@noop {} {\  (\bibinfo {year} {2024})},\ \Eprint
  {http://arxiv.org/abs/2405.02884} {arXiv:2405.02884 [gr-qc]} \BibitemShut
  {NoStop}%
\bibitem [{\citenamefont {Junqueira}\ \emph {et~al.}(2017)\citenamefont
  {Junqueira}, \citenamefont {Pereira}, \citenamefont {Sadovski}, \citenamefont
  {Santos}, \citenamefont {Sobreiro},\ and\ \citenamefont
  {Tomaz}}]{Junqueira:2016hlu}%
  \BibitemOpen
  \bibfield  {author} {\bibinfo {author} {\bibfnamefont {O.~C.}\ \bibnamefont
  {Junqueira}}, \bibinfo {author} {\bibfnamefont {A.~D.}\ \bibnamefont
  {Pereira}}, \bibinfo {author} {\bibfnamefont {G.}~\bibnamefont {Sadovski}},
  \bibinfo {author} {\bibfnamefont {T.~R.~S.}\ \bibnamefont {Santos}}, \bibinfo
  {author} {\bibfnamefont {R.~F.}\ \bibnamefont {Sobreiro}}, \ and\ \bibinfo
  {author} {\bibfnamefont {A.~A.}\ \bibnamefont {Tomaz}},\ }\href {\doibase
  10.1140/epjc/s10052-017-4820-y} {\bibfield  {journal} {\bibinfo  {journal}
  {Eur. Phys. J. C}\ }\textbf {\bibinfo {volume} {77}},\ \bibinfo {pages} {249}
  (\bibinfo {year} {2017})},\ \Eprint {http://arxiv.org/abs/1612.05590}
  {arXiv:1612.05590 [hep-th]} \BibitemShut {NoStop}%
\bibitem [{\citenamefont {Woit}(2021)}]{Woit:2021bmb}%
  \BibitemOpen
  \bibfield  {author} {\bibinfo {author} {\bibfnamefont {P.}~\bibnamefont
  {Woit}},\ }\href@noop {} {\  (\bibinfo {year} {2021})},\ \Eprint
  {http://arxiv.org/abs/2104.05099} {arXiv:2104.05099 [hep-th]} \BibitemShut
  {NoStop}%
\bibitem [{\citenamefont {Barbero~G.}(1995)}]{BarberoG:1994eia}%
  \BibitemOpen
  \bibfield  {author} {\bibinfo {author} {\bibfnamefont {J.~F.}\ \bibnamefont
  {Barbero~G.}},\ }\href {\doibase 10.1103/PhysRevD.51.5507} {\bibfield
  {journal} {\bibinfo  {journal} {Phys. Rev. D}\ }\textbf {\bibinfo {volume}
  {51}},\ \bibinfo {pages} {5507} (\bibinfo {year} {1995})},\ \Eprint
  {http://arxiv.org/abs/gr-qc/9410014} {arXiv:gr-qc/9410014} \BibitemShut
  {NoStop}%
\bibitem [{\citenamefont {Holst}(1996)}]{Holst:1995pc}%
  \BibitemOpen
  \bibfield  {author} {\bibinfo {author} {\bibfnamefont {S.}~\bibnamefont
  {Holst}},\ }\href {\doibase 10.1103/PhysRevD.53.5966} {\bibfield  {journal}
  {\bibinfo  {journal} {Phys. Rev. D}\ }\textbf {\bibinfo {volume} {53}},\
  \bibinfo {pages} {5966} (\bibinfo {year} {1996})},\ \Eprint
  {http://arxiv.org/abs/gr-qc/9511026} {arXiv:gr-qc/9511026} \BibitemShut
  {NoStop}%
\bibitem [{\citenamefont {Unruh}(2014)}]{Unruh:2014aka}%
  \BibitemOpen
  \bibfield  {author} {\bibinfo {author} {\bibfnamefont {W.~G.}\ \bibnamefont
  {Unruh}},\ }\href@noop {} {\  (\bibinfo {year} {2014})},\ \Eprint
  {http://arxiv.org/abs/1401.3393} {arXiv:1401.3393 [gr-qc]} \BibitemShut
  {NoStop}%
\bibitem [{\citenamefont {Lemos}\ and\ \citenamefont
  {Silva}(2021)}]{Lemos:2020qxk}%
  \BibitemOpen
  \bibfield  {author} {\bibinfo {author} {\bibfnamefont {J.~P.~S.}\
  \bibnamefont {Lemos}}\ and\ \bibinfo {author} {\bibfnamefont {D.~L. F.~G.}\
  \bibnamefont {Silva}},\ }\href {\doibase 10.1016/j.aop.2021.168497}
  {\bibfield  {journal} {\bibinfo  {journal} {Annals Phys.}\ }\textbf {\bibinfo
  {volume} {430}},\ \bibinfo {pages} {168497} (\bibinfo {year} {2021})},\
  \Eprint {http://arxiv.org/abs/2005.14211} {arXiv:2005.14211 [gr-qc]}
  \BibitemShut {NoStop}%
\bibitem [{\citenamefont {Hohmann}(2020)}]{Hohmann:2019fvf}%
  \BibitemOpen
  \bibfield  {author} {\bibinfo {author} {\bibfnamefont {M.}~\bibnamefont
  {Hohmann}},\ }\href {\doibase 10.3390/sym12030453} {\bibfield  {journal}
  {\bibinfo  {journal} {Symmetry}\ }\textbf {\bibinfo {volume} {12}},\ \bibinfo
  {pages} {453} (\bibinfo {year} {2020})},\ \Eprint
  {http://arxiv.org/abs/1912.12906} {arXiv:1912.12906 [math-ph]} \BibitemShut
  {NoStop}%
\bibitem [{\citenamefont {Majumdar}(1947)}]{PhysRev.72.390}%
  \BibitemOpen
  \bibfield  {author} {\bibinfo {author} {\bibfnamefont {S.~D.}\ \bibnamefont
  {Majumdar}},\ }\href {\doibase 10.1103/PhysRev.72.390} {\bibfield  {journal}
  {\bibinfo  {journal} {Phys. Rev.}\ }\textbf {\bibinfo {volume} {72}},\
  \bibinfo {pages} {390} (\bibinfo {year} {1947})}\BibitemShut {NoStop}%
\bibitem [{\citenamefont {Ace\~na}\ \emph {et~al.}(2024)\citenamefont
  {Ace\~na}, \citenamefont {Guntsche},\ and\ \citenamefont
  {de~Austria}}]{Acena:2023yly}%
  \BibitemOpen
  \bibfield  {author} {\bibinfo {author} {\bibfnamefont {A.}~\bibnamefont
  {Ace\~na}}, \bibinfo {author} {\bibfnamefont {B.~C.}\ \bibnamefont
  {Guntsche}}, \ and\ \bibinfo {author} {\bibfnamefont {I.~G.}\ \bibnamefont
  {de~Austria}},\ }\href {\doibase 10.31349/RevMexFis.70.030701} {\bibfield
  {journal} {\bibinfo  {journal} {Rev. Mex. Fis.}\ }\textbf {\bibinfo {volume}
  {70}},\ \bibinfo {pages} {030701} (\bibinfo {year} {2024})},\ \Eprint
  {http://arxiv.org/abs/2310.06575} {arXiv:2310.06575 [gr-qc]} \BibitemShut
  {NoStop}%
\bibitem [{\citenamefont {Gorji}\ \emph {et~al.}(2020)\citenamefont {Gorji},
  \citenamefont {Allahyari}, \citenamefont {Khodadi},\ and\ \citenamefont
  {Firouzjahi}}]{Gorji:2020ten}%
  \BibitemOpen
  \bibfield  {author} {\bibinfo {author} {\bibfnamefont {M.~A.}\ \bibnamefont
  {Gorji}}, \bibinfo {author} {\bibfnamefont {A.}~\bibnamefont {Allahyari}},
  \bibinfo {author} {\bibfnamefont {M.}~\bibnamefont {Khodadi}}, \ and\
  \bibinfo {author} {\bibfnamefont {H.}~\bibnamefont {Firouzjahi}},\ }\href
  {\doibase 10.1103/PhysRevD.101.124060} {\bibfield  {journal} {\bibinfo
  {journal} {Phys. Rev. D}\ }\textbf {\bibinfo {volume} {101}},\ \bibinfo
  {pages} {124060} (\bibinfo {year} {2020})},\ \Eprint
  {http://arxiv.org/abs/1912.04636} {arXiv:1912.04636 [gr-qc]} \BibitemShut
  {NoStop}%
\bibitem [{\citenamefont {Maeda}\ and\ \citenamefont
  {Maeda}(2012)}]{Maeda:2012tu}%
  \BibitemOpen
  \bibfield  {author} {\bibinfo {author} {\bibfnamefont {H.}~\bibnamefont
  {Maeda}}\ and\ \bibinfo {author} {\bibfnamefont {K.-i.}\ \bibnamefont
  {Maeda}},\ }\href {\doibase 10.1103/PhysRevD.86.124045} {\bibfield  {journal}
  {\bibinfo  {journal} {Phys. Rev. D}\ }\textbf {\bibinfo {volume} {86}},\
  \bibinfo {pages} {124045} (\bibinfo {year} {2012})},\ \Eprint
  {http://arxiv.org/abs/1208.5777} {arXiv:1208.5777 [gr-qc]} \BibitemShut
  {NoStop}%
\bibitem [{\citenamefont {Sotiriou}(2015)}]{Sotiriou:2015pka}%
  \BibitemOpen
  \bibfield  {author} {\bibinfo {author} {\bibfnamefont {T.~P.}\ \bibnamefont
  {Sotiriou}},\ }\href {\doibase 10.1088/0264-9381/32/21/214002} {\bibfield
  {journal} {\bibinfo  {journal} {Class. Quant. Grav.}\ }\textbf {\bibinfo
  {volume} {32}},\ \bibinfo {pages} {214002} (\bibinfo {year} {2015})},\
  \Eprint {http://arxiv.org/abs/1505.00248} {arXiv:1505.00248 [gr-qc]}
  \BibitemShut {NoStop}%
\bibitem [{\citenamefont {Gallagher}\ \emph {et~al.}(2022)\citenamefont
  {Gallagher}, \citenamefont {Koivisto},\ and\ \citenamefont
  {Marzola}}]{Gallagher:2022kvv}%
  \BibitemOpen
  \bibfield  {author} {\bibinfo {author} {\bibfnamefont {P.}~\bibnamefont
  {Gallagher}}, \bibinfo {author} {\bibfnamefont {T.}~\bibnamefont {Koivisto}},
  \ and\ \bibinfo {author} {\bibfnamefont {L.}~\bibnamefont {Marzola}},\ }\href
  {\doibase 10.1103/PhysRevD.105.125010} {\bibfield  {journal} {\bibinfo
  {journal} {Phys. Rev. D}\ }\textbf {\bibinfo {volume} {105}},\ \bibinfo
  {pages} {125010} (\bibinfo {year} {2022})},\ \Eprint
  {http://arxiv.org/abs/2202.05657} {arXiv:2202.05657 [hep-th]} \BibitemShut
  {NoStop}%
\bibitem [{\citenamefont {Gallagher}(2024)}]{Gallagher:2024haq}%
  \BibitemOpen
  \bibfield  {author} {\bibinfo {author} {\bibfnamefont {P.}~\bibnamefont
  {Gallagher}},\ }\href@noop {} {\  (\bibinfo {year} {2024})},\ \Eprint
  {http://arxiv.org/abs/2403.02578} {arXiv:2403.02578 [hep-th]} \BibitemShut
  {NoStop}%
\bibitem [{\citenamefont {Isham}\ \emph {et~al.}(1971)\citenamefont {Isham},
  \citenamefont {Salam},\ and\ \citenamefont {Strathdee}}]{Isham:1971gm}%
  \BibitemOpen
  \bibfield  {author} {\bibinfo {author} {\bibfnamefont {C.~J.}\ \bibnamefont
  {Isham}}, \bibinfo {author} {\bibfnamefont {A.}~\bibnamefont {Salam}}, \ and\
  \bibinfo {author} {\bibfnamefont {J.~A.}\ \bibnamefont {Strathdee}},\ }\href
  {\doibase 10.1103/PhysRevD.3.867} {\bibfield  {journal} {\bibinfo  {journal}
  {Phys. Rev. D}\ }\textbf {\bibinfo {volume} {3}},\ \bibinfo {pages} {867}
  (\bibinfo {year} {1971})}\BibitemShut {NoStop}%
\bibitem [{\citenamefont {Israelit}\ and\ \citenamefont
  {Rosen}(1989)}]{Israelit:1986ez}%
  \BibitemOpen
  \bibfield  {author} {\bibinfo {author} {\bibfnamefont {M.}~\bibnamefont
  {Israelit}}\ and\ \bibinfo {author} {\bibfnamefont {N.}~\bibnamefont
  {Rosen}},\ }\href {\doibase 10.1007/BF00737765} {\bibfield  {journal}
  {\bibinfo  {journal} {Found. Phys.}\ }\textbf {\bibinfo {volume} {19}},\
  \bibinfo {pages} {33} (\bibinfo {year} {1989})}\BibitemShut {NoStop}%
\bibitem [{\citenamefont {Beltran~Jimenez}\ \emph {et~al.}(2012)\citenamefont
  {Beltran~Jimenez}, \citenamefont {Golovnev}, \citenamefont {Karciauskas},\
  and\ \citenamefont {Koivisto}}]{BeltranJimenez:2012sz}%
  \BibitemOpen
  \bibfield  {author} {\bibinfo {author} {\bibfnamefont {J.}~\bibnamefont
  {Beltran~Jimenez}}, \bibinfo {author} {\bibfnamefont {A.}~\bibnamefont
  {Golovnev}}, \bibinfo {author} {\bibfnamefont {M.}~\bibnamefont
  {Karciauskas}}, \ and\ \bibinfo {author} {\bibfnamefont {T.~S.}\ \bibnamefont
  {Koivisto}},\ }\href {\doibase 10.1103/PhysRevD.86.084024} {\bibfield
  {journal} {\bibinfo  {journal} {Phys. Rev. D}\ }\textbf {\bibinfo {volume}
  {86}},\ \bibinfo {pages} {084024} (\bibinfo {year} {2012})},\ \Eprint
  {http://arxiv.org/abs/1201.4018} {arXiv:1201.4018 [gr-qc]} \BibitemShut
  {NoStop}%
\bibitem [{\citenamefont {Schmidt-May}\ and\ \citenamefont {von
  Strauss}(2016)}]{Schmidt-May:2015vnx}%
  \BibitemOpen
  \bibfield  {author} {\bibinfo {author} {\bibfnamefont {A.}~\bibnamefont
  {Schmidt-May}}\ and\ \bibinfo {author} {\bibfnamefont {M.}~\bibnamefont {von
  Strauss}},\ }\href {\doibase 10.1088/1751-8113/49/18/183001} {\bibfield
  {journal} {\bibinfo  {journal} {J. Phys. A}\ }\textbf {\bibinfo {volume}
  {49}},\ \bibinfo {pages} {183001} (\bibinfo {year} {2016})},\ \Eprint
  {http://arxiv.org/abs/1512.00021} {arXiv:1512.00021 [hep-th]} \BibitemShut
  {NoStop}%
\bibitem [{\citenamefont {Castro}(2017)}]{Castro:2016rvj}%
  \BibitemOpen
  \bibfield  {author} {\bibinfo {author} {\bibfnamefont {C.}~\bibnamefont
  {Castro}},\ }\href {\doibase 10.1007/s00006-016-0702-x} {\bibfield  {journal}
  {\bibinfo  {journal} {Adv. Appl. Clifford Algebras}\ }\textbf {\bibinfo
  {volume} {27}},\ \bibinfo {pages} {1031} (\bibinfo {year}
  {2017})}\BibitemShut {NoStop}%
\bibitem [{\citenamefont {Krssak}(2017)}]{Krssak:2017nlv}%
  \BibitemOpen
  \bibfield  {author} {\bibinfo {author} {\bibfnamefont {M.}~\bibnamefont
  {Krssak}},\ }\href@noop {} {\  (\bibinfo {year} {2017})},\ \Eprint
  {http://arxiv.org/abs/1705.01072} {arXiv:1705.01072 [gr-qc]} \BibitemShut
  {NoStop}%
\bibitem [{\citenamefont {Blixt}\ \emph {et~al.}(2023)\citenamefont {Blixt},
  \citenamefont {Hohmann}, \citenamefont {Koivisto},\ and\ \citenamefont
  {Marzola}}]{Blixt:2023qbg}%
  \BibitemOpen
  \bibfield  {author} {\bibinfo {author} {\bibfnamefont {D.}~\bibnamefont
  {Blixt}}, \bibinfo {author} {\bibfnamefont {M.}~\bibnamefont {Hohmann}},
  \bibinfo {author} {\bibfnamefont {T.}~\bibnamefont {Koivisto}}, \ and\
  \bibinfo {author} {\bibfnamefont {L.}~\bibnamefont {Marzola}},\ }\href
  {\doibase 10.1140/epjc/s10052-023-12247-7} {\bibfield  {journal} {\bibinfo
  {journal} {Eur. Phys. J. C}\ }\textbf {\bibinfo {volume} {83}},\ \bibinfo
  {pages} {1120} (\bibinfo {year} {2023})},\ \Eprint
  {http://arxiv.org/abs/2305.03504} {arXiv:2305.03504 [gr-qc]} \BibitemShut
  {NoStop}%
\bibitem [{\citenamefont {Faraoni}\ and\ \citenamefont
  {Vachon}(2020)}]{Faraoni:2020ehi}%
  \BibitemOpen
  \bibfield  {author} {\bibinfo {author} {\bibfnamefont {V.}~\bibnamefont
  {Faraoni}}\ and\ \bibinfo {author} {\bibfnamefont {G.}~\bibnamefont
  {Vachon}},\ }\href {\doibase 10.1140/epjc/s10052-020-8345-4} {\bibfield
  {journal} {\bibinfo  {journal} {Eur. Phys. J. C}\ }\textbf {\bibinfo {volume}
  {80}},\ \bibinfo {pages} {771} (\bibinfo {year} {2020})},\ \Eprint
  {http://arxiv.org/abs/2006.10827} {arXiv:2006.10827 [gr-qc]} \BibitemShut
  {NoStop}%
\bibitem [{\citenamefont {Herrero}\ and\ \citenamefont
  {Morales-Lladosa}(2010)}]{Herrero:2010yt}%
  \BibitemOpen
  \bibfield  {author} {\bibinfo {author} {\bibfnamefont {A.}~\bibnamefont
  {Herrero}}\ and\ \bibinfo {author} {\bibfnamefont {J.~A.}\ \bibnamefont
  {Morales-Lladosa}},\ }\href {\doibase 10.1088/0264-9381/27/17/175007}
  {\bibfield  {journal} {\bibinfo  {journal} {Class. Quant. Grav.}\ }\textbf
  {\bibinfo {volume} {27}},\ \bibinfo {pages} {175007} (\bibinfo {year}
  {2010})},\ \Eprint {http://arxiv.org/abs/1006.3182} {arXiv:1006.3182 [gr-qc]}
  \BibitemShut {NoStop}%
\bibitem [{\citenamefont {Fazzini}\ \emph {et~al.}(2023)\citenamefont
  {Fazzini}, \citenamefont {Rovelli},\ and\ \citenamefont
  {Soltani}}]{Fazzini:2023scu}%
  \BibitemOpen
  \bibfield  {author} {\bibinfo {author} {\bibfnamefont {F.}~\bibnamefont
  {Fazzini}}, \bibinfo {author} {\bibfnamefont {C.}~\bibnamefont {Rovelli}}, \
  and\ \bibinfo {author} {\bibfnamefont {F.}~\bibnamefont {Soltani}},\ }\href
  {\doibase 10.1103/PhysRevD.108.044009} {\bibfield  {journal} {\bibinfo
  {journal} {Phys. Rev. D}\ }\textbf {\bibinfo {volume} {108}},\ \bibinfo
  {pages} {044009} (\bibinfo {year} {2023})},\ \Eprint
  {http://arxiv.org/abs/2307.07797} {arXiv:2307.07797 [gr-qc]} \BibitemShut
  {NoStop}%
\bibitem [{\citenamefont {Gomes}\ \emph {et~al.}(2023)\citenamefont {Gomes},
  \citenamefont {Beltr\'an~Jim\'enez},\ and\ \citenamefont
  {Koivisto}}]{Gomes:2022vrc}%
  \BibitemOpen
  \bibfield  {author} {\bibinfo {author} {\bibfnamefont {D.~A.}\ \bibnamefont
  {Gomes}}, \bibinfo {author} {\bibfnamefont {J.}~\bibnamefont
  {Beltr\'an~Jim\'enez}}, \ and\ \bibinfo {author} {\bibfnamefont {T.~S.}\
  \bibnamefont {Koivisto}},\ }\href {\doibase 10.1103/PhysRevD.107.024044}
  {\bibfield  {journal} {\bibinfo  {journal} {Phys. Rev. D}\ }\textbf {\bibinfo
  {volume} {107}},\ \bibinfo {pages} {024044} (\bibinfo {year} {2023})},\
  \Eprint {http://arxiv.org/abs/2205.09716} {arXiv:2205.09716 [gr-qc]}
  \BibitemShut {NoStop}%
\bibitem [{\citenamefont {Capozziello}\ \emph {et~al.}(2024)\citenamefont
  {Capozziello}, \citenamefont {De~Bianchi},\ and\ \citenamefont
  {Battista}}]{Capozziello:2024ucm}%
  \BibitemOpen
  \bibfield  {author} {\bibinfo {author} {\bibfnamefont {S.}~\bibnamefont
  {Capozziello}}, \bibinfo {author} {\bibfnamefont {S.}~\bibnamefont
  {De~Bianchi}}, \ and\ \bibinfo {author} {\bibfnamefont {E.}~\bibnamefont
  {Battista}},\ }\href {\doibase 10.1103/PhysRevD.109.104060} {\bibfield
  {journal} {\bibinfo  {journal} {Phys. Rev. D}\ }\textbf {\bibinfo {volume}
  {109}},\ \bibinfo {pages} {104060} (\bibinfo {year} {2024})},\ \Eprint
  {http://arxiv.org/abs/2404.17267} {arXiv:2404.17267 [gr-qc]} \BibitemShut
  {NoStop}%
\bibitem [{\citenamefont {Khodadi}\ \emph {et~al.}(2024)\citenamefont
  {Khodadi}, \citenamefont {Vagnozzi},\ and\ \citenamefont
  {Firouzjaee}}]{Khodadi:2024ubi}%
  \BibitemOpen
  \bibfield  {author} {\bibinfo {author} {\bibfnamefont {M.}~\bibnamefont
  {Khodadi}}, \bibinfo {author} {\bibfnamefont {S.}~\bibnamefont {Vagnozzi}}, \
  and\ \bibinfo {author} {\bibfnamefont {J.~T.}\ \bibnamefont {Firouzjaee}},\
  }\href@noop {} {\  (\bibinfo {year} {2024})},\ \Eprint
  {http://arxiv.org/abs/2408.03241} {arXiv:2408.03241 [gr-qc]} \BibitemShut
  {NoStop}%
\bibitem [{\citenamefont {Chen}\ \emph {et~al.}(2018)\citenamefont {Chen},
  \citenamefont {Bouhmadi-L\'opez},\ and\ \citenamefont {Chen}}]{Chen:2017ify}%
  \BibitemOpen
  \bibfield  {author} {\bibinfo {author} {\bibfnamefont {C.-Y.}\ \bibnamefont
  {Chen}}, \bibinfo {author} {\bibfnamefont {M.}~\bibnamefont
  {Bouhmadi-L\'opez}}, \ and\ \bibinfo {author} {\bibfnamefont
  {P.}~\bibnamefont {Chen}},\ }\href {\doibase 10.1140/epjc/s10052-018-5556-z}
  {\bibfield  {journal} {\bibinfo  {journal} {Eur. Phys. J. C}\ }\textbf
  {\bibinfo {volume} {78}},\ \bibinfo {pages} {59} (\bibinfo {year} {2018})},\
  \Eprint {http://arxiv.org/abs/1710.10638} {arXiv:1710.10638 [gr-qc]}
  \BibitemShut {NoStop}%
\bibitem [{\citenamefont {Nashed}\ \emph {et~al.}(2019)\citenamefont {Nashed},
  \citenamefont {El~Hanafy},\ and\ \citenamefont {Bamba}}]{Nashed:2018qag}%
  \BibitemOpen
  \bibfield  {author} {\bibinfo {author} {\bibfnamefont {G.~G.~L.}\
  \bibnamefont {Nashed}}, \bibinfo {author} {\bibfnamefont {W.}~\bibnamefont
  {El~Hanafy}}, \ and\ \bibinfo {author} {\bibfnamefont {K.}~\bibnamefont
  {Bamba}},\ }\href {\doibase 10.1088/1475-7516/2019/01/058} {\bibfield
  {journal} {\bibinfo  {journal} {JCAP}\ }\textbf {\bibinfo {volume} {01}},\
  \bibinfo {pages} {058} (\bibinfo {year} {2019})},\ \Eprint
  {http://arxiv.org/abs/1809.02289} {arXiv:1809.02289 [gr-qc]} \BibitemShut
  {NoStop}%
\bibitem [{\citenamefont {Chamseddine}\ \emph {et~al.}(2020)\citenamefont
  {Chamseddine}, \citenamefont {Mukhanov},\ and\ \citenamefont
  {Russ}}]{Chamseddine:2019fog}%
  \BibitemOpen
  \bibfield  {author} {\bibinfo {author} {\bibfnamefont {A.~H.}\ \bibnamefont
  {Chamseddine}}, \bibinfo {author} {\bibfnamefont {V.}~\bibnamefont
  {Mukhanov}}, \ and\ \bibinfo {author} {\bibfnamefont {T.~B.}\ \bibnamefont
  {Russ}},\ }\href {\doibase 10.1002/prop.201900103} {\bibfield  {journal}
  {\bibinfo  {journal} {Fortsch. Phys.}\ }\textbf {\bibinfo {volume} {68}},\
  \bibinfo {pages} {1900103} (\bibinfo {year} {2020})},\ \Eprint
  {http://arxiv.org/abs/1912.03162} {arXiv:1912.03162 [hep-th]} \BibitemShut
  {NoStop}%
\bibitem [{\citenamefont {Nojiri}\ and\ \citenamefont
  {Odintsov}(2024)}]{Nojiri:2024txy}%
  \BibitemOpen
  \bibfield  {author} {\bibinfo {author} {\bibfnamefont {S.}~\bibnamefont
  {Nojiri}}\ and\ \bibinfo {author} {\bibfnamefont {S.~D.}\ \bibnamefont
  {Odintsov}},\ }\href@noop {} {\  (\bibinfo {year} {2024})},\ \Eprint
  {http://arxiv.org/abs/2408.05668} {arXiv:2408.05668 [gr-qc]} \BibitemShut
  {NoStop}%
\bibitem [{\citenamefont {Han}\ and\ \citenamefont {Liu}(2024)}]{Han:2022rsx}%
  \BibitemOpen
  \bibfield  {author} {\bibinfo {author} {\bibfnamefont {M.}~\bibnamefont
  {Han}}\ and\ \bibinfo {author} {\bibfnamefont {H.}~\bibnamefont {Liu}},\
  }\href {\doibase 10.1103/PhysRevD.109.084033} {\bibfield  {journal} {\bibinfo
   {journal} {Phys. Rev. D}\ }\textbf {\bibinfo {volume} {109}},\ \bibinfo
  {pages} {084033} (\bibinfo {year} {2024})},\ \Eprint
  {http://arxiv.org/abs/2212.04605} {arXiv:2212.04605 [gr-qc]} \BibitemShut
  {NoStop}%
\bibitem [{\citenamefont {Giesel}\ \emph {et~al.}(2024)\citenamefont {Giesel},
  \citenamefont {Liu}, \citenamefont {Singh},\ and\ \citenamefont
  {Weigl}}]{Giesel:2024mps}%
  \BibitemOpen
  \bibfield  {author} {\bibinfo {author} {\bibfnamefont {K.}~\bibnamefont
  {Giesel}}, \bibinfo {author} {\bibfnamefont {H.}~\bibnamefont {Liu}},
  \bibinfo {author} {\bibfnamefont {P.}~\bibnamefont {Singh}}, \ and\ \bibinfo
  {author} {\bibfnamefont {S.~A.}\ \bibnamefont {Weigl}},\ }\href@noop {} {\
  (\bibinfo {year} {2024})},\ \Eprint {http://arxiv.org/abs/2405.03554}
  {arXiv:2405.03554 [gr-qc]} \BibitemShut {NoStop}%
\bibitem [{\citenamefont {Wiesendanger}(2018)}]{Wiesendanger:2018dzw}%
  \BibitemOpen
  \bibfield  {author} {\bibinfo {author} {\bibfnamefont {C.}~\bibnamefont
  {Wiesendanger}},\ }\href {\doibase 10.1088/1361-6382/ab04e9} {\bibfield
  {journal} {\bibinfo  {journal} {Class. Quant. Grav.}\ }\textbf {\bibinfo
  {volume} {36}},\ \bibinfo {pages} {065015} (\bibinfo {year} {2018})},\
  \Eprint {http://arxiv.org/abs/1806.05037} {arXiv:1806.05037 [gr-qc]}
  \BibitemShut {NoStop}%
\bibitem [{\citenamefont {Wiesendanger}(2024)}]{Wiesendanger:2024dnp}%
  \BibitemOpen
  \bibfield  {author} {\bibinfo {author} {\bibfnamefont {C.}~\bibnamefont
  {Wiesendanger}},\ }\href@noop {} {\  (\bibinfo {year} {2024})},\ \Eprint
  {http://arxiv.org/abs/2405.03719} {arXiv:2405.03719 [gr-qc]} \BibitemShut
  {NoStop}%
\bibitem [{\citenamefont {Wiesendanger}(2020)}]{Wiesendanger:2020lwa}%
  \BibitemOpen
  \bibfield  {author} {\bibinfo {author} {\bibfnamefont {C.}~\bibnamefont
  {Wiesendanger}},\ }\href {\doibase 10.1088/1361-6382/aba80b} {\bibfield
  {journal} {\bibinfo  {journal} {Class. Quant. Grav.}\ }\textbf {\bibinfo
  {volume} {37}},\ \bibinfo {pages} {195029} (\bibinfo {year}
  {2020})}\BibitemShut {NoStop}%
\bibitem [{\citenamefont {Gielen}\ and\ \citenamefont
  {Wise}(2013)}]{Gielen:2012fz}%
  \BibitemOpen
  \bibfield  {author} {\bibinfo {author} {\bibfnamefont {S.}~\bibnamefont
  {Gielen}}\ and\ \bibinfo {author} {\bibfnamefont {D.~K.}\ \bibnamefont
  {Wise}},\ }\href {\doibase 10.1063/1.4802878} {\bibfield  {journal} {\bibinfo
   {journal} {J. Math. Phys.}\ }\textbf {\bibinfo {volume} {54}},\ \bibinfo
  {pages} {052501} (\bibinfo {year} {2013})},\ \Eprint
  {http://arxiv.org/abs/1210.0019} {arXiv:1210.0019 [gr-qc]} \BibitemShut
  {NoStop}%
\bibitem [{\citenamefont {Doran}(2000)}]{Doran:1999gb}%
  \BibitemOpen
  \bibfield  {author} {\bibinfo {author} {\bibfnamefont {C.}~\bibnamefont
  {Doran}},\ }\href {\doibase 10.1103/PhysRevD.61.067503} {\bibfield  {journal}
  {\bibinfo  {journal} {Phys. Rev. D}\ }\textbf {\bibinfo {volume} {61}},\
  \bibinfo {pages} {067503} (\bibinfo {year} {2000})},\ \Eprint
  {http://arxiv.org/abs/gr-qc/9910099} {arXiv:gr-qc/9910099} \BibitemShut
  {NoStop}%
\bibitem [{\citenamefont {Baines}(2021)}]{Baines:2021jbj}%
  \BibitemOpen
  \bibfield  {author} {\bibinfo {author} {\bibfnamefont {J.}~\bibnamefont
  {Baines}},\ }\href@noop {} {\  (\bibinfo {year} {2021})},\ \Eprint
  {http://arxiv.org/abs/2104.10332} {arXiv:2104.10332 [gr-qc]} \BibitemShut
  {NoStop}%
\bibitem [{\citenamefont {M\"akinen}\ \emph {et~al.}(2019)\citenamefont
  {M\"akinen}, \citenamefont {Dmitriev}, \citenamefont {Nissinen},
  \citenamefont {Rysti}, \citenamefont {Volovik}, \citenamefont {Yudin},
  \citenamefont {Zhang},\ and\ \citenamefont {Eltsov}}]{Makinen:2018ltj}%
  \BibitemOpen
  \bibfield  {author} {\bibinfo {author} {\bibfnamefont {J.~T.}\ \bibnamefont
  {M\"akinen}}, \bibinfo {author} {\bibfnamefont {V.~V.}\ \bibnamefont
  {Dmitriev}}, \bibinfo {author} {\bibfnamefont {J.}~\bibnamefont {Nissinen}},
  \bibinfo {author} {\bibfnamefont {J.}~\bibnamefont {Rysti}}, \bibinfo
  {author} {\bibfnamefont {G.~E.}\ \bibnamefont {Volovik}}, \bibinfo {author}
  {\bibfnamefont {A.~N.}\ \bibnamefont {Yudin}}, \bibinfo {author}
  {\bibfnamefont {K.}~\bibnamefont {Zhang}}, \ and\ \bibinfo {author}
  {\bibfnamefont {V.~B.}\ \bibnamefont {Eltsov}},\ }\href {\doibase
  10.1038/s41467-018-08204-8} {\bibfield  {journal} {\bibinfo  {journal}
  {Nature Commun.}\ }\textbf {\bibinfo {volume} {10}},\ \bibinfo {pages} {237}
  (\bibinfo {year} {2019})},\ \Eprint {http://arxiv.org/abs/1807.04328}
  {arXiv:1807.04328 [cond-mat.other]} \BibitemShut {NoStop}%
\bibitem [{\citenamefont {Volovik}(2021)}]{Volovik:2020lwn}%
  \BibitemOpen
  \bibfield  {author} {\bibinfo {author} {\bibfnamefont {G.~E.}\ \bibnamefont
  {Volovik}},\ }\href {\doibase 10.1007/s10909-020-02538-8} {\bibfield
  {journal} {\bibinfo  {journal} {J. Low Temp. Phys.}\ }\textbf {\bibinfo
  {volume} {202}},\ \bibinfo {pages} {11} (\bibinfo {year} {2021})},\ \Eprint
  {http://arxiv.org/abs/2008.04682} {arXiv:2008.04682 [cond-mat.other]}
  \BibitemShut {NoStop}%
\bibitem [{\citenamefont {Hayward}(2006)}]{Hayward:2005gi}%
  \BibitemOpen
  \bibfield  {author} {\bibinfo {author} {\bibfnamefont {S.~A.}\ \bibnamefont
  {Hayward}},\ }\href {\doibase 10.1103/PhysRevLett.96.031103} {\bibfield
  {journal} {\bibinfo  {journal} {Phys. Rev. Lett.}\ }\textbf {\bibinfo
  {volume} {96}},\ \bibinfo {pages} {031103} (\bibinfo {year} {2006})},\
  \Eprint {http://arxiv.org/abs/gr-qc/0506126} {arXiv:gr-qc/0506126}
  \BibitemShut {NoStop}%
\bibitem [{\citenamefont {Koivisto}\ \emph {et~al.}(2019)\citenamefont
  {Koivisto}, \citenamefont {Hohmann},\ and\ \citenamefont
  {Z\l{}o\'snik}}]{Koivisto:2019ejt}%
  \BibitemOpen
  \bibfield  {author} {\bibinfo {author} {\bibfnamefont {T.}~\bibnamefont
  {Koivisto}}, \bibinfo {author} {\bibfnamefont {M.}~\bibnamefont {Hohmann}}, \
  and\ \bibinfo {author} {\bibfnamefont {T.}~\bibnamefont {Z\l{}o\'snik}},\
  }\href {\doibase 10.3390/universe5070168} {\bibfield  {journal} {\bibinfo
  {journal} {Universe}\ }\textbf {\bibinfo {volume} {5}},\ \bibinfo {pages}
  {168} (\bibinfo {year} {2019})},\ \Eprint {http://arxiv.org/abs/1905.02967}
  {arXiv:1905.02967 [gr-qc]} \BibitemShut {NoStop}%
\bibitem [{\citenamefont {Kruskal}(1960)}]{PhysRev.119.1743}%
  \BibitemOpen
  \bibfield  {author} {\bibinfo {author} {\bibfnamefont {M.~D.}\ \bibnamefont
  {Kruskal}},\ }\href {\doibase 10.1103/PhysRev.119.1743} {\bibfield  {journal}
  {\bibinfo  {journal} {Phys. Rev.}\ }\textbf {\bibinfo {volume} {119}},\
  \bibinfo {pages} {1743} (\bibinfo {year} {1960})}\BibitemShut {NoStop}%
\bibitem [{\citenamefont {Finch}(2015)}]{Finch:2012vli}%
  \BibitemOpen
  \bibfield  {author} {\bibinfo {author} {\bibfnamefont {T.~K.}\ \bibnamefont
  {Finch}},\ }\href {\doibase 10.1007/s10714-015-1891-7} {\bibfield  {journal}
  {\bibinfo  {journal} {Gen. Rel. Grav.}\ }\textbf {\bibinfo {volume} {47}},\
  \bibinfo {pages} {56} (\bibinfo {year} {2015})},\ \Eprint
  {http://arxiv.org/abs/1211.4337} {arXiv:1211.4337 [gr-qc]} \BibitemShut
  {NoStop}%
\bibitem [{\citenamefont {MacLaurin}(2018)}]{MacLaurin:2018aze}%
  \BibitemOpen
  \bibfield  {author} {\bibinfo {author} {\bibfnamefont {C.}~\bibnamefont
  {MacLaurin}},\ }in\ \href {\doibase 10.1007/978-3-030-18061-4_9} {\emph
  {\bibinfo {booktitle} {{Domoschool - the International Alpine School in
  Mathematics and Physics}: {Einstein Equations: Physical and Mathematical
  aspects of General Relativity}}}}\ (\bibinfo {year} {2018})\ pp.\ \bibinfo
  {pages} {267--287},\ \Eprint {http://arxiv.org/abs/1911.05988}
  {arXiv:1911.05988 [gr-qc]} \BibitemShut {NoStop}%
\bibitem [{\citenamefont {Martel}\ and\ \citenamefont
  {Poisson}(2001)}]{Martel:2000rn}%
  \BibitemOpen
  \bibfield  {author} {\bibinfo {author} {\bibfnamefont {K.}~\bibnamefont
  {Martel}}\ and\ \bibinfo {author} {\bibfnamefont {E.}~\bibnamefont
  {Poisson}},\ }\href {\doibase 10.1119/1.1336836} {\bibfield  {journal}
  {\bibinfo  {journal} {Am. J. Phys.}\ }\textbf {\bibinfo {volume} {69}},\
  \bibinfo {pages} {476} (\bibinfo {year} {2001})},\ \Eprint
  {http://arxiv.org/abs/gr-qc/0001069} {arXiv:gr-qc/0001069} \BibitemShut
  {NoStop}%
\bibitem [{\citenamefont {Gautreau}\ and\ \citenamefont
  {Hoffmann}(1978)}]{Gautreau:1978zz}%
  \BibitemOpen
  \bibfield  {author} {\bibinfo {author} {\bibfnamefont {R.}~\bibnamefont
  {Gautreau}}\ and\ \bibinfo {author} {\bibfnamefont {B.}~\bibnamefont
  {Hoffmann}},\ }\href {\doibase 10.1103/PhysRevD.17.2552} {\bibfield
  {journal} {\bibinfo  {journal} {Phys. Rev. D}\ }\textbf {\bibinfo {volume}
  {17}},\ \bibinfo {pages} {2552} (\bibinfo {year} {1978})}\BibitemShut
  {NoStop}%
\bibitem [{\citenamefont {Emtsova}\ \emph {et~al.}(2021)\citenamefont
  {Emtsova}, \citenamefont {Kr\v{s}\v{s}\'ak}, \citenamefont {Petrov},\ and\
  \citenamefont {Toporensky}}]{Emtsova:2021ehh}%
  \BibitemOpen
  \bibfield  {author} {\bibinfo {author} {\bibfnamefont {E.~D.}\ \bibnamefont
  {Emtsova}}, \bibinfo {author} {\bibfnamefont {M.}~\bibnamefont
  {Kr\v{s}\v{s}\'ak}}, \bibinfo {author} {\bibfnamefont {A.~N.}\ \bibnamefont
  {Petrov}}, \ and\ \bibinfo {author} {\bibfnamefont {A.~V.}\ \bibnamefont
  {Toporensky}},\ }\href {\doibase 10.1140/epjc/s10052-021-09505-x} {\bibfield
  {journal} {\bibinfo  {journal} {Eur. Phys. J. C}\ }\textbf {\bibinfo {volume}
  {81}},\ \bibinfo {pages} {743} (\bibinfo {year} {2021})},\ \Eprint
  {http://arxiv.org/abs/2105.13312} {arXiv:2105.13312 [gr-qc]} \BibitemShut
  {NoStop}%
\bibitem [{\citenamefont {Kiefer}\ and\ \citenamefont
  {Mohaddes}(2023)}]{Kiefer:2023zxt}%
  \BibitemOpen
  \bibfield  {author} {\bibinfo {author} {\bibfnamefont {C.}~\bibnamefont
  {Kiefer}}\ and\ \bibinfo {author} {\bibfnamefont {H.}~\bibnamefont
  {Mohaddes}},\ }\href {\doibase 10.1103/PhysRevD.107.126006} {\bibfield
  {journal} {\bibinfo  {journal} {Phys. Rev. D}\ }\textbf {\bibinfo {volume}
  {107}},\ \bibinfo {pages} {126006} (\bibinfo {year} {2023})},\ \Eprint
  {http://arxiv.org/abs/2303.17924} {arXiv:2303.17924 [gr-qc]} \BibitemShut
  {NoStop}%
\bibitem [{\citenamefont {Chowdhury}(2024)}]{Chowdhury:2024rja}%
  \BibitemOpen
  \bibfield  {author} {\bibinfo {author} {\bibfnamefont {A.}~\bibnamefont
  {Chowdhury}},\ }\href@noop {} {\  (\bibinfo {year} {2024})},\ \Eprint
  {http://arxiv.org/abs/2404.15948} {arXiv:2404.15948 [gr-qc]} \BibitemShut
  {NoStop}%
\bibitem [{\citenamefont {Thiemann}(2024)}]{Thiemann:2024nmy}%
  \BibitemOpen
  \bibfield  {author} {\bibinfo {author} {\bibfnamefont {T.}~\bibnamefont
  {Thiemann}},\ }\href@noop {} {\  (\bibinfo {year} {2024})},\ \Eprint
  {http://arxiv.org/abs/2404.18956} {arXiv:2404.18956 [gr-qc]} \BibitemShut
  {NoStop}%
\end{thebibliography}%

\section{Enhanced symmetry}  
\label{extrasym}

The formula for the change of curvature
\be
\bR^{ab} \rightarrow \bR^{ab} + \bDiff\bX^{ab} + \bX^a{}_c\wedge\bX^c{}_b\,,
\ee
under generic distortion 
\be
\bomega^{ab} \rightarrow \bomega^{ab} + \bX^{ab}\,,
\ee
applied to the case
\be \label{thecase}
\bX^{ab} = \xi^{ab}\bdiff\phi^2\,,
\ee
simplifies to
\be 
\bR^{ab} \rightarrow \bR^{ab} + \bDiff\xi^{ab}\wedge\bdiff\phi^2\,, \label{simple}
\ee
due to $\bdiff^2 \phi^2 = 0$ and $\bdiff\phi^2\wedge\bdiff\phi^2 = 0$. 

We would like to show that the the 4-form 
\be \label{4form}
\bL_{g_{4}} =  \star\lp \bDiff\phi^a\wedge\bDiff\phi^b\rp\wedge\bR_{ab} = \bDiff\phi^a\wedge\bDiff\phi^b\wedge\star\bR_{ab}\,, 
\ee
changes by a boundary term when the connection is distorted by (\ref{thecase}). We obtain that under (\ref{thecase})
\be \label{change2}
\star \lp \bDiff\phi^a\wedge\bDiff\phi^b\rp \rightarrow \star \lp \bDiff\phi^a\wedge\bDiff\phi^b\rp + \epsilon^{ab}{}_{de}\xi^d{}_c \phi^c\bdiff\phi^2\wedge\bDiff\phi^e\,. 
\ee
Taking into account (\ref{simple}) and (\ref{change2}), the 4-form (\ref{4form}) changes under (\ref{thecase}) as
\be
\delta \bL_{g_{4}} = \epsilon_{abde}\lp -\xi^d{}_c\phi^c\bDiff\phi^e\wedge\bR^{ab} + \frac{1}{2}\bDiff\phi^d\wedge\bDiff\phi^e\wedge\bDiff\xi^{ab}\rp\wedge\bdiff\phi^2\,.   
\ee
This can be matched with the exact form 
\be
\bdiff\lp \bDiff\phi^a\wedge\bDiff\phi^b\star\xi_{ab}\wedge\bdiff\phi^2\rp = \epsilon_{abcd}\lp  \bR^a{}_e\phi^e\wedge\bDiff\phi^b \xi^{cd}
+ \frac{1}{2}\bDiff\phi^a\wedge\bDiff\phi^b\wedge\bDiff\xi^{cd}\rp\wedge\bdiff\phi^2\,,
\ee
by noting that
\be \label{condition}
\epsilon_{abcd}\phi^e\lp \bR^a{}_e\xi^{cd} - \bR^{ad}\xi^c{}_e\rp\wedge\bDiff\phi^b\wedge\bdiff\phi^2 = 2\epsilon_{abcd}\phi_e\bR^{a[d}\xi^{e]c}\wedge\bDiff\phi^b\wedge\bdiff\phi^2 = 0\,. 
\ee
The latter equality follows from that $\epsilon_{abcd}\bR^{a[d}\xi^{e]c} = \delta^e_b\bY$, for some 2-form $\bY$. 
To check this, one can by inspection be convinced that the contraction $\epsilon_{abcd}\bR^{a[d}\xi^{e]c}$ vanishes if the free indices are not equal, $b \neq e$. 
Thus we have that  $2\epsilon_{abcd}\phi_e\bR^{a[d}\xi^{e]c}\wedge\bDiff\phi^b\wedge\bdiff\phi^2 = 2\bY \phi_b\bDiff\phi^b\wedge\bdiff\phi^2 = \bY\wedge\bdiff\phi^2\wedge\bdiff\phi^2 = 0$. 

Using the above results, it is now easy to see that the modification of the 4-form 
\be
\bL_{g_{3}} = \bDiff\phi^a\wedge\bDiff\phi^b\wedge\bR_{ab}\,, 
\ee
under the distortion (\ref{thecase}) is not a pure boundary term. Instead, we obtain
\be
\delta \bL_{g_{3}} = 4\bR_{ab}\xi^a{}_c\phi^{[b}\wedge\bDiff\phi^{c]}\wedge\bdiff\phi^2\,. 
\ee 
Summarising the result, the action (\ref{Iphi}) changes as
\be
\delta I_{(2)} = 4 g_3 \int  \bR_{ab}\xi^a{}_c\phi^{[b}\wedge\bDiff\phi^{c]}\wedge\bdiff\phi^2 + g_4\oint \bDiff\phi^a\wedge\bDiff\phi^b\star\xi_{ab}\wedge\bdiff\phi^2\,,
\ee
which is consistent with Ref.\cite{Nikjoo:2023flm}.
It is perhaps interesting to note that the 4-form
\be \label{uptoa}
\bL =  -\epsilon_{abcd}\phi^a\phi^e\bR^b{}_e\wedge\bR^{cd}\,, 
\ee
which is equivalent to (\ref{4form}) up to a boundary term, transforms trivially,
\be
\bL \rightarrow \bL -\epsilon_{abcd}\phi^a\phi_e\lp \xi^{be}\bR^{cd} + \xi^{cd}\bR^{be}\rp\wedge\bdiff\phi^2 = \bL - 2\epsilon_{abcd}\phi^a\phi_e\xi^{b[d}\bR^{e]c}\wedge\bdiff\phi^2 = \bL\,,
\ee
where the last step follows again from that $\epsilon_{abcd}\bR^{a[d}\xi^{e]c} = \delta^e_b\bY$. Thus, whilst the $g_4$-theory given by (\ref{4form}) possesses the exact shift symmetry but inexact distortion symmetry, the $\tilde{g}_4$-theory given by (\ref{uptoa}) has the exact distortion symmetry but inexact shift symmetry. 

Finally, we check the invariance of the quartic action (\ref{Ilambda}). 
\ba
\delta I_{(4)} & = & 4\lambda\epsilon_{abcd}\xi^{a}{}_e\phi^e\bdiff\phi^2\wedge\bDiff\phi^b\wedge\bDiff\phi^c\wedge\bDiff\phi^d \nn \\ &.= & 
24\lambda\xi^a{}_b\phi^b\bdiff\phi^2\wedge\star\bDiff\phi_a = 48\lambda\xi^a{}_b\phi^b\phi_c\bDiff\phi^c\wedge\star\bDiff\phi_a = -48\lambda\xi^a{}_b\phi^b\phi_c\delta^c_a\star {\bf 1} = -48\lambda\xi_{ab}\phi^a\phi^b\star {\bf 1} = 0\,. 
\ea

\section{Spherically symmetric geodesics}
\label{geodesics}

The line element (\ref{sline})
\be \label{sline2}
\bdiff s^2 = -f^2(r)\bdiff t^2 + g^2(r)\bdiff r^2 + r^2\lp \bdiff \theta^2 + \sin^2{\theta}\bdiff \varphi^2\rp\,,
\ee  
ia associated with the Christoffel symbols
\ba
\mathring{\Gamma}^t{}_{tr} & = & \frac{f'(r)}{f(r)}\,, \quad \mathring{\Gamma}^r{}_{tt} = \frac{f'(r) f(r)}{g^2(r)}\,, \quad 
\mathring{\Gamma}^r{}_{rr} = \frac{g'(r)}{g(r)}\,, \nn \\  \mathring{\Gamma}^r{}_{\theta\theta} 
& = & \frac{\mathring{\Gamma}^r{}_{\varphi\varphi}}{\sin^2\theta}  = -\frac{r}{g^2(r)}\,, \quad 
\mathring{\Gamma}^\theta{}_{\theta r}  = \mathring{\Gamma}^\varphi{}_{\varphi r}  = \frac{1}{r}\,, \quad
 \mathring{\Gamma}^\theta{}_{\varphi \varphi}  = -\cos\theta\sin\theta\,, \quad  
 \mathring{\Gamma}^\varphi{}_{\varphi \theta}  = \cot\theta\,.  
\ea
Due to the spherical symmetry, we can choose to consider movement on the equivatorial plane $\theta=\pi/2$. The three nontrivial equations for geodesic parallel transport with an affine parameter $\lambda$ are then
\bs
\ba
t''(\lambda) + 2\frac{f'(r)}{f(r}t'(\lambda)r'(\lambda) & = & 0\,, \\
r''(\lambda) + \frac{f'(r)f(r)}{g^2(r)}(t'(\lambda))^2 + \frac{g'(r)}{g(r)}(r'(\lambda))^2 & = & 0\,, \\
\varphi''(\lambda) + \frac{1}{r}\varphi'(\lambda)r'(\lambda) & = & 0\,.
\ea
\es
The property of Killing vectors $k$ can be expressed in terms of the corresponding Killing 1-forms $\bk = \flat k$ s.t. $\bk(x'(\lambda)) = x'(\lambda)\lrcorner\bk$ is a constant. For the metric (\ref{sline2}), there exists a time-like Killing vector $k_e=\partial_t$ with the Killing 1-form $\bk_e = -f^2\bdiff t$ reflecting a conserved energy $e$, as well as a space-like Killing vector $k_L = \partial_\varphi$ with the Killing 1-form $\bk_L = r^2\sin^2\theta\bdiff\varphi$ reflecting a conservation of an angular momentum $L$. We parameterise these conserved quantities as
\bs
\label{constants}
\ba
\bk_E(x'(\lambda)) & = & -f^2t'(\lambda) = -e\,,  \label{energy} \\
\bk_L(x'(\lambda)) & = & r^2\varphi'(\lambda) = L\,.
\ea
\es
Yet, we can use the fact that due to metric compatibility, 
\bs
\be \label{epsilon}
\lp\bDiff\tau^a\otimes\bDiff\tau_a\rp(x'(\lambda),x'(\lambda)) = -f^2(r)\lp t'(\lambda)\rp^2 + g^2(r)\lp r'(\lambda)\rp^2 + r^2\lp (\theta'(\lambda))^2 + \sin^2{\theta}(\varphi'(\lambda))^2\rp = \epsilon\,,
\ee
is some constant $\epsilon$. Choosing the proper time $\tau$ of a massive particle as the affine parameter $\lambda=\tau$ we have that $\epsilon =-1$, i.e.
\be
 -f^2(r)\dot{t}^2 + g^2(r)\dot{r}^2 + r^2\lp \dot{\theta}^2 + \sin^2{\theta}\dot{\varphi}^2\rp = -1\,,
\ee
\es
Combining the information from (\ref{constants}) and (\ref{epsilon}) , we find that a geodesic trajectory is described by the 4-vector
\bs
\be \label{trajectory}
x'(\lambda) = \frac{e}{f^2(r)}\partial_t \mp \sqrt{\frac{e^2}{[f(r)g(r)]^2} + \frac{\epsilon - r^{-2}L}{g^2(r)}}\partial_r + \frac{L}{r^2}\partial_\varphi\,,
\ee
or alternatively, the 1-form
\be
\flat x'(\lambda) =  -e\bdiff t \mp \sqrt{\frac{e^2 g^2(r)}{f^2(r)} + \lp \epsilon - \frac{L}{r^2}\rp g^2(r)}\bdiff r - L\bdiff\varphi\,.  
\ee
\es
Considering the proper time $\tau$ as the affine parameter $\lambda=\tau$ we obtain the proper 4-velocity $\dot{x}$ from (\ref{trajectory}) by setting $\epsilon=-1$. Note that for $g=f^{-1}$ (\ref{trajectory}) then coincides precisely with (\ref{observer1}) for purely radial geodesics $L=0$. Let us we denote $\bu = \flat \dot{x}$,
\be
\bu = -e\bdiff t \mp \sqrt{\frac{e^2 g^2(r)}{f^2(r)} - \lp 1 + \frac{L}{r^2}\rp g^2(r)}\bdiff r - L\bdiff\varphi\,.
\ee 
The relevance of this to the Lorentz gauge theory in the synchronous phase is that we can recognise the $\bu$ as an exact form, $\bu = -\bdiff T$ for some function $T$. We can easily see that the conjectured function has the form
\be \label{conjectured}
T = e(t-t_0)  \pm \int  \sqrt{\frac{e^2}{f^2(r)} - \lp 1 + \frac{L}{r^2}\rp}g(r)\bdiff r  + L(\varphi-\varphi_0)\,.    
\ee
We have demonstrated that in the case $L=0$ there exists solutions to the Lorentz gauge theory s.t. the khronon $\tau^a = \delta^a_0 T$. Then the 1-form $\bu$ is the 
zeroth component of the Lorentz frame i.e. $\bu=\bbe_0=\bDiff\tau_0 = -\bdiff T$.   

There exist different types of radial time-like geodesics, depending on the value of $e$. These have been called the {\it rain} ($e=1$), {\it hail} ($e>1$) and {\it drip} ($e<1$) geodesics \cite{taylor}. 
\begin{itemize}
\item The case $e>1$ describes the hyperbolic, or  {\it hail} geodesics of massive objects emerging from infinity $r = \infty$ with a finite initial velocity $v_\infty$.  The relation (\ref{energy}) tells us that the specific energy i.e. the energy per the mass of the object is related to the $v_\infty$ by the relativistic $\gamma$-factor s.t. $e=1/\sqrt{1-v_\infty^2}$, since we can assume that $f(\infty)=1$.  
\item According to the above, the case $e=1$ corresponds to particles starting at rest from infinity, $v_\infty=0$. The proper time along these parabolic, or {\it rain} geodesics is the usual Lema{\^i}tre time coordinate.
\item The case $e<1$ describes the elliptic or {\it drip} geodesics of objects which start at rest from a finite radius $r_\text{max}$. The corresponding coordinates become imaginary above the $r>r_\text{max}$, and in this sense cover only a part of the manifold $r>0$. In general,  $r_\text{max}$ is solved from $F( r_\text{max})=0$, which for the case of Schwarzschild black hole gives $r_\text{max}=r_S/(1-e^2)$.  
\item The limit $e\rightarrow \infty$ describes null geodesics, and thus recovers the Eddington-Finkelstein time. This limit is not captured by our scheme (\ref{conjectured}), but becomes accessible via the simple rescaling $T \rightarrow T/e$ (which yields the ``Lake-Martel-Poisson family'' of coordinates versus the ``proper time family'' in Finch's terminology \cite{Finch:2012vli}).       
\item The geodesics with $e \leq 0$ emerge from the white hole region of an analytic extension of the manifold. Trajectories of negative energy particles $e<0$ were called {\it mist} geodesics, and they can be divided into drip mist $-1<e<0$, rain mist $e=-1$ and hail mist $e<-1$ in analogy to the positive energy cases \cite{MacLaurin:2018aze}. 
\end{itemize}
A generalisation of the Painlev{\'e}-Gullstrand coordinates in the case $e>1$ was introduced in Ref.\cite{Martel:2000rn}. A case corresponding to the situation $e>1$ had been discussed earlier in Ref.\cite{Gautreau:1978zz}. The united picture was systematically constructed in Ref.\cite{Finch:2012vli}, and the analytic continuation was considered in Ref.\cite{MacLaurin:2018aze}. The applications are many and diverse, some recent examples including Refs.\cite{Emtsova:2021ehh,Kiefer:2023zxt,Chowdhury:2024rja,Thiemann:2024nmy}. 

\end{document}